%% file: main.tex
\documentclass{article}

\usepackage[final]{aaj2026}
\usepackage[numbers]{natbib}

\usepackage[utf8]{inputenc} 
\usepackage[T1]{fontenc}    
\usepackage{hyperref}       
\usepackage{url}            
\usepackage{booktabs}       
\usepackage{amsfonts}       
\usepackage{nicefrac}       
\usepackage{microtype}      
\usepackage{xcolor}         

\usepackage{graphicx}
\usepackage{natbib}
\usepackage{amsmath}
\usepackage{amssymb}
\usepackage{multirow}
\usepackage{multicol}
\usepackage{subcaption}
\usepackage{gensymb}
\usepackage{float}
\usepackage[shortcuts,acronym]{glossaries}
\usepackage{import}
\usepackage{mathtools}
\usepackage{txfonts}
\usepackage{pxfonts}
\usepackage{cancel}
\usepackage[utf8]{inputenc}
\usepackage{placeins}
\usepackage{tikz}
\usetikzlibrary{tikzmark}

\title{Co-simulation-Based Safety-Assessment of Black-Box Maritime Cyber-Physical Systems}

\author{%
  Andreas R. G.~Sitorus\thanks{\includegraphics[height=1em]{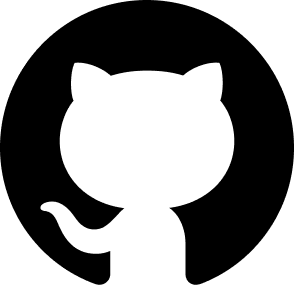}\url : {https://github.com/AndreasKing-Goks/ship-in-transit-FMUs}.}\\
  Department of Engineering Cybernetics\\
  Norwegian Institute of Science and Technology\\
  Trondheim, Norway \\
  \texttt{andreas.r.g.sitorus@ntnu.no} \\
\And
 Melih~Akdağ\\
  Det Norske Veritas\\
  Trondheim, Norway \\
  \texttt{melih.akdag@dnv.com} \\
\And
 Hoàng Anh~Tràn\\
  Department of Engineering Cybernetics\\
  Norwegian Institute of Science and Technology\\
  Trondheim, Norway \\
  \texttt{hoang.a.tran@ntnu.no} \\
\And
 Etienne~Purcell\\
  SINTEF Nordvest\\
  Ålesund, Norway \\
  \texttt{etienne.purcell@sintef.no} \\
}

\begin{document}

\newacronym{lhs}{LHS}{\textit{Latin Hypercube Sampling}}
\newacronym{ast}{AST}{\textit{Adaptive Stress-Testing}}
\newacronym{ebast}{EB-AST}{\textit{Event-Based Adaptive Stress-Testing}}
\newacronym{ebast2}{EB-ASTv2}{\textit{Event-Based Adaptive Stress-Testing v2}}
\newacronym{rl}{RL}{\textit{Reinforcement Learning}}
\newacronym{drl}{DRL}{\textit{Deep Reinforcement Learning}}
\newacronym{mlp}{MLP}{\textit{Multi-Layer Perceptron}}
\newacronym{rnn}{RNN}{\textit{Recurrent Neural Network}}
\newacronym{lstm}{LSTM}{\textit{Long Short-Term Memory}}
\newacronym{mdp}{MDP}{\textit{Markov Decision Process}}
\newacronym{pomdp}{POMDP}{\textit{Partially Observable Markov Decision Process}}
\newacronym{sbt}{SBT}{\textit{Simulation-Based Testing}}
\newacronym{sac}{SAC}{\textit{Soft Actor-Critic}}
\newacronym{kl}{KL}{\textit{Kullback-Leibler}}
\newacronym{cpi}{CPI}{\textit{Conservative Policy Iteration}}
\newacronym{trpo}{TRPO}{\textit{Trust Region Policy Optimization}}
\newacronym{gae}{GAE}{\textit{General Advantage Estimation}}
\newacronym{ppo}{PPO}{\textit{Proximal Policy Optimization}}
\newacronym{rppo}{R-PPO}{\textit{Recurrent Proximal Policy Optimization}}
\newacronym{mcts}{MCTS}{\textit{Monte Carlo Tree Search}}
\newacronym{mcts-sa}{MCTS-SA}{\textit{Monte Carlo Tree Search-Seed Action}}
\newacronym{bo}{BO}{\textit{Bayesian Optimization}}
\newacronym{dp}{DP}{\textit{Dynamic Programming}}
\newacronym{cps}{CPS}{\textit{Cyber-Physical System}}
\newacronym{ipr}{IPR}{\textit{Intellectual Property Right}}
\newacronym{hla}{HLA}{\textit{High-Level Architecture}}
\newacronym{fmi}{FMI}{\textit{Functional Mock-up Interface}}
\newacronym{fmu}{FMU}{\textit{Functional Mock-up Unit}}
\newacronym{3dof}{3-DoF}{\textit{3-Degree of Freedom}}
\newacronym{los}{LOS}{\textit{Line-of-Sight}}
\newacronym{mso}{MSO}{\textit{Machinery System Operational}}
\newacronym{pto}{PTO}{\textit{Power Take Out}}
\newacronym{mec}{MEC}{\textit{Mechanical}}
\newacronym{pti}{PTI}{\textit{Power Take In}}
\newacronym{hsg}{HSG}{\textit{Hybrid Shaft Generator}}
\newacronym{colav}{COLAV}{\textit{Collision-Avoidance}}
\newacronym{cpa}{CPA}{\textit{Closest Point of Approach}}
\newacronym{tcpa}{$t_\text{CPA}$}{\textit{Time to Closest Point of Approach}}
\newacronym{dcpa}{$D_\text{CPA}$}{\textit{Distance to Closest Point of Approach}}
\newacronym{osp}{OSP}{\textit{Open Simulation Platform}}
\newacronym{roa}{RoA}{\textit{Radius of Acceptance}}
\newacronym{iws}{IWS}{\textit{Intermediate Waypoint Sampling}}
\newacronym{iw}{IW}{\textit{Intermediate Waypoint}}
\newacronym{stg}{STG}{\textit{Ship Traffic Generator}}
\newacronym{ntnu}{NTNU}{\textit{Norwegian University of Science and Technology}}
\newacronym{dnv}{DNV}{\textit{Det Norske Veritas}}
\newacronym{sog}{SoG}{\textit{Speed over Ground}}
\newacronym{cog}{CoG}{\textit{Course over Ground}}

\maketitle

\begin{abstract}
Modern maritime cyber-physical systems consist of tightly interconnected subsystems ranging from multiple engineering domains, making exhaustive safety verification impractical. Simulation-based testing offers a scalable alternative, while co-simulation enables heterogeneous subsystem models, potentially developed by different stakeholders, to be integrated without requiring access to their internal implementations. Adaptive Stress-Testing further improves testing efficiency by formulating failure discovery as a sequential decision-making problem, guiding the search toward the most likely failure scenarios, and giving the counter safety examples using reinforcement learning. 

However, black-box co-simulation may provide only partial access to the system state required by conventional Markov decision process formulations. To address this limitation, we proposed Event-Based Adaptive Stress-Testing version 2, a co-simulation-based safety-assessment method formulated as a partially observable Markov decision process. This method employs Recurrent Proximal Policy Optimization to exploit observation histories when selecting adversarial actions. The method is evaluated on a ship-in-transit operation in which a collision-avoidance system is stress-tested against multi-adversarial target ships. An extended Intermediate Waypoint Sampling mechanism enables independent multi-adversary trajectory control and supports the generation of realistic collision-inducing scenarios. The experiment demonstrate that the learned stress-testing policy generalizes across different encounter cases, can be further improved through continued training, and can be made effective when the collision-avoidance system under test is replaced by another version. These results demonstrate the method's potential for scalable black-box safety assessment of complex maritime cyber-physical systems.
\end{abstract}


\import{./Sections}{Section_1.tex}
\import{./Sections}{Section_2.tex}
\import{./Sections}{Section_3.tex}
\import{./Sections}{Section_4.tex}

\import{./Sections}{Section_5.tex}

\import{./Sections}{Section_6.tex}
\import{./Sections}{Section_7.tex}


\bibliography{EB-ASTv2_Journal_Paper.bib}


\appendix
\vspace{-1em}
\section{Appendix / supplemental material}

\subsection{Interception Heading Prediction}\label{sec:interception_heading_prediction}
In order to compute the interception-guidance reward term, a procedure has been developed to determine whether a target ship may intercept the own ship  at every decision event. If interception is deemed feasible, it then predicts the interception heading $\psi^\mathrm{ic}_t$ accordingly.

Let us define own ship positions ($\boldsymbol{p}_\mathrm{OS}\in\mathbb{R}^2$), target ship positions ($\boldsymbol{p}_\mathrm{TS}\in\mathbb{R}^2$), own ship velocities ($\boldsymbol{v}_\mathrm{OS}\in\mathbb{R}^2$), target ship forward speed ($u_\mathrm{TS}\in\mathbb{R}^1$), and target ship directional vector unit ($\boldsymbol{\mu}_\mathrm{TS}\in\mathbb{R}^2$). An \textit{interception} can be framed from the kinematic perspective as
\begin{equation}
    \boldsymbol{p}_\mathrm{OS} + \boldsymbol{v}_\mathrm{OS}t = \boldsymbol{p}_\mathrm{TS} + u_\mathrm{TS}\boldsymbol{\mu}_\mathrm{TS}t.
\end{equation}
Refactor this by moving $\boldsymbol{\mu_\mathrm{TS}}$ to left side, and we get
\begin{equation}
    \boldsymbol{\mu}_\mathrm{TS} = \frac{\boldsymbol{r}+\boldsymbol{v}_\mathrm{OS}}{u_\mathrm{TS}t},
\end{equation}
where $\boldsymbol{r}=\boldsymbol{p}_\mathrm{OS}-\boldsymbol{p}_\mathrm{TS}$ following Equation \ref{eq:rel_pos}. By applying $\|\boldsymbol{\mu}_\mathrm{TS}|=1$, we ended up with
\begin{equation}
    \|u_\mathrm{TS}t\| = \|\boldsymbol{r}+\boldsymbol{v}_\mathrm{OS}\|.
\end{equation}
By squaring both sides, and arranging the equation by moving all terms to the left side, we get
\begin{equation}\label{eq:intercept_qudratic}
     \left(\boldsymbol{v}_\mathrm{OS}\cdot\boldsymbol{v}_\mathrm{OS}-u_\mathrm{TS}^2 \right)t^2+
     2\left(\boldsymbol{r}\cdot\boldsymbol{v}_\mathrm{OS}\right)t + 
     \left(\boldsymbol{r}\cdot\boldsymbol{r}\right) =0.
\end{equation}
Above is the \textit{quadratic formula} that follows $\mathbf{A}t^2+\mathbf{B}t+\mathbf{C}=0$. Observe now that we have
\begin{align}
    \mathbf{A} &= \boldsymbol{v}_\mathrm{OS}\cdot\boldsymbol{v}_\mathrm{OS}-u_\mathrm{TS}^2\\
    \mathbf{B} &= 2\left(\boldsymbol{r}\cdot\boldsymbol{v}_\mathrm{OS}\right)\\
    \mathbf{C} &= \boldsymbol{r}\cdot\boldsymbol{r}.
\end{align}
From these, we can learn several things about the own-ship--target-ship kinematics. When $\mathbf{A}\approx0$ means that both own ship and target ship travels at a similar speed, hence the only enabling factor for feasible interception are their geometrical positions. We can also compute the \textit{discriminant} as
\begin{equation}
    \mathbf{disc} = \mathbf{B}^2 - 4\mathbf{A}\mathbf{C},
\end{equation}
where it determines whether interception is \textit{infeasible} when $\mathbf{disc}<0$. If the discriminant is positive, solution exists and the \textit{roots} of Equation \ref{eq:intercept_qudratic} can solved by computing
\begin{align}
    t_1 &=\frac{-\mathbf{B}+\sqrt{\mathbf{disc}}}{2\mathbf{A}} \\
    t_2 &=\frac{-\mathbf{B}-\sqrt{\mathbf{disc}}}{2\mathbf{A}}.
\end{align}
These terms are all of the possible times-to-intercept. Given the threshold $t_\text{predict limit}$, interception is only feasible when at least one of the root fulfill the condition $0\leq t_\mathrm{intercept} \leq t_\text{predict limit}$. Beyond that, the interception either occurred in the past or is too far in the future.

Using the solution $t_\mathrm{intercept}$, target ship velocity can now be computed by
\begin{equation}
    \boldsymbol{\nu}_\mathrm{TS} = 
    \frac{\boldsymbol{r}+\boldsymbol{\nu}_\mathrm{OS}t_\mathrm{intercept}}{t_\mathrm{intercept}},
\end{equation}
where the interception heading $\psi^\mathrm{ic}_t$ can be obtained by solving
\begin{equation}
    \psi^\mathrm{ic}_t = \frac{\boldsymbol{\nu}_\mathrm{TS}}{\|\boldsymbol{\nu}_\mathrm{TS}\|}.
\end{equation}

\subsection{\ac{stg} parameters}\label{sec:stg_parameters}
\begin{table}[ht] 
\centering 
\caption{\ac{stg} parameters} 
\label{tab:stg_parameters} 
    \begin{tabular}{lc} 
        \toprule 
        \textbf{Parameter} & \textbf{Value} \\
        \midrule 
        $\theta_{13}$ criteria                  & $67.5^\circ$ \\
        $\theta_{14}$ criteria                  & $5.0^\circ$ \\
        $\theta_{15}$ criteria                  & $5.0^\circ$ \\
        $\theta_{15}$                           & $\theta_{15}\in\{112.5^\circ, 247.5^\circ\}$ \\
        $v_{\text{rel}}$ head on                & $v_\text{rel}\in[0.5, 1.5]$ \\
        $v_{\text{rel}}$ overtaking stand-on    & $v_\text{rel}\in[1.5, 2.0]$ \\
        $v_{\text{rel}}$ overtaking give-way    & $v_\text{rel}\in[0.25, 0.75]$ \\
        $v_{\text{rel}}$ crossing stand-on      & $v_\text{rel}\in[0.5, 1.5]$ \\
        $v_{\text{rel}}$ crossing give-way      & $v_\text{rel}\in[0.5, 1.5]$ \\
        Situation length                        & $90.0$ minutes \\
        Maximum meeting distance                & $5.0$ nautical miles \\
        Situation develop time                  & $60.0$ minutes \\
        Vector time                             & $t_\text{vector}\in[40.0, 60.0]$ minutes \\
        Own ship \ac{sog}                       & $6.0$ knots\\
        Own ship \ac{cog}                       & $\psi_\text{sog}\in\{0.0^\circ, 15.0^\circ, \cdots,  345.0^\circ, 360^\circ\}$ \\
        Margin $\delta$                         & $2.0^\circ$\\
        $\beta$ head on                         & $\beta \in [(-\theta_{14}+\delta), (\theta_{14}-\delta) ]$\\
        $\beta$ overtaking give-way             & $\beta \in [(-\theta_{13}+\delta), (\theta_{13}-\delta) ]$\\
        $\beta$ overtaking stand-on starboard   & $\beta \in [(\theta_{15, \mathrm{min}}+\delta), (180.0^\circ-\delta )]$\\
        $\beta$ overtaking stand-on port        & $\beta \in [(-180.0^\circ+\delta), (-\theta_{15, \mathrm{min}}-\delta) ]$\\
        $\beta$ crossing give-way               & $\beta \in [(\theta_{15, \mathrm{crit}}+\delta), (\theta_{15, \mathrm{min}}-\delta) ]$\\
        $\beta$ crossing stand-on               & $\beta \in [(-\theta_{15, \mathrm{min}}+\delta), (\theta_{15, \mathrm{crit}}-\delta) ]$\\
        Navigation status                       & Under way using engine \\
        \bottomrule 
    \end{tabular} 
\end{table}


\end{document}

%% file: Sections/Section_1.tex
\section{Introduction}

Advances in maritime automation are allowing increasingly autonomous and functionally integrated vessel operations. Modern maritime systems may consist of tightly coupled propulsion, navigation, machinery, energy-management, sensing, and control subsystems, together with intelligent modules such as collision avoidance. Their interaction increasingly gives rise to complex maritime \acp{cps}, whose underlying subsystems may also be developed by different stakeholders. Co-simulation provides an enabling technology for integrating reusable and interchangeable subsystem models into a common virtual prototype, supporting rapid system development and system-level simulation \cite{sadjina_distributed_2019}.

The increasing integration of these subsystems also makes safety assessment more challenging. Component-level verification alone provides limited evidence of system-level behavior because failures may emerge from interactions between each correctly functioning subsystems. Consequently, complex maritime \acp{cps} require testing approaches that considers all subsystem interactions under given operational conditions.

Simulation-based testing provides a scalable ways for evaluating such systems without exposing physical assets to potentially hazardous scenarios. However, its effectiveness strongly depends on the selection of the \textit{test cases}. Test scenarios are traditionally selected manually using expert knowledge, operational experience, and risk analysis. On the other hand, systematic experimental-design approaches such as factorial designs, fractional factorial designs, \ac{lhs}, and Sobol sequences can be used to sample the input space and characterize system variability \cite{montgomery_design_2017,saltelli_global_2008}. Nevertheless, with  a large number of interacting variables and possible environmental conditions exhaustive explorations become computationally infeasible. Automatic search mechanisms that direct simulations toward safety-critical regions are therefore desirable.

\ac{ast} addresses this problem by formulating failure-scenario discovery as a sequential decision-making problem and using an optimization or \ac{rl} agent to search for likely failure trajectories. Different search strategies have been investigated, including \ac{mcts} \cite{lee_adaptive_2015,laursen_evolutionary_2025}, the Backward Algorithm \cite{koren_finding_2021}, Go-Explore \cite{koren_adaptive_2020}, and deep \ac{rl}-based approaches \cite{hjelmeland_identification_2022,sitorus_identifying_2025}. These methods have demonstrated the capability of \ac{ast} to stress-test safety-critical systems in domains including automotive, aerospace, and, more recently, maritime applications.

However, applying \ac{ast} directly to distributed co-simulation introduces an additional challenge. Individual subsystem models may be accessible only through their exposed interfaces, while their complete internal states remain unavailable to the stress-testing agent. Therefore, the agent must make decisions from incomplete observations and infer information relevant to the underlying system state from its interaction history. Under these conditions, the fully observable \ac{mdp} formulation is insufficient, and the stress-testing problem should be represented as a \ac{pomdp}. This motivates an \ac{ast} formulation that is compatible with black-box co-simulation interfaces by having it capable of retaining temporal information.

\subsection{Our Approach and Contributions}

This article presents \ac{ebast2}, an extension of \ac{ast} for simulation-based failure discovery in partially observable maritime co-simulation. The method is evaluated using a ship-in-transit scenario in which an \ac{rl} agent controls multiple adversarial target ships to search for collision-inducing scenarios against an own ship equipped with a \ac{colav} system.

The principal contributions of this work are:
\begin{enumerate}
    \item \textbf{Partial-observability-aware stress testing.}
    \ac{ebast2} employs \ac{rppo} as its primary policy-optimization method. The recurrent policy retains information from previous observations and actions, enabling the agent to account for temporal information that is not available from the current co-simulation observation alone. The trained recurrent policy's performance will also be compared with its non-recurrent policy alternatives.

    \item \textbf{Multi-adversary scenario generation.}
    The framework introduces multiple adversarial control feature to enhance its failure discovery potential. An event-activity masking mechanism allows individual target ships to trigger waypoint-sampling events independently while remaining compatible with a joint policy action space.

    \item \textbf{Reward-guided generation of plausible adversarial trajectories.}
    An improved reward formulation is introduced to guide the target ships toward interception opportunities while discouraging abrupt trajectory changes. This formulation balances collision-chasing tendencies with realistic target ship behavior.

    \item \textbf{Generalization across traffic configurations and systems under test.}
    The policy is trained across a diverse collection of traffic configurations to improve its generalization capability. The modular co-simulation architecture further allows the trained stress-testing policy to be applied to different \ac{colav} implementations by replacing the corresponding \ac{fmu}, enabling comparative safety assessment without redesigning the complete testing framework.
\end{enumerate}

Through these contributions, \ac{ebast2} aims to provide a reusable and scalable safety-assessment method for discovering safety-critical scenarios in partially observable maritime \acp{cps}, while exploiting the modularity offered by co-simulation.

%% file: Sections/Section_2.tex
\section{Preliminaries and Backgrounds}


    




\subsection{Co-simulation}

Co-simulation technology is a simulation technique that is based on the construction of systems from loosely-coupled stand-alone models and the simulation across different subsystems. Co-simulation offers the ability for separate subsystems initialization, pre-processing, time-integration, and post-processing with different specialized tools \cite{sadjina_distributed_2019}. Co-simulation also enables the use of \textit{black-box} models \cite{skjong_virtual_2018}, where the models is compiled into a low-level machine code that hides the internal algorithm from the users. These enables rapid and accurate model development that make use of already available models/software while at the same time protecting the \ac{ipr} associated to the subsystems.
\begin{figure}[h!]
\begin{center}
    \includegraphics[width=.5\textwidth]{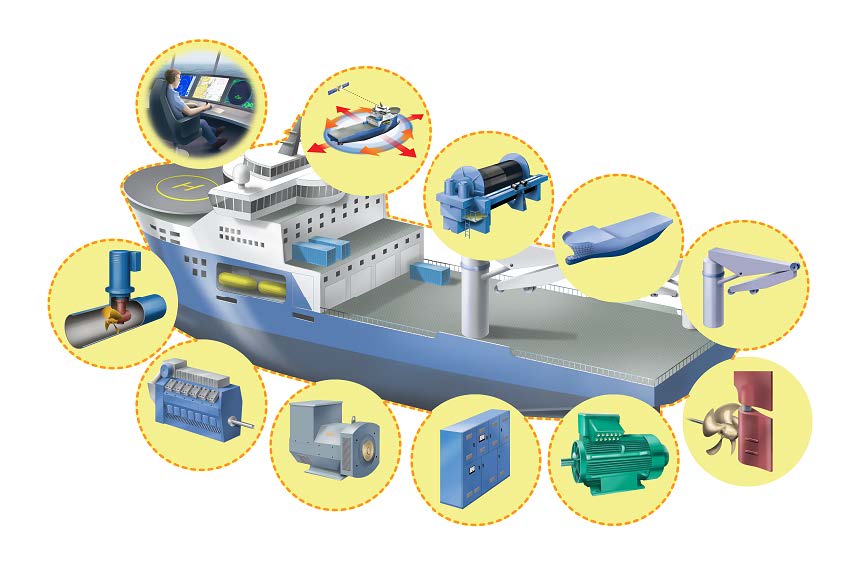}
    \caption{Illustration of real maritime system modeled as co-simulation. It contains different tools and models, but they are interconnected with each other, run independently, and used in parallel to form a full-system simulation \cite{sadjina_distributed_2019}.}
    \label{fig:co-simulation_settings}
\end{center}
\end{figure}

The subsystems in a co-simulation exchange information through their inputs and outputs. This exchange is based on the time-ordered values computed by the individual simulators during the simulation  at discrete \textit{communication points} $t_i \in \{t_0, t_1, \ldots, t_N\}$ \cite{schweiger_empirical_2019}. Information exchange between subsystems is performed only at these communication points and can be expressed through interface constraint equations \cite{sadjina_distributed_2019}
\begin{equation}
    \mathbf{u}(t_i) = \mathbf{L}\mathbf{y}(t_i),
\end{equation}
where $\mathbf{u}$ and $\mathbf{y}$ denote the subsystem input and output vectors, respectively, and $\mathbf{L}$ is a connection graph matrix that maps subsystem outputs to subsystem inputs. The communication points are separated by a \textit{macro step size} $\Delta t_i$, such that $t_{i+1} = t_i + \Delta t_i$. In practice, the connection structure represented by $\mathbf{L}$ is specified through an \textit{I/O configuration}, while an \textit{orchestrator} algorithm is responsible for executing the co-simulation. 

To enable interoperability among different subsystems and simulation tools, co-simulation relies on standardized interfaces or communication protocols. In this work we extensively used the \ac{fmi} standards, which has gained increasing attention in the engineering community in recent years \cite{blockwitz_functional_2012}. \ac{fmi} specifies that models should be packaged as \acp{fmu}, which are archive files that contain the model code along with its metadata and documentation. This standard is maintained and developed by the Modelica Association. Co-simulation has reached a decent maturity level and has been used across many industries, such as automotive industries \cite{krammer_system_2014, winter_simulation_2015, nalic_development_2019}, aviation industries \cite{sogandares_stone_2002, zhu_real-time_2024, zhao_Co-simulation_2023}, and recently maritime industries \cite{perabo_multiphysics_2024, skjong_virtual_2018, nagel_holistic_2016, hassani_virtual_2016, chu_virtual_2018, paquin_modern_2009}.

\subsection{Failures and Failure Analysis}\label{sec:Failure_analysis}
Failure and failure mode are the two fundamental concepts in system reliability analysis. This section is based on Rausand \cite{rausand_reliability_2014} and summarizes the concepts relevant to failure analysis.

 A \textit{failure} occurs when an \textit{item} is no longer able to perform one or more of its required functions. An item may be an individual component, a channel, a subsystem, or a safety loop. Hence, failure is the \textit{event} that takes place when a required function is terminated. A failure may originate from an \textit{error}, which defined as a ``discrepancy between a computed, observed or measured value or condition and the true, specified or theoretically correct value or condition''.
 
 Failure commonly followed by a \textit{failed state}. The condition responsible for, or associated with, this state is referred to as a \textit{fault}, where it is characterized by the inability of an item to perform a required function. A fault therefore is a \textit{state} that may have its origin in a random failure event or in a \textit{deficiency} related to the item, its location, or its application.

Finally, a \textit{failure mode} is a description of a fault. It tells \textit{how} a fault manifests itself. A failure mode lead to one or more consequences, referred to as \textit{failure effects}. However, these effects can vary considerably in severity. Thus, the main motivation for identifying failure effects is to distinguish the more severe failure modes from less severe ones.

\begin{figure}[h!]
\begin{center}
    \includegraphics[width=.7\textwidth]{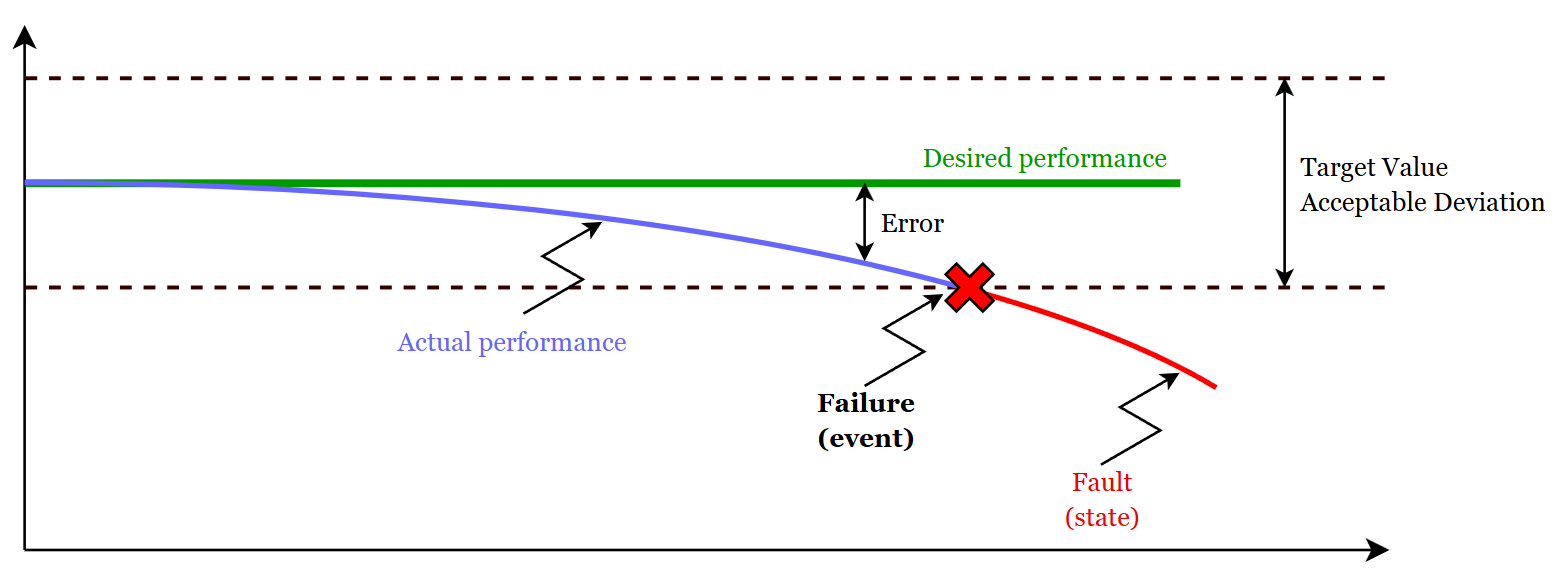}
    \caption{Error, failure, and fault \cite{rausand_reliability_2014}.}
    \label{fig:error_failure_fault}
\end{center}
\end{figure}

\subsection{Simulation-Based Safety-Assessment of Black-Box Cyber-Physical Systems} \label{sec:CPS_assessment} 

We first define a system model, denoted by $\mathcal{M}$, together with a parameter space $\mathcal{P}$, an input space $\mathcal{I}$, and a property $\Psi$ that the system is required to satisfy. The behavior of the model for a parameter value $p\in\mathcal{P}$ and an input $i\in\mathcal{I}$ is denoted by $\Phi(\mathcal{M},p,i)$. In general, the input $i$ may be time-dependent. The set of all possible behaviors of $\mathcal{M}$ over the parameter space $\mathcal{P}$ and input space $\mathcal{I}$ is denoted by $\Phi(\mathcal{M},\mathcal{P},\mathcal{I})$ \cite{kapinski_simulation-based_2016}.

\textit{\textbf{Testing}} is the task of evaluating whether the behaviors of the model $\mathcal{M}$, for selected subsets $\hat{\mathcal{P}}\subseteq\mathcal{P}$ and $\hat{\mathcal{I}}\subseteq\mathcal{I}$, satisfy the property $\Psi$ \cite{kapinski_simulation-based_2016}. Assuming that $\hat{\mathcal{P}}$ and $\hat{\mathcal{I}}$ are finite, testing can be expressed as:
\begin{equation}
\Phi(\mathcal{M},\hat{\mathcal{P}},\hat{\mathcal{I}}) \vDash \Psi.
\end{equation}

\textit{\textbf{Verification}} is the task of proving that all possible behaviors of the model $\mathcal{M}$ satisfy the property $\Psi$ \cite{kapinski_simulation-based_2016}. It therefore provides a formal guarantee of rightness or correctness over the specified parameter and input spaces \cite{sargent_verification_2010, balci_verification_2003}. Verification can be expressed as:
\begin{equation}
\Phi(\mathcal{M},\mathcal{P},\mathcal{I}) \vDash \Psi.
\end{equation}

In this research, we focus on \acp{cps}, which are common in maritime systems because they integrate physical, digital, and networked components \cite{torben_automatic_2023}. \acp{cps} are difficult to verify exhaustively because they are complex, hybrid, and often \textit{black-box}. A system is considered a \textit{black box} when its internal model $\mathcal{M}$ is unknown or too complex to support explicit reasoning \cite{corso_survey_2021}, or when it is deliberately concealed from external parties, for example to protect proprietary information.
\begin{figure}[h!]
\begin{center}
    \includegraphics[width=.8\linewidth]{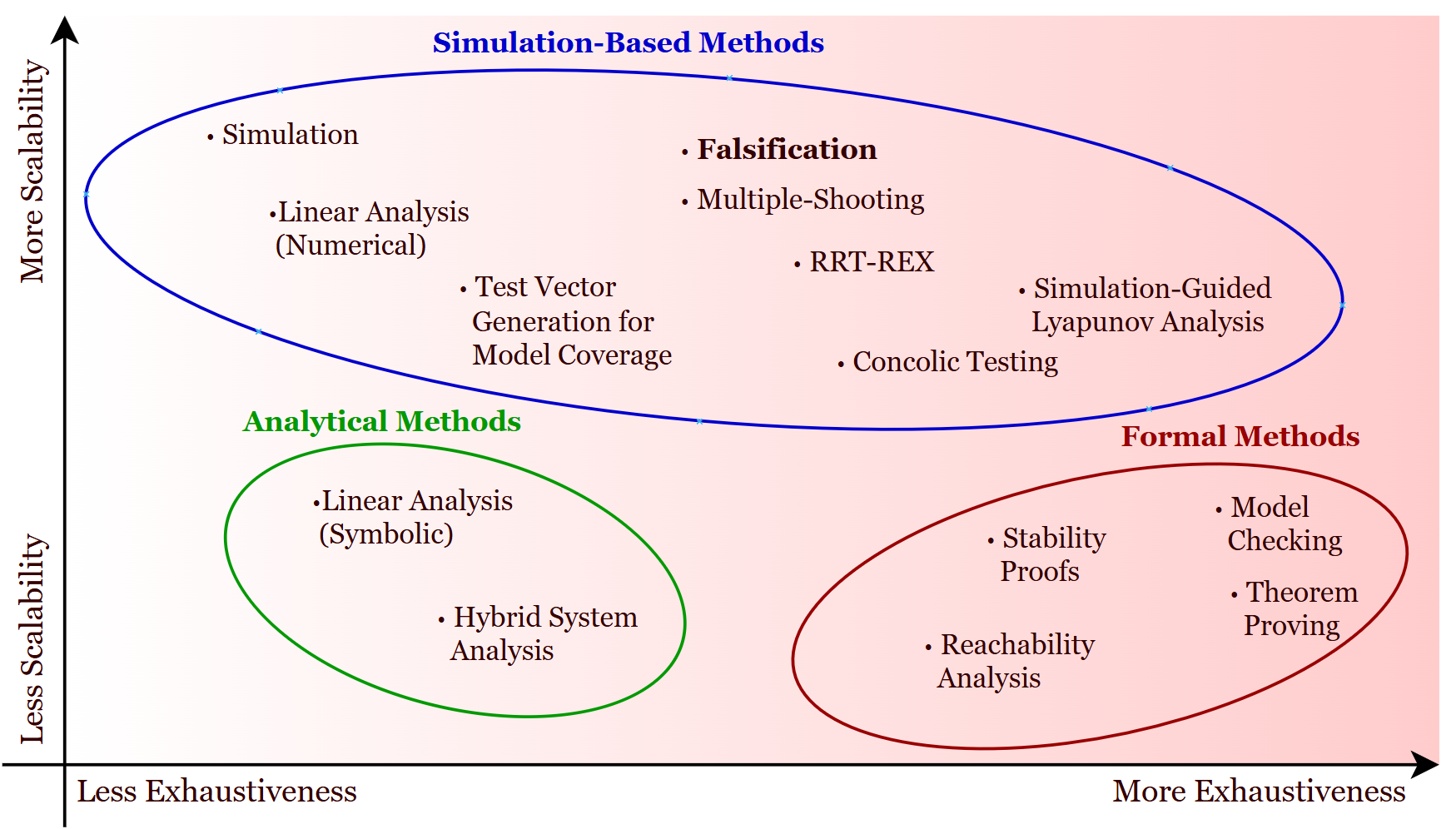}
    \caption{Spectrum of verification methods for \acp{cps} \cite{kapinski_simulation-based_2016}.}
    \label{fig:Spectrum_VM}
\end{center}
\end{figure}

Kapinski \textit{et al.} \cite{kapinski_simulation-based_2016} provide an overview on existing system verification method for \acp{cps}, illustrated in Figure \ref{fig:Spectrum_VM}. The spectrum shows the trade-off between \textit{scalability} and \textit{exhaustiveness} of each method. \textit{Simulation-based testing} is scalable but non-exhaustive. \textit{Formal verification} is exhaustive but less scalable. In maritime autonomy, Torben \textit{et al.} \cite{torben_automatic_2023} show that simulation-based testing is attractive but requires automated test-case selection due to the large scenario space. Thus, \textit{Falsification} is useful because it searches for counterexamples rather than proving universal correctness.

\begin{figure}[h!]
\begin{center}
    \includegraphics[width=\textwidth]{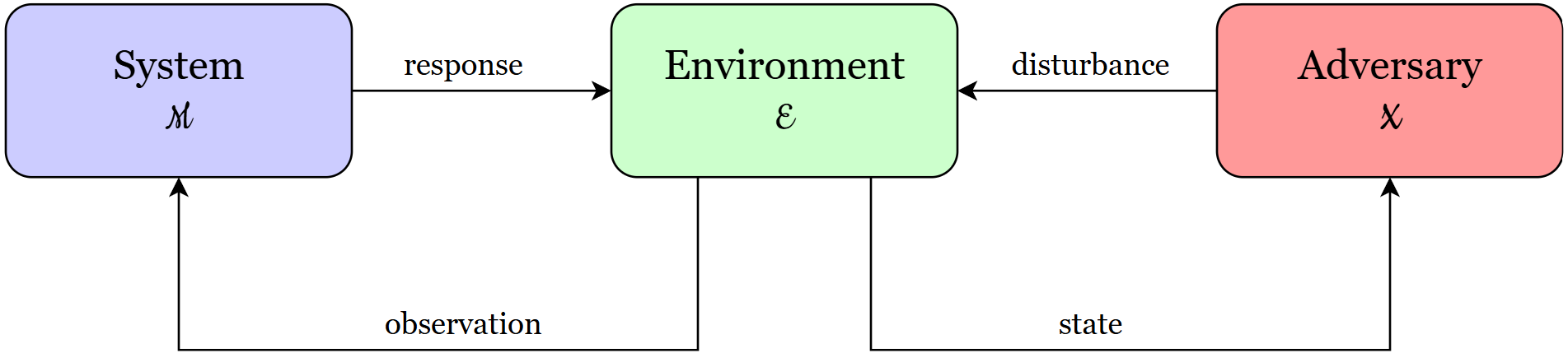}
    \caption{The safety validation problem \cite{corso_survey_2021}.}
    \label{fig:safety_validation}
\end{center}
\end{figure}

After introducing the relevant definitions, we describe a method known as \textit{safety validation}, introduced by Corso \textit{et al.} \cite{corso_survey_2021}. Safety validation, as illustrated in Figure \ref{fig:safety_validation}, is the process of ensuring the correct and safe operation of a system within its environment $\mathcal{E}$. The desired safety property are expressed in a specification language, and simulation is typically used to identify violations of these properties, which are also referred to as \textit{failures}. System failures may be caused by disturbances from the adversary $\mathcal{X}$, and a model of these disturbances can be used to estimate the probability of failure. A system is deemed safe if no failure is found during a finite exploration or if the estimated failure probability remains below an acceptable threshold. Accordingly, safety validation encompasses three main objectives:

\begin{enumerate}
    \item \textit{\textbf{Falsification}}, which is the task of finding a parameter value and an input for which the resulting system behavior violates the property $\Psi$ \cite{kapinski_simulation-based_2016}. In other words, the objective is to find a counterexample $(p,i)$ such that:
    \begin{equation}
        p\in\mathcal{P}, \quad i\in\mathcal{I}, \quad
        \Phi(\mathcal{M},p,i) \nvDash \Psi.
    \end{equation}
    
    Equivalently, the set of all counterexamples may be written as:
    \begin{equation}
        \{(p,i)\in\mathcal{P}\times\mathcal{I}
        \mid \Phi(\mathcal{M},p,i)\nvDash\Psi\}.
    \end{equation}
    
    \item \textit{\textbf{Most-Likely Failure Analysis}}, which is the task to find the counterexample with maximum likelihood:
    \begin{equation}
        \arg\max_{p,i}\Pr(p,i) \quad\text{such that}\quad \Phi(\mathcal{M},p,i)\nvDash\Psi.
    \end{equation}
    
    \item \textit{\textbf{Failure Probability Estimation}}, which is the task to compute the probability that a safety property will be violated. Failure probability is given by the expectation of observing failure under the disturbance model
    \begin{equation}
        {\Pr}_{failure} = \mathbb{E}\left[\mathbf{1} \{\Phi(\mathcal{M},p,i)\nvDash\Psi\} \right].
    \end{equation}
    Where,
    \begin{equation}
        \mathbf{1} \{\Phi(\mathcal{M},p,i)\nvDash\Psi\} =
        \begin{cases}
            &1,\quad\text{if }\Phi(\mathcal{M},p,i)\nvDash\Psi \\
            &0,\quad\text{if }\Phi(\mathcal{M},p,i)\vDash\Psi.
        \end{cases}
    \end{equation}
\end{enumerate}

In the verification literature, falsification is often discussed as part of simulation-based verification because it searches for counterexamples to formalized requirements. In the black-box safety-validation literature, the same activity is interpreted as one goal of safety validation, because the discovered counterexamples reveal unsafe behavior in an operational environment. In this work, falsification is therefore treated as a simulation-based safety-assessment activity: it does not provide a formal proof of correctness, but it supports a scalable verification via counterexamples and validation by identifying requirement-violating trajectories.

\subsection{Reinforcement Learning Preliminaries}\label{sec:RL_preliminaries}

Section \ref{sec:RL_preliminaries} provides a brief overview of Markov decision processes, largely following the presentation in Sutton and Barto \cite{sutton_reinforcement_2018}.

\subsubsection{Markov Decision Process}\label{sec:mdp}

\ac{mdp} is the classical of formalization of sequential decision making and serves as the foundation of many modern \ac{rl} algorithms.  \ac{mdp} is intended as an intuitive framework of the problem of learning through interaction to achieve goal. The framework defines two entities: the ``learner'' which will be often refer to as the \textit{agent}, and the \textit{environment} which consists of everything outside the agent that it interacts with. Together, they create the \textit{agent-environment interface} where both entities interact continually: The agent selects actions and the environment responds to it, transitioning it to a new situations or \textit{states} to the agent. The environments then assigns a specific reward signals attribute to this transition.

 \ac{mdp} can be formally defined as a tuple $\langle S, A, T, R \rangle$. The agent and environment interact for every discrete time steps, $t=[0,1,2,3,\ldots]$. At each time step $t$, the agent receives some representation of the environment state, $S_t \in \mathcal{S}$. Then based on the states, the agent choose an action , $A_t \in \mathcal{A}(S_t)$. On the next timestep, in part as a consequence of its action, the agent will find itself in the new state $S_{t+1}$ and receive the numerical reward, $R_{t+1} \in \mathcal{R}(S_t, A_t) \subset \mathbb{R}$. The continuous process of this thereby give rise to a sequence of
\begin{equation*}
    S_0, A_0, R_1, S_1, A_1, R_2, S_2, A_2, R_3, S_3, \ldots.
\end{equation*}
In a \textit{finite} \ac{mdp}, the sets of states, actions, and rewards $\mathcal{S,A,R}$ all have finite number elements. In this case, the next obtained state $s'\in\mathcal{S}$ and reward $r'\in\mathcal{R}$ when starting from a known current state $s\in\mathcal{S}$ while taking action $a\in\mathcal{A}$ is random and based on a probability distribution dependent only on the preceding state and action.
\begin{equation}\label{eq:mdp_transition}
    T(s',r\;\mid\;s,a)\coloneqq \Pr\{S_t=s', R_t=r\;\mid\;S_{t-1}=s, A_{t-1}=a\},
\end{equation}
for all $s',s\in\mathcal{S}$; $r\in\mathcal{R}$; $a\in\mathcal{A}(s)$ and the \textit{transition function} $T$ defines the dynamics of the \ac{mdp}. Due to this property, a state-action transition not just affect the immediate rewards of one transition but also the future rewards through the subsequent agent-environment interactions. This transition model also satisfies the \textit{Markov Property}, meaning that, given the current state, future state transitions are independent of past agent-environment interactions. Thus, the state must contain all information from the interaction history that is relevant for predicting future dynamics.

\subsubsection{Partially Observable Markov Decision Process}\label{sec:pomdp}

Many real-world problems do not satisfy the assumptions of the \ac{mdp} framework because the learning agent can only partially observe the state of the environment at each time step \cite{shani_survey_2013}. In other words, some components of the state may be inherently hidden from the agent, while others may be observed inaccurately due to environmental noise. Such problem can be modeled as a \ac{pomdp} \cite{astrom_optimal_1965}, which is formally defined as a tuple 
\begin{equation}
    \langle S, A, T, R, \Omega , \mathcal{O}, b_o \rangle.
\end{equation}
The $\langle S, A, T, R \rangle$ follows the \ac{mdp} in Section \ref{sec:mdp}. $\Omega$ is the observation space that describe the possible observations, $\mathcal{O}$ is an observation model where 
\begin{equation}
    \mathcal{O}(a, s', o) = \Pr(o_{t+1}\mid a_t, s_{t+1})
\end{equation} 
is the probability of observing $o$ given that the agent has reached state $s'$ after executing action $a$. 

One option to address \ac{pomdp} challenge is to use the agent \textit{history} as its internal state. This history is composed of previous agent-environment interactions, denoted as
\begin{equation}\label{eq:history_eq}
    h_t = \langle a_0, o_1, a_1 , o_2, \ldots, a_{t-1}, o_t \rangle.
\end{equation}
The complete history $h_t$ can be used as an internal Markov state, but it is infeasible for the reason $h_t$ will continually grows over time \cite{shani_survey_2013}. To tackle this problem, the concept of a \textit{belief state} is introduced  in the context of nondeterministic and partially observable planning problems \cite{russell_artificial_2021}. Belief state is the set of actual states in which the agent \textit{may} be in. In \ac{pomdp}, the belief state $b$ is generalized into a \textit{probability distribution} over the possible states. Following Shani \textit{et al.} \cite{shani_survey_2013}, the belief state is conditioned on the agent's history, and can be expressed as
\begin{equation}\label{eq:belief_state}
    b_o(s)=\Pr(s\mid h),
\end{equation}
where $b(s)$ represents the probability that the agent is in state $s$ after observing history $h$.

\subsubsection{Rewards, Returns and Episodes}

In \ac{rl}, the reward function is formulated to capture the desired objective of the problem. Accordingly, the agent seeks to maximize the cumulative reward over the course of its interaction with the environment, rather than prioritizing immediate reward maximization. If the sequence of rewards for every time step $t$ is denoted as $R_{t+1}, R_{t+2}, R_{t+3}, \ldots$, we can then formalized the \textit{expected return} $G_t$ of that sequence. Formally, the expected return is what the agent aims to maximize. The simplest form of return can be written as
\begin{equation}\label{eq:return_non_inf}
    G_t \coloneqq R_{t+1}+R_{t+2}+R_{t+3}+\ldots+R_T,
\end{equation}
where $T$ is a \textit{final time step}. This approach is sensible in the case when there is a natural notion of final time step, that is when the agent-environment interactions break naturally under a specific condition. This specific condition is generally called as \textit{terminal state}. Furthermore, the overall sequence from the initial until terminal state is called \textit{episodes}.

However, in many case the agent-environment interactions go infinitely rather than having a definitive end, which could cause the return to become unbounded.  To address this, the \textit{discounting} concept is introduced to discount the future rewards, emphasizing the importance of the immediate rewards while reducing the influence of the future rewards and preventing the return computation to explode. It is written as
\begin{equation}\label{eq:return_inf}
    G_t \coloneqq R_{t+1} + \gamma R_{t+2} + \gamma^2 R_{t+3} + \ldots = \sum^{\infty}_{k=0}\gamma^k R_{t+k+1},
\end{equation}
where $\gamma$ is called \textit{discount rate} and bounded with $0\leq \gamma \leq 1$. As $\gamma$ approaches $1$, the greater the emphasis placed on future rewards, and \textit{vice versa}, thereby defining the trade-offs between immediate rewards and long-term rewards.


\subsubsection{Policies and Value Functions}

A \textit{policy} is formally defined as a mapping from states to probabilities of selecting each possible action. In a stochastic policy, the probability of selecting action $a$ in state $s$ is denoted by $\pi(a\mid s)$. Thus, when the agent follows policy $\pi$, the action $A_t$ at time $t$ is sampled from $\pi(\cdot\mid S_t)$.

The \textit{value function}, $v_{\pi}(s)$, is a function that estimate the \textit{value} of the agent being in a state $s$ under policy $\pi$. Similarly, \textit{action-value function}, $q_{\pi}(s,a)$, also estimate the \textit{value} of the agent taking one action $a$ in a state $s$ while following policy $\pi$. The \textit{value} here is defined in terms of the expected returns $G_t$.

Solving \ac{rl} problem means finding the policy that maximizes the returns collection over the episodes. In that context,there is always at least one policy that is better or equal to all other policies which can be measured by the value function associated to that policy. In essence, $\pi \geq \pi'$ if and only if $v_\pi(s) \geq v_{\pi'}(s)$ for all $s\in\mathcal{S}$. This policy is called the \textit{optimal policy}, denotes as $\pi^*$. They also share the same state-value function and action-value function, namely the \textit{optimal state-value function} $v^*(s)$ and \textit{optimal action-value function} $q^*(s,a)$ respectively. It can be defined as
\begin{align}\label{eq:optimal_val_func}
    v^*(s) \coloneqq\;& \max_\pi(v_\pi(s)), \quad\quad \text{for all } s\in\mathcal{S}, \\
    q^*(s,a) \coloneqq\;& \max_\pi(q_\pi(s,a)), \quad \text{for all }s\in\mathcal{S}\text{ and }a\in\mathcal{A}.
\end{align}

\subsubsection{Policy Evaluation and Policy Improvement}
Many \ac{rl} algorithms can be interpreted as alternating, explicitly or implicitly, between two operations: \textit{policy evaluation} and \textit{policy improvement}. Policy evaluation estimates how good the current policy is, usually through a value function or action-value function. Policy improvement then modifies the policy toward actions that are expected to yield higher return. These two sub-processes alternate iteratively until they converge into the optimal policy.  Each \ac{rl} algorithm differs on how they evaluate and improve the policy, some may have an edge than the others depend on the circumstances. 

\subsubsection{Policy Gradient, Advantage, and Actor-Critic Idea}
One could also consider to use a \textit{parameterized policy} that can select actions without requiring a value functions. Let the policy parameter vector be denoted as ${\theta}\in\mathbb{R}^n$. Using this parameter, we parameterized the policy into $\pi_\theta$ which can be written as
\begin{equation}\label{eq:parameterized_policy}
   \pi_\theta(a\;\mid \;s) = \pi(a\;\mid \;s,\theta) = \Pr\{A_t=a \;\mid \; S_t = s, \theta_t = \theta\}. 
\end{equation}
The policy parameter vector can then be learned by optimizing a scalar performance measure $J(\pi_\theta)$, which quantifies the expected performance of the policy. The value functions may be included in $J(\pi_\theta)$ to aid for the policy parameter learning. Since the objective is to maximize $J(\pi_\theta)$, the policy parameters can be updated using \textit{stochastic gradient ascent}
\begin{equation}\label{eq:policy_param_update}
    \theta_{t+1}=\theta + \alpha\widehat{\nabla_\theta J(\pi_{\theta_t})},
\end{equation}
where $\alpha$ is the \textit{learning rate} and $\widehat{\nabla_\theta J(\pi_{\theta_t})}\in\mathbb{R}^n$ is a \textit{stochastic estimate} whose expectation approximates the gradient of the performance measure with respect to the policy parameters $\theta_t$. Methods that update the policy in this manner are known as \textit{policy gradient methods}. Based on Marbach and Tsitsiklis in \cite{marbach_simulation-based_1998} and (Sutton \textit{et al.},) in \cite{sutton_policy_1999}, the gradient of the scalar function $J(\pi_\theta)$ with respect to the policy parameters can be formalized as:
\begin{equation}\label{eq:policy_gradient_obj}
    \nabla_\theta J(\pi_\theta) \mathbb{E}_{(s_t, a_t) \sim {\pi_\theta}} 
    \left[\frac{\nabla_\theta \pi_\theta(a_t\;\mid \;s_t)}{\pi_\theta(a_t\;\mid \;s_t)} q_\pi(s_t, a_t) \right].
\end{equation}
This gradient increases the likelihood of high returns-action sampling, however it introduces high variances because the gradient is approximated using sampled estimates from trajectories collected by the policy. A common strategy as demonstrated in (\cite{baird_reinforcement_1994}, \cite{greensmith_variance_2001}, \cite{suau_breaking_2025}) can reduce variance by subtracting the state-value function from the action-value function, such that
\begin{equation}\label{eq:advantage_func}
    A_\pi(s_t, a_t) \coloneqq q_\pi(s_t, a_t) - v_\pi(s_t).
\end{equation}
This adjustment is referred to as the \textit{advantage function}, and it describes how much better is action $a_t$ than the average action under the current policy at state $s_t$. This adjustment does not bias the gradient as well because the value function is independent of the action.

The \textit{actor-critic} architecture is a method that learns approximations to both of policy learning and value estimation. The \textit{actor} is a policy network that proposes actions given a state, while the \textit{critic} consists of value function networks that estimate the expected return of actions taken in given states. This separation allows the actor to be updated using gradients provided by the critic, leading to more stable learning than purely policy-gradient or value-based methods.

\subsection{Long Short-Term Memory Neural Network}

\ac{lstm} is an advanced variant of the \ac{rnn} that is specifically tailored for learning \textit{long-term dependencies}. \ac{lstm} was first introduced by Hochreiter and Schmidhuber \cite{hochreiter_long_1997}, and now it is widely use and works remarkably well on many variety of problems.
\begin{figure}[h!]
\begin{center}
    \includegraphics[width=.6\textwidth]{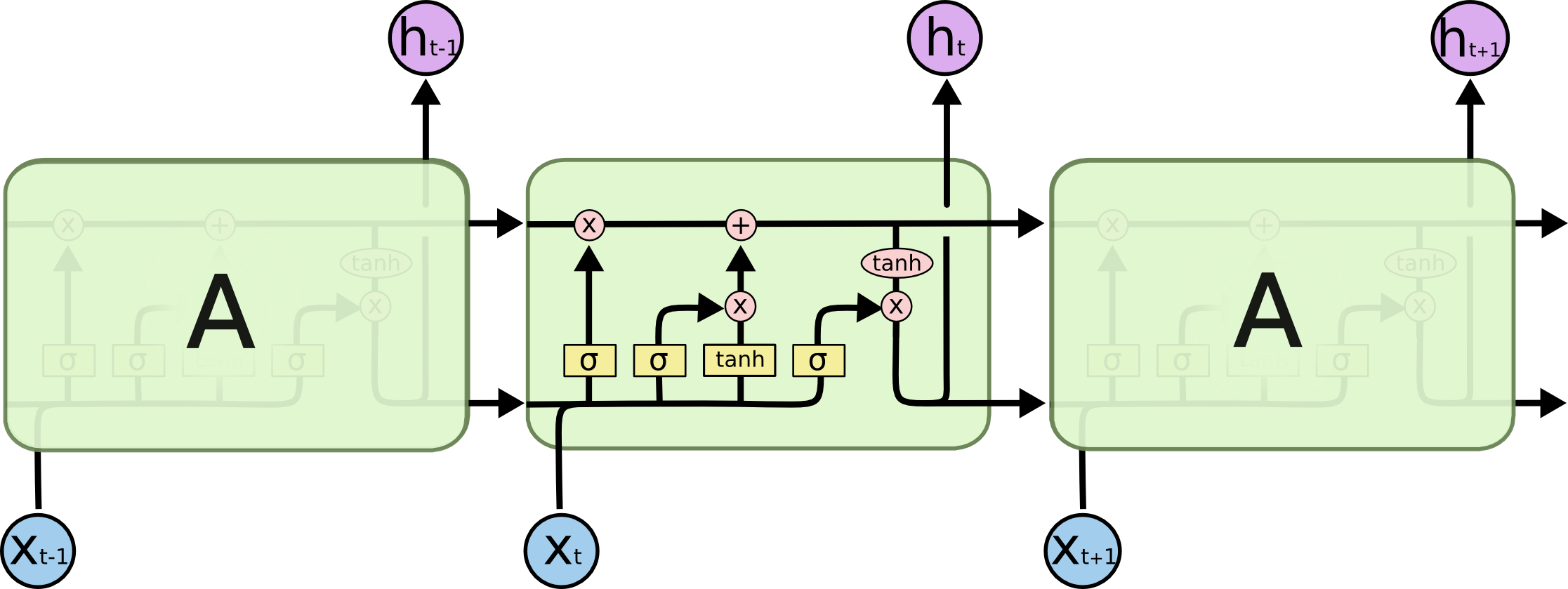}
    \caption{Long Short-Term Memory Networks \cite{olah_understanding_2015}.}
    \label{fig:lstm}
\end{center}
\end{figure}

\ac{lstm} addresses the limitations of standard \ac{rnn}, where it often struggles to learn long-term dependencies due to the \textit{vanishing gradient} and \textit{exploding gradient} problems \cite{schmidt_recurrent_2019}. \ac{lstm} solves this by introducing a separate memory pathway known as the \textit{cell state}. As shown in Figure \ref{fig:lstm}, the cell state is represented by the horizontal connection running through the top of the structure. The flow of information into and out of the cell state is controlled by several \textit{gates}, which are the \textit{forget gate}, \textit{input gate}, and \textit{output gate}. These gates use sigmoid activation $\sigma$ and $\tanh$ functions to regulates how much information should be retained, added, or exposed. Through this mechanism, \ac{lstm} is better able to retain useful long-range information while filtering out irrelevant information.

\subsection{Reinforcement Learning Cores for Event-Based Adaptive Stress-Testing}

\subsubsection{Proximal Policy Optimization and Recurrent Extension}

\ac{ppo} is a family of policy-gradient methods that optimizes a surrogate objective using stochastic gradient ascent. Introduced by Schulman \textit{et al.} \cite{schulman_proximal_2017}, \ac{ppo} was designed to retain some of the data efficiency and reliable performance associated with \ac{trpo}, while avoiding the complexity of second-order constrained optimization. This is achieved through a clipped probability-ratio objective, which discourages excessively large policy updates and improves training stability.

Let $r_t(\theta)$ denotes the probability ratio between the updated policy and the policy used to collect the data
\begin{equation}
    r_t(\theta)=\frac{\pi_\theta(a_t\mid s_t)}{\pi_{\theta_{\mathrm{old}}}(a_t\mid s_t)}.
\end{equation}
When $\theta=\theta_{\mathrm{old}}$, the ratio satisfies $r_t(\theta_{\mathrm{old}})=1$. Values greater than one indicate that the updated policy assigns higher probability to action $a_t$ in state $s_t$ while values less than one indicate that the updated policy assigns lower probability to that action.


\ac{ppo} avoids the explicit constrained \ac{kl} divergence optimization of \ac{trpo} by modifying the surrogate objective with a clipping operation
\begin{equation}\label{eq:ppo_obj_func}
    J^{\mathrm{CLIP}}(\theta)=\mathbb{E}_t\left[\min\left(r_t(\theta) A_t;\mathrm{clip}\left(r_t(\theta),1-\epsilon,1+\epsilon\right) A_t\right)\right],
\end{equation}
where $\epsilon$ is a hyperparameter that determines the clipping range. The clipping mechanism limits the incentive for the updated policy to move too far from the old policy by maintaining the balance between the probability ratio $r_t(\theta)$ and the advantage value $A_t$ while hard-constraining the objective with a predetermined threshold $\epsilon$. In this way, \ac{ppo} approximates the stabilizing effect of \ac{trpo} algorithm while using a simpler first-order optimization procedure.


The objective in advantage function estimators can also use value function loss formulated using squared-error loss, denoted as $J^{VF}$, making it follows the \textit{actor-critic} formalism
\begin{equation}
    J^{\mathrm{VF}}_t(\theta) = \left(v_\theta(s_t) - v^{target}_t \right)^2.
\end{equation}
The objective can also be further improved by introducing \textit{entropy bonus}, $\mathcal{H(\pi)}$ maximization, which encourages exploration. Combining this terms, we can obtain a more robust \ac{ppo} formulation
\begin{equation}\label{eq:robust_ppo_obj}
    J^{\mathrm{CLIP+VF+}\mathcal{H}}(\theta) = \mathbb{E}_t\left[J^{\mathrm{CLIP}}_t(\theta) - k_1J^{\mathrm{VF}}_t(\theta) + k_2\mathcal{H}(\pi(\cdot\mid s_t)) \right],
\end{equation}
where $k_1$ and $k_2$ control the strength of the value-function and entropy-bonus terms.

While the standard \ac{ppo} is designed for fully observable \acp{mdp}, many real-life problems provide only for partial observation of the underlying state. Standard \ac{ppo} generally employs feedforward \ac{mlp} policy and value networks, therefore basing each decision on the current observation. \ac{rppo} extends \ac{ppo} by leveraging recurrent architecture into the actor and critic networks, enabling the agent to retains memory that can provide information on the history of the current task that is not present in the agent's most recent observation \cite{pleines_generalization_2022}. In this research, \ac{lstm} networks are chosen to represent the policy $\pi_\theta$ and the value function $v_\theta$ while preserving the same optimization objectives as standard \ac{ppo} (Equation \ref{eq:robust_ppo_obj}).

\subsubsection{Soft Actor-Critic}

\ac{sac} is introduced in \cite{haarnoja_soft_2018} by Haarnoja \textit{et al}. as an off-policy actor-critic algorithm based on the \textit{maximum entropy RL framework}. For the standard \ac{rl} algorithm, the goal is to learn the policy $\pi$ to maximizes the returns $G_t$. On the other hand, the maximum entropy objective generalizes the standard \ac{rl} objectives by augmenting it with an \textit{entropy term}, such that the optimal policy additionally maximize its entropy at each visited state:
\begin{equation}
    \pi^* = \arg \max_\pi\sum_t \mathbb{E}_{(s_t,a_t)\sim\rho_\pi}\left[r(s_t,a_t) + \alpha\mathcal{H}(\pi(\cdot\;\mid \;s_t)) \right],
\end{equation}
where $\alpha$ is the temperature parameter that determines the strength of the entropy term $\mathcal{H}$. \ac{sac} algorithm incorporates three key ingredients. First, an \textit{actor-critic architecture} consisting of a policy network $\pi_\theta(a_t \mid s_t)$ (actor) and a value function network $Q_\phi(s_t, a_t)$ (critic), parameterized by $\theta$ and $\phi$, respectively. Second, an \textit{off-policy formulation} that enables reuse of previously collected data through \textit{memory buffer} for efficiency. Third, an \textit{entropy maximization process} to encourage stability and exploration.

\ac{sac} is trained on the \textit{soft policy iteration}, which includes \textit{soft policy evaluation} and \textit{soft policy improvement}. The main goal of this framework is to simultaneously maximize expected \textit{return} and \textit{entropy} to encourage exploration while still discouraging clearly bad actions. Intuitively, it is analogous to succeed at the task while acting randomly to the policy entropy. We encourage reader to visit Haarnoja \textit{et al.} works \cite{haarnoja_soft_2018}, where they provide the details about the loss functions and derivations of the \ac{sac} algorithm.

\subsection{Adaptive Stress-Testing}

\ac{ast} is a simulation-based falsification method that aims not only to find a scenario that results in a \textit{failure}, but also to maximize the  \textit{probability} of the \textit{failure modes} \cite{lee_adaptive_2020}. One most notable implementation of \ac{ast} was carried by Lee \textit{et al.} \cite{lee_adaptive_2015}, where \ac{ast} was introduced as a \textit{scalable} and \textit{black-box} compatible failure discovery method to stress-test a prototype air collision avoidance system by identifying the most likely trajectories that leads to mid-air collision.
\begin{figure}[h!]
\begin{center}
    \includegraphics[width=.6\textwidth]{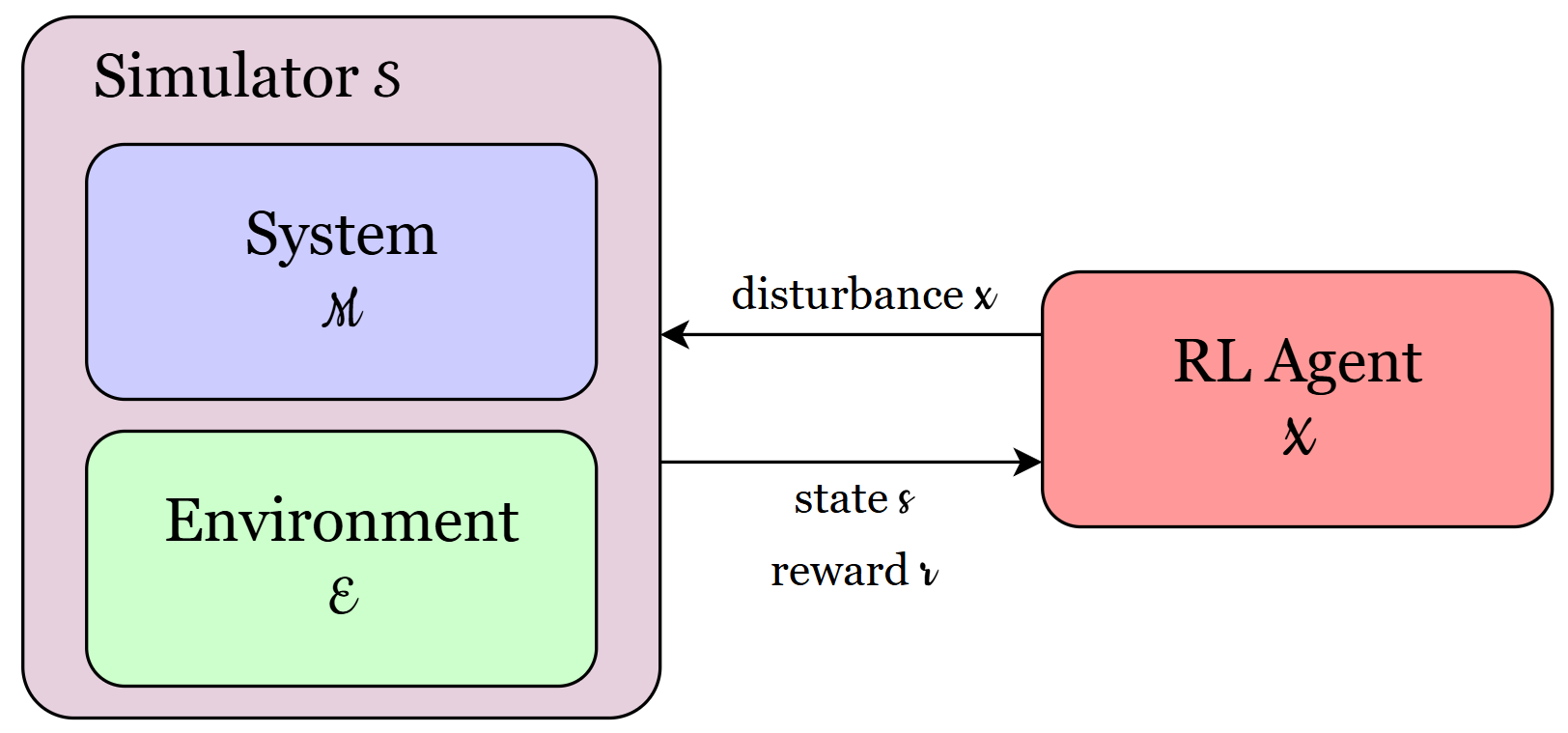}
    \caption{Adaptive Stress-Testing illustration.}
    \label{fig:AST_illustration}
\end{center}
\end{figure}

Borrowing the terminologies from Section \ref{sec:CPS_assessment}, we define a \textit{simulator} denoted as $\mathcal{S}$ that contains the \textit{system under test} (or just \textit{system}) $\mathcal{M}$ interacting with the \textit{environment} $\mathcal{E}$. The system with state $\mu \in \mathcal{M}$ takes a response $\varrho \in \mathrm{P}$ based on observation $o \in \Omega$ of the environment state $z \in \mathcal{Z}$. The simulator state $s \in \mathcal{S}$ is the stacked system and environment states $\left[\mu, z \right]$, The system interacts with the environment over discrete timestep $t \in \left[0, \cdots, t_{\text{end}} \right]$. As conventions, we denote the variable at time $t$ using subscript $t$ and we also denote the sequence of variable over a range of time steps with colon range $t_{\text{start}}:t_{\text{end}}$. The state and action of the system depend on the environment observations, and are modeled as:
\begin{equation}
    \mu_{t+1}, \varrho_t = \mathcal{M}(o_{0:t}).
\end{equation}

The environment state and observation evolve over time depending on the actions of the system and \textit{disturbances} $x \in \mathcal{X} \subseteq \mathbb{R}^n$, where $n$ is the dimensionality of the disturbances. The disturbances here can be depicted as any stochastic variables that can affect the environment, hence these terms are very broad and need to be specified. For example, the can be weather loads, sensor noises, or adversarial actors wandering inside the simulation (e.g. pedestrian, target ships, other vehicles)
\begin{equation}
    z_{t+1}, o_{t+1} = \mathcal{E}(\varrho_{0:t}, x_{0:t}).
\end{equation}

The disturbances are assumed to be conditionally sampled from a probability density $\Pr(x \mid s)$, which may be specified using expert knowledge or estimated from relevant data. Let $\Upsilon \subseteq \mathcal{S}$ denote the desired event space, and let $\neg\Upsilon \subseteq \mathcal{S}$ denote its complement, corresponding to undesired events. There also exists states that cause the simulation to be terminated which is referred to as \textit{terminal state}. In this work, we focus on failure, i.e., terminal states satisfying $s \in \Upsilon$. They signify the end of one episode, and they can belong both in $\Upsilon$ (e.g. ship collision) and $\neg\Upsilon$ (e.g. ship reaches end trajectory). The objective of \ac{ast} is to find the \textit{most likely failure modes} by maximizing the \textit{reward function}. The objective can be formulated as:
\begin{equation}\label{eq:AST_objective}
     \max_{x_{0:t_{\text{end}}}} \quad
     \prod_{t=0}^{t_{\text{end}}-1} \Pr(x_t \mid s_t)
     \quad
     \text{subject to} \quad
     s_{t_{\text{end}}}\in\Upsilon .
\end{equation}

\ac{ast} frames failure discovery as a reinforcement learning problem, shown in Figure \ref{fig:AST_illustration}, where an adversarial agent selects disturbances $x_t$ that challenge the system under test $\mathcal{M}$. Meanwhile the simulator evolves the system state and reports whether a failure has occurred. To avoid confusion, \ac{ast} frames the \textit{disturbance sampling} as the equivalent of \textit{action sampling} on typical \ac{rl} methods. Depending on the extent to which the simulator exposes its internal state, \ac{ast} can be categorized into two types:

\subsubsection{Full Observability}
If the simulator state is fully observable, we can formulate the \ac{ast} problem as an \ac{mdp} as demonstrated in Section \ref{sec:mdp}. In this case, the agent can selects a disturbance $x_t$ from the policy $\pi$ given the true state $s_t$,
\begin{equation}
    x_t \sim \pi(\cdot \mid s_t),
\end{equation}
where the general reward function design by Lee \textit{et al.} \cite{lee_adaptive_2020} is given by:
\begin{equation}\label{eq:reward_function_mdp}
    \mathcal{R}(s, x) = 
    \begin{cases}
        R_{\Upsilon} \quad &s\text{ is terminal and }\text{if } s\in\Upsilon \\
        -d \quad &s\text{ is terminal and }\text{if } s\in\neg\Upsilon\\
        \log\left(\Pr(x \mid s)\right) \quad &\text{otherwise }.
    \end{cases}
\end{equation}
The first term of Equation \ref{eq:reward_function_mdp} provides a non-negative reward signal, $R_{\Upsilon}$, when the trajectory terminates due to a failure. If the trajectory terminates without encountering a failure, the second term of Equation \ref{eq:reward_function_mdp} penalizes the agent by assigning a \textit{miss distance}, $d$, which measures how close the simulator came to a failure event. This term can significantly accelerate the search by enabling the agent to differentiate the relative desirability of non-failure trajectories \cite{lee_adaptive_2020}.

The final term of Equation \ref{eq:reward_function_mdp} seeks to maximize the \ac{ast} objective defined in Equation \ref{eq:AST_objective} by rewarding the log-likelihood of each state transition. Recall that unlikely disturbances are associated with lower probabilities, whereas likely disturbances have probabilities closer to one. Consequently, the log-likelihood of an unlikely transition approaches $-\infty$, while that of a highly likely transition approaches $0$. This reward formulation therefore penalizes unlikely transitions while favoring more probable ones, thereby guiding the agent toward discovering failure trajectories that are as likely as possible.

\subsubsection{Partial Observability}
Not all simulator supports fully observable states. Some simulators are given as a software binary that maintained its internal dynamics hidden over function calls \cite{lee_adaptive_2020}. Example of this are co-simulation. Even though the underlying process can be modeled as a \ac{mdp}, the agent only sees part of simulator states $s$ in form of observations $o$. 
\begin{align}
    o_t &\sim \mathcal{O}(\cdot \mid s_t), \\
    x_t &\sim \pi(\cdot \mid o_t).
\end{align}

This is a problem for typical \ac{mdp} solvers because the same observation may correspond to multiple underlying hidden states. Consequently, a policy conditioned solely on the current observation $o_t$ may select suboptimal actions, as it lacks sufficient information to infer the true state of the environment. To address this issue, the partially observable \ac{ast} problem is naturally formulated as a \ac{pomdp} (see Section \ref{sec:pomdp}), where sampled disturbances are conditioned on the observation history $h_t$ (see Equation \ref{eq:history_eq}) rather than only the current observation:
\begin{equation}
x_t \sim \pi(\cdot \mid h_t).
\end{equation}

Based on our literature review, we identified two main approaches for handling partial observability in \ac{ast}:

\begin{enumerate}
\item \textbf{History-based tree search.} This approach employs algorithms that maintain and revisit the action-observation history during planning, allowing the agent to reason over past information when selecting future disturbances. This direction was pioneered by Lee \textit{et al.} \cite{lee_adaptive_2020} through the introduction of the \ac{mcts-sa} algorithm.

\item \textbf{Recurrent policy gradient methods.} This approach augments policy gradient algorithms with \ac{rnn} architectures, such as \ac{lstm}, instead of conventional \ac{mlp}s. The recurrent layers enable the policy to learn temporal dependencies from the observation history, therefore forming an implicit approximation of the underlying belief state. One example is the work of Koren and Kochenderfer \cite{koren_efficient_2019}, who demonstrated the application of \ac{drl} algorithms with \ac{lstm} layers for partially observable \ac{ast}.
\end{enumerate}

As for the reward function designs, it mirrors the form and purpose those in Equation \ref{eq:reward_function_mdp} \cite{lee_adaptive_2020}:
\begin{equation}\label{eq:reward_function_pomdp}
    \mathcal{R}(\rho, f, d, \tau) = 
    \begin{cases}
        R_{\Upsilon} \quad &\text{if } \tau \text{ and } f \\
        -d \quad &\text{if } \tau \text{ and } \neg f\\
        \log\rho \quad &\text{otherwise },
    \end{cases}
\end{equation}
where the simulator \textit{indirectly} (due to agent's limited access to the actual simulator states) provides the transition likelihood $\rho = \Pr(x \mid s)$, a boolean indicating whether failure $f = (s \in \Upsilon)$ has occurred, and the miss distance $d$. Additionally, $\tau$ represents a boolean from a function call that explains whether a terminal state has been reached based on the actual simulator states.

\subsection{Event-Based Adaptive Stress-Testing}

\ac{ebast} was first introduced as an extension of the conventional \ac{ast} framework by revisiting its action sampling strategy, and was extensively evaluated for stress-testing maritime collision avoidance systems \cite{sitorus_identifying_2025}. In conventional \ac{colav} \ac{ast}, the adversary is modeled as an \ac{rl} agent that directly controls the target ship as a disturbance by sampling low-level control actions corresponding to its \textit{surge}, \textit{sway}, and \textit{yaw} accelerations. These actions are updated at every fixed \ac{rl} decision interval throughout the simulation.

One representative example is the work of Hjelmeland \textit{et al.} \cite{hjelmeland_identification_2022}, which demonstrated that this formulation is effective in discovering collision-inducing trajectories under several reward function designs. Despite its promising performance, this approach exhibits several limitations. Since the adversary updates its control actions at a higher frequency, the resulting trajectories may become overly discontinuous and operationally unrealistic. This effect is particularly evident in scenarios where the target ship continuously steers toward the own ship or performs excessively aggressive maneuvers during near-collision encounters. 
\begin{figure}[h!]
\begin{center}
    \includegraphics[width=.42\textwidth]{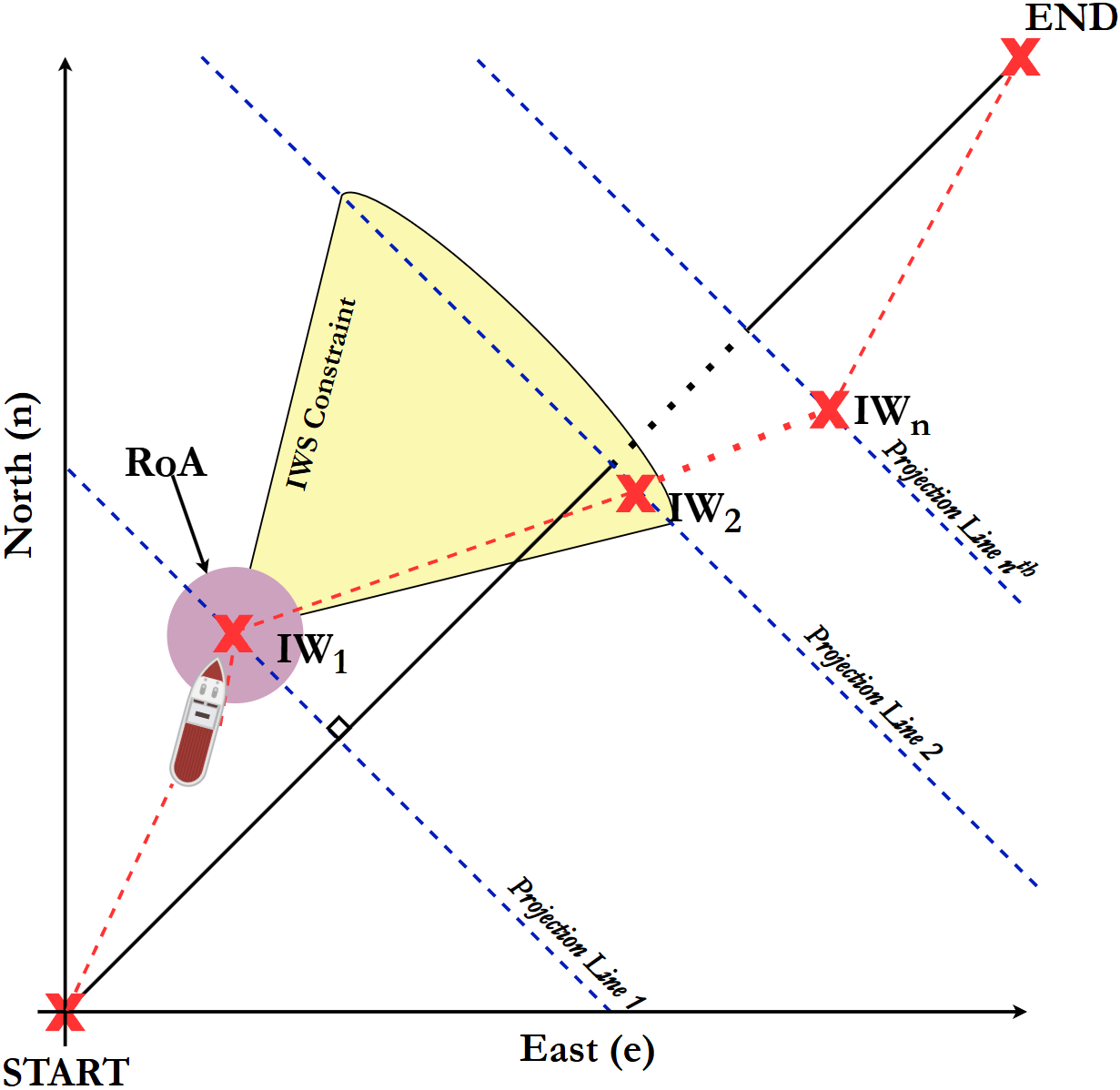}
    \caption{Intermediate waypoint sampling process in \ac{ebast}.}
    \label{fig:IWS_v1}
\end{center}
\end{figure}

\ac{ebast} considers a typical ship-in-transit scenario in which the own ship executes its nominal voyage while an RL agent controls an adversarial target ship. The objective of the adversary is to interfere with the own ship's mission and induce collision scenarios. Each of ship asset initially has their own \textit{baseline} trajectory (which is a straight trajectory between \textit{start} to \textit{end} waypoint). The main takeaway is that the target ship is permitted to sample new \textit{waypoints} in between the start and end of its trajectory to follow, mid simulation. These new waypoints is called \acp{iw}, and an \ac{iw} can be sampled only after the target ship enters the current \ac{iw}'s \ac{roa}. Consequently, \ac{iw} updates are triggered by \textit{navigation events} rather than fixed time intervals, giving rise to the term \textit{event-based}. This waypoint-generation mechanism, referred to as \ac{iws} (illustrated in Figure \ref{fig:IWS_v1}), and constitutes the key characteristics of the \ac{ebast} framework. Additionally, \ac{iws} is constrained such that every newly sampled waypoint advances the vessel toward its destination, preventing unrealistic backward maneuvers or oscillatory behavior. The overall \ac{ast} process is then constructed as \ac{mdp} and then optimized using \ac{sac} algorithm. The objective of \ac{ebast} is to discover critical waypoint sequences that induce collision while preserving realistic target-ship trajectories through the waypoint sampling constraints \cite{sitorus_identifying_2025}.

The proposed \ac{ebast} framework however was limited to identifying the near-collision events \cite{sitorus_identifying_2025} and did not uncover actual collision events. The authors suggest that introducing multiple adversarial vessels may place the own ship's \ac{colav} system under greater operational stress, which might increasing the likelihood of genuine collision events.

%% file: Sections/Section_3.tex
\section{Ship-in-Transit Co-simulation Setup}\label{sec:sit_co-simulation_setup}

\subsection{Ship Model}\label{sec:Ship_Model}
\begin{figure}[h!]
\begin{center}
    \includegraphics[width=.6\textwidth]{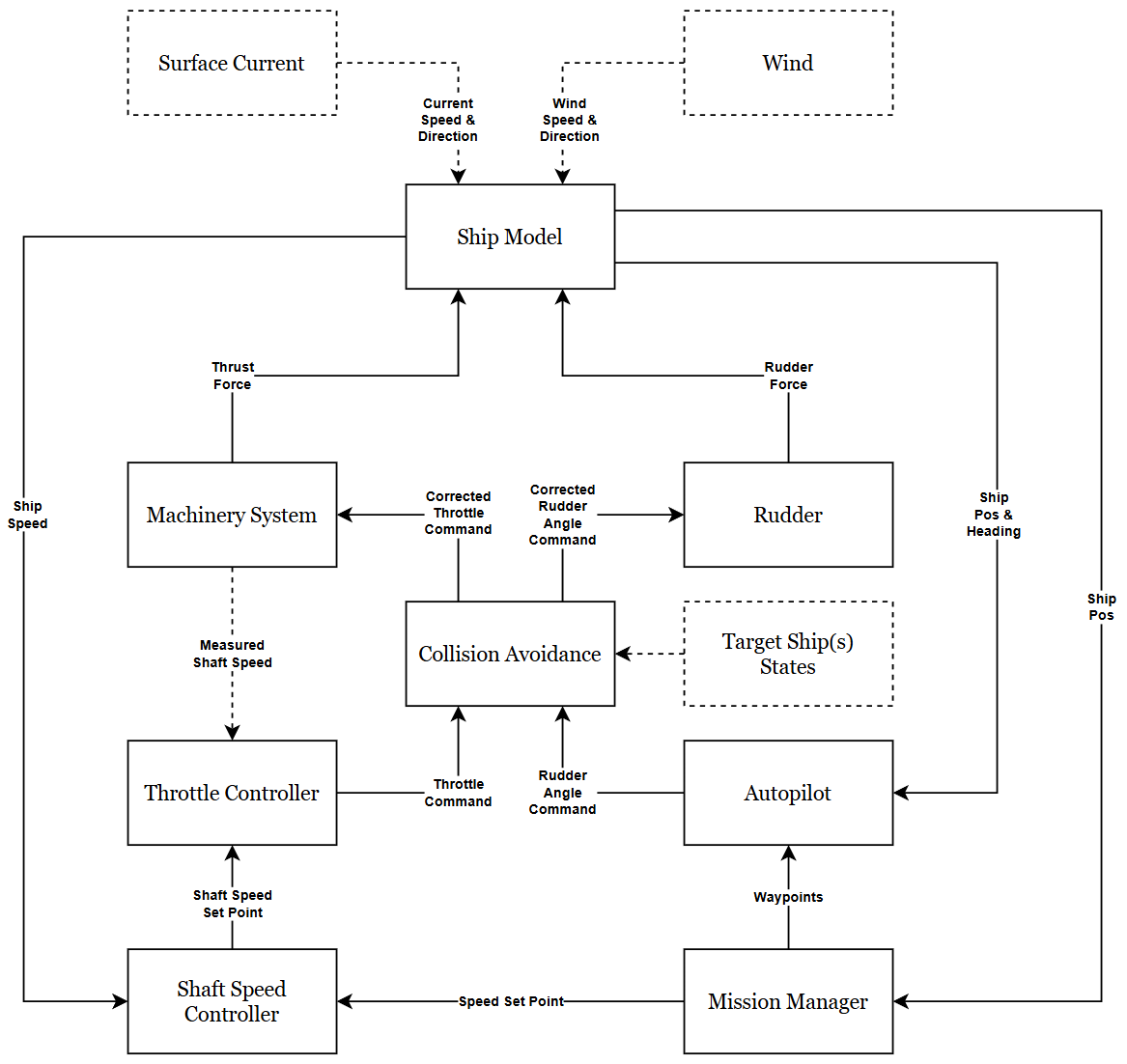}
    \caption{\acp{fmu} connection diagram for the ship simulation.}
    \label{fig:SiT_FMUs}
\end{center}
\end{figure}
Figure \ref{fig:SiT_FMUs} shows the interconnection structure of the \acp{fmu} used in the ship simulation framework. In this research, this structure is specifically used to model the own ship in order to include the system complexity for the stress testing process. Each block in the diagram is implemented as a separate \ac{fmu}, and exchange data through their defined input and output variables. To orchestrate the communication between each block, we use a Python library called \texttt{libcosimpy} \cite{steinsto_libcosimpy_2026}, which is a Python wrapper for co-simulation library in C++, \texttt{libcosim} \cite{noauthor_libcosim_2026}, both are developed by \ac{osp} \cite{noauthor_open_nodate}.

The \textit{ship model} \ac{fmu} represents the vessel hull dynamics. It receives the \textit{thrust force} generated by the \textit{machinery-system} \ac{fmu} and the \textit{rudder force} generated by the \textit{rudder} \ac{fmu}. Environmental disturbances are modeled as two \ac{fmu} blocks: \textit{surface current} and \textit{wind}. which provides \textit{the speed and direction} of the surface current and wind to ship model \ac{fmu}.

The nominal actuator commands are generated by the \textit{autopilot} and the \textit{speed-control} subsystems and are subsequently processed by the \ac{colav} \ac{fmu}. When no collision threat are detected, the \textit{nominal throttle and and rudder commands} are passed directly to the machinery-system and rudder \ac{fmu}, respectively. When an avoidance action is required, \ac{colav} system modifies these command and output the corrected nominal commands.

The autopilot \ac{fmu} generates the nominal rudder-angle command using \ac{los} guidance scheme. The \textit{mission-manager} \ac{fmu} selects and supplies the active waypoint segment, consisting of the previous and next waypoints, according to the current vessel position received from the ship model \ac{fmu}. Using this waypoints, the autopilot determines the desired heading and calculates the corresponding rudder command.

The forward speed control is handled by a cascaded structure consisting of the \textit{shaft-speed controller} and \textit{throttle-controller} \ac{fmu}. The shaft-speed controller compares the actual vessel speed from the ship model \ac{fmu} and the predefined speed set point from the mission manager \ac{fmu}. It then generates the desired shaft-speed setpoint that will be used by the throttle-controller to compare with the measured shaft-speed retrieved from the machinery-system \ac{fmu}. From this, it will output the nominal throttle command, closing the loop.

\subsection{Simplified Ship Model}
\begin{figure}[h!]
\begin{center}
    \includegraphics[width=.6\textwidth]{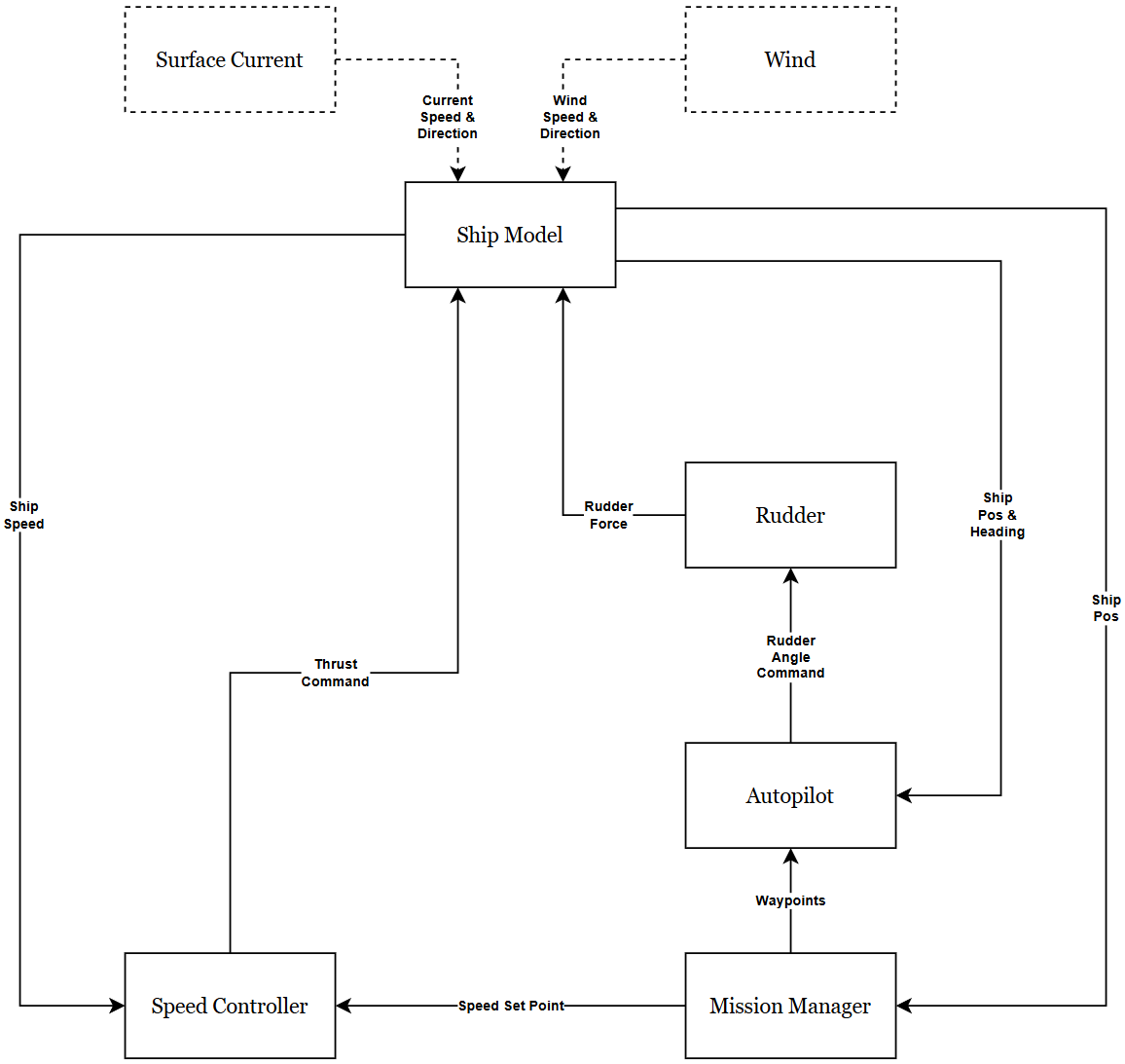}
    \caption{\acp{fmu} connection diagram for the simplified ship simulation.}
    \label{fig:Simplified_SiT_FMUs}
\end{center}
\end{figure}
Figure \ref{fig:Simplified_SiT_FMUs} shows the interconnection structure of the \acp{fmu} used in the simplified ship simulation framework. Like the previous Ship Model structure in Section \ref{sec:Ship_Model}, all blocks in the diagram are implemented as a separate \acp{fmu}, and are orchestrated using the same \texttt{libcosimpy} Python library.  In this research, a simplified model is used for the target ships because their role as adversarial actors does not require the same level of system complexity unlike the own ship. Therefore, we stripped off the machinery-system \ac{fmu} along with the cascading speed-control subsystem and replace it with a simple PI \textit{speed controller}. Additionally, we also exclude the \ac{colav} system \ac{fmu}, which enables a direct thrust and rudder command input into the ship model \ac{fmu}.

\subsection{Simple Collision-Avoidance System}\label{sec:simple_colav}
\begin{figure}[h!]
\begin{center}\includegraphics[width=.8\textwidth]{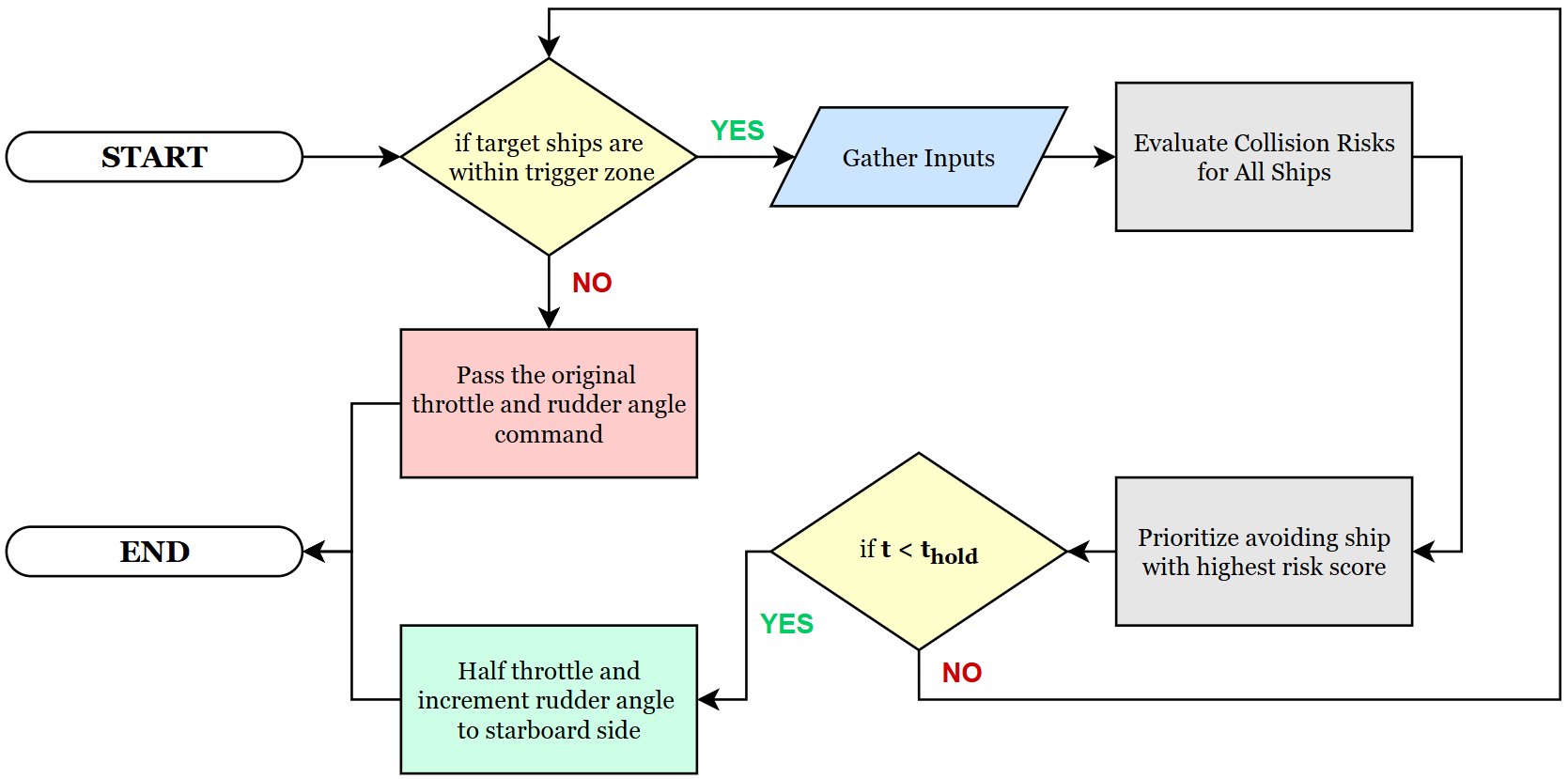}
    \caption{Process flow diagram of the simple \ac{colav} system to stress-test.}
    \label{fig:simple_colav}
\end{center}
\end{figure}
The simple \ac{colav} implementation follows a relatively simple decision strategy, as shown in Figure \ref{fig:simple_colav}. When an avoidance maneuver is required, the system commands a half-throttle and a fixed-angle turn to starboard with the intention of passing the selected target ship on the own ship's port side. This intention is arbitrary and serves as a demonstration only. The avoidance command will considered once the system assess own ship's collision risk. At each assessment interval, it evaluates the collision risk associated with every target ship and selects the ship presenting the highest risk. The collision risk score computation is based on the \ac{cpa} metrics, and \textit{closing speed} value.

Let the relative north-east position vector and the relative north-east velocity between the own ship and the target ship be defined as $\boldsymbol{r}$ and $\boldsymbol{\nu}$ respectively,
\begin{align}
    \boldsymbol{r} &= \boldsymbol{p}_{\text{OS}} - \boldsymbol{p}_{\text{TS}},\quad \boldsymbol{p}:[N,E]^\top\in\mathbb{R}^2 \label{eq:rel_pos} \\
    \boldsymbol{\nu} &= \boldsymbol{\nu}_{\text{OS}} - \boldsymbol{\nu}_{\text{TS}},\quad \boldsymbol{\nu}:[u,v]^\top\in\mathbb{R}^2.\label{eq:rel_vel}
\end{align}
Given the own ship north-east position $p_\text{OS}$ and heading $\psi_\text{OS}$, we compute each target ship's relative bearing to own ship $\beta$
\begin{equation}
    \beta = \arctan{(E_\text{OS} - E_\text{TS}, N_\text{OS} - N_\text{TS})} - \psi_\text{OS},
\end{equation}
and the \textit{instantaneous separation} distance
\begin{equation}
    D = \|\boldsymbol{r}\|.
\end{equation}
Differentiating $D$ with respect to time, yields the closing speed
\begin{equation}
    \dot{D} = \frac{\boldsymbol{r}^\top \boldsymbol{\nu}}{\|\boldsymbol{r}\|}.
\end{equation}
\textit{Closing speed} explains whether two ships are closing in on each other when $\dot{D}$ is negative or moving apart when $\dot{D}$ is positive.

We also introduce the concept of the \ac{cpa}, which is widely used in the definition and evaluation of collision-threat parameters and criteria \cite{lenart_analysis_2015}. The \ac{cpa} denotes the predicted point at which the distance between the own ship and another object, (can be as a target ship or an obstacle), reaches its minimum value \cite{errico_methodology_2019}. Two quantities associated with the \ac{cpa} are the \ac{tcpa} and \ac{dcpa}
\begin{align}
t_{\text{CPA}} &= -\frac{\boldsymbol{r}^\top \boldsymbol{\nu}}{\boldsymbol{\nu}^\top \boldsymbol{\nu}}\quad, \\
D_{\text{CPA}} &= \|\boldsymbol{r} + \boldsymbol{\nu} t_{\text{CPA}}\|.
\end{align}

During the simulation, the own ship will continuously compute the \ac{tcpa}, \ac{dcpa}, and $\dot{D}$ of each target ship. Given the \textit{collision zone radius} $D_\text{collision}$ and the \textit{safe look-ahead time} $t_\text{look-ahead}$, the target ship is considered as threat when its $t_\text{CPA}>0$, $t_\text{CPA}<t_\text{look-ahead}$, $D_\text{CPA} < D_\text{collision}$ and $\dot{D}<0$. In the case of multiple target ships are considered as threats, the own ship's \ac{colav} system will compute The \textit{collision risk score} $\Xi$ and prioritize the target ships that has the \textit{highest risk score}. $\Xi$ is formulated as:
\begin{equation}
    \Xi = \frac{\lambda_{t_\text{CPA}}}{t_\text{CPA}} + \frac{\lambda_{D_\text{CPA}}}{D_\text{CPA}}- \lambda_{\dot{D}}\dot{D}
\end{equation}

Where $\lambda_{t_\text{CPA}}$, $\lambda_{D_\text{CPA}}$, $\lambda_{\dot{D}}$ are all coefficients which control the strength of each terms. The \ac{colav} system can respond to only one target ship during each decision interval. After the interval has elapsed, the $\Xi$ are reassessed and the target ship with the highest $\Xi$ is selected.

\subsection{Ship Traffic Generator}
\ac{stg} is a tool developed by \ac{dnv} to generate a structured set of ship encounter scenarios for verifying automated collision and grounding avoidance systems \cite{pedersen_ship_nodate}. To generate an encounter scenario involving the own ship, the desired situation must first be specified. This desired situation consists of several user-defined parameters:
\begin{enumerate}
    \item \textbf{\textit{Initial Position}}. The initial position of the own ship, given by latitude and longitude coordinates in decimal degrees.
    
    \item \textbf{\ac{sog}}. The speed of the own ship relative to the ground, given in knots.
    
    \item \textbf{\ac{cog}}. The nominal mission course of the own ship, given in degrees using a north-referenced clockwise convention. For example, a northbound trajectory has a \ac{cog} of $0^\circ$.
    
    \item \textbf{\textit{Initial Heading}}. The initial heading angle of the own ship, given in degrees.
    
    \item \textbf{\textit{Navigational Status}}. The navigational status assigned to the simulated vessel.
    
    \item \textbf{\textit{Vector Time}}. The time, given in minutes from the start of the simulation, at which the desired encounter situation is expected to occur within a specified range from the own ship.
    
    \item \textbf{\textit{Relative Bearing}} $\beta$. Which is the angle between the own ship heading and the target ship directions, given in degrees.
    
    \item \textbf{\textit{Relative Speed}}. The speed of the target ship relative to the own ship, expressed as a fraction of the own ship speed. A value of $1$ indicates that the target ship travels at the same speed as the own ship, while values below or above $1$ indicate lower or higher relative target-ship speeds, respectively. The minimum value is $0$.
\end{enumerate}

\begin{figure}[htbp]
    \centering
    \begin{minipage}{0.48\textwidth}
        \centering
        \includegraphics[width=\textwidth]{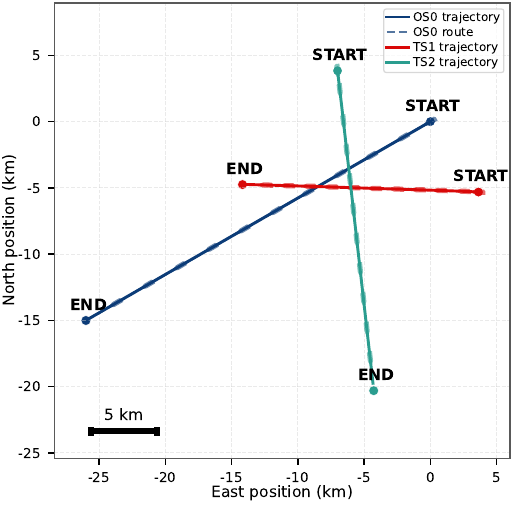}
        \label{fig:40_scenario}
    \end{minipage}
    \hfill
    \begin{minipage}{0.48\textwidth}
        \centering
        \includegraphics[width=\textwidth]{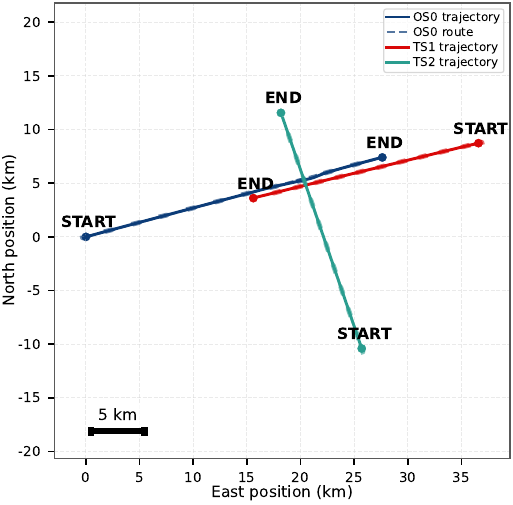}
        \label{fig:50_scenario}
    \end{minipage}
    \caption{Examples of the generated traffic scenarios by \ac{stg}.}
    \label{fig:STG_examples}
\end{figure}

The desired situation is typically specified in a JSON file, which serves as input to the \ac{stg}. The corresponding output is also provided in JSON format and contains relevant vessel settings, such as waypoints and \ac{sog}, for all simulated vessels. These settings can subsequently be translated and implemented in the Ship-in-Transit co-simulation framework to generate the desired encounter scenarios. Examples of the generated traffic scenarios are shown in Figure \ref{fig:STG_examples}.

In this work, \ac{stg} is used as an external scenario-generation tool to construct the initial ship encounter configurations used in the simulation study. The complete encounter generation algorithm is not explained here, since the present work does not modify the internal \ac{stg} procedure. Instead, we summarize the input parameters relevant to our study and describe how the generated encounter cases are integrated into the proposed \ac{ebast2} environment. A detailed description of the \ac{stg} algorithm can be found in Pedersen \textit{et al.}  paper \cite{pedersen_generating_2023}.

%% file: Sections/Section_4.tex
\section{Case Study: Collision-Avoidance Stress-Testing in a Ship-in-Transit Co-simulation}

This case study considers a ship-in-transit operation in which the functionality and failure behavior of a shipboard \ac{colav} system are assessed through simulation-based stress-testing. We are aiming to develop a stress-testing method that automatically searches for scenarios that expose limitations in the \ac{colav} system and lead to collision events. Identifying such scenarios can provide insight into the mechanisms underlying the failures and support subsequent improvements to the \ac{colav} algorithm.

The simulated traffic scenario consists of three vessels. The first vessel is designated as the \textit{own ship} and is equipped with the \ac{colav} system under test. The remaining two vessels are designated as \textit{target ships} and are distinguished by numerical identifiers. From the perspective of the own ship, these target ships represent surrounding traffic that must be monitored and avoided during the voyage.

Each vessel is initially assigned a nominal mission defined by two waypoints: a \textit{start} waypoint $w_{\text{start}}$ and an \textit{end} waypoint $w_{\text{end}}$, where all of them attempt to complete their respective missions using their onboard speed and heading controllers. For the own ship, these waypoints define a straight-line route that the vessel is required to follow. Conversely, for each \ac{iw}-enabled target ship belongs to set $\mathcal{I}_t$, let $j$ denotes the target ship identifier, $\mathcal{T}$ denotes the set of all \textit{feasible trajectories} $\zeta$ as a \textit{finite waypoint sequence} connecting $w_{\text{start}}$ to $w_{\text{end}}$. 
\begin{equation}
    \zeta^j = \left(w_{\text{start}}^j, w_0^j, w_1^j, \cdots, w_m^j, w_{\text{end}}^j\right) \in \left(\mathbb{R}^2\right)^{m+3}, \quad
    w_m = (N_m, E_m)
\end{equation}
\begin{equation}
    \mathcal{T}_{\leq n_w^j}^j = \bigcup^{n_w^j}_{m=0}\left\{\left(w_0^j, w_1^j, \cdots, w_{m+3}^j\right)
    \in \left(\mathbb{R}^2\right)^{m+3}
    \mid
    w_0^j=w_{\text{start}}^j, w_{m+3}^j=w_{\text{end}}^j, \zeta^j \text{ is feasible}\right\}. 
\end{equation}
\begin{equation}
    \mathcal{T}_{\leq n_w}^j = \mathcal{T}_0^j \cup \mathcal{T}_1^j \cup \cdots \mathcal{T}_{n_w^j}^j
\end{equation}
where $j$ is the number of adversarial target ships up to $n_\text{TS}$, ($w_0, w_1, \cdots, w_m$) are the \acp{iw}, and $n_w^j$ is the predetermined maximum number of \acp{iw} associated to each target ship $j$. We also define the distance between two consecutive waypoints $L(\zeta^j)$ as
\begin{equation}
    L(\zeta^j) = \sum_{i=0}^m \left\|w_{i+1}^j - w_i^j\right\|,
\end{equation}
where among the trajectories in $\mathcal{T}_{\leq n_w}^j$, there exists $\zeta^{\min,j}$ that has the minimum path length.
\begin{equation}
    \zeta^{\min,j} = \arg\min_{\zeta^j \in \mathcal{T}_{\leq n_w^j}}L(\zeta^j).
\end{equation}
In the absence of obstacles or additional constraints, $\zeta^{\min *,j} \in \mathcal{T}_0^j$ is the \textit{straight-line route} connecting $w_{\text{start}}^j$ and $w_{\text{end}}^j$.

The nominal missions are generated using the \ac{stg}. Depending on the specified \textit{encounter type}, the straight-line route $\zeta^{\min *,j}$ may intersect or overlap in a manner that produces head-on, crossing, or overtaking encounters between the own ship and the target ships. The complete set of initial waypoints and vessel states used to initialize an episode is referred to as a \textit{spawn request} or \textit{spawn case}.

A \textit{scenario} \(\mathcal{C}\) is then defined as the combination of the own ship trajectory $\zeta^0 \in \mathcal{T}^0$ and the target ships' trajectories within a single simulation:
\begin{equation}
\mathcal{C}=\left(\zeta^0,\zeta^1,\zeta^2,\ldots,\zeta^{n_{\mathrm{TS}}}\right).
\end{equation}
Accordingly, the overall scenario space is given by
\begin{equation}
\mathbb{C}
=
\mathcal{T}^0
\times
\prod_{j=1}^{n_{\mathrm{TS}}}
\mathcal{T}_{\leq n_w^j}^{j},
\end{equation}
where the Cartesian product indicates that one trajectory is selected for each target ship.

During a nominal encounter, the own ship relies on its onboard \ac{colav} system to avoid the target ships. However, the simplified \ac{colav} system used in this stress-testing experiment has a known limitation: it cannot simultaneously account for all target ships when multiple collision threats are present (see Section \ref{sec:simple_colav}). Instead, collision-avoidance actions are determined according to a sequential prioritization scheme. This behavior introduces a potential vulnerability in encounters involving multiple simultaneous threats.

The objective of the stress-testing method then is to identify the subset of scenarios \(\mathbb{C}^* \subseteq \mathbb{C}\) that exposes limitations of the onboard \ac{colav} system and results in a collision between the own ship and at least one target ship. This subset is defined as
\begin{equation}
\mathbb{C}^*=
\left\{\mathcal{C}\in\mathbb{C}\mid\exists j\in\{1,\ldots,n_{\mathrm{TS}}\}:\zeta^j \in\mathcal{T}_{\leq n_w^j}^{j*}
\right\}.
\end{equation}

Here, $\mathcal{T}_{\leq n_w^j}^{j*}$ is defined as the set of collision-inducing trajectories for each target ship
\begin{equation}
\mathcal{T}_{\leq n_w^j}^{j*}=
\left\{\zeta^j \in \mathcal{T}_{\leq n_w^j}^{j}\mid\Upsilon(\zeta^0,\zeta^j)\right\},
\end{equation}
where \(\Upsilon(\zeta^0,\zeta^j)\) indicates that the trajectory \(\zeta^j\) results in a collision involving target ship \(j\) and the own ship. Thus, a scenario belongs to \(\mathbb{C}^*\) if at least one of the target-ship trajectory is collision-inducing; the trajectories of the remaining target ships are not required to result in collisions. To identify such scenarios, the stress-testing method searches for waypoint sequences between \(w_{\text{start}}\) and \(w_{\text{end}}\) along each target ship's nominal route.

In addition to identifying failures for individual initial conditions, the stress-testing method is designed to produce a \textit{scenario generator} that can adapt to a diverse range of spawn cases. To achieve this, the stress-testing process is formulated as a \ac{rl} problem, in which the agent's policy serves as the scenario generator. A finite collection of traffic configurations is generated using the \ac{stg}, stored in a \textit{spawn-request bank}, and sampled during training. These configurations expose the agent to variations in vessel positions, headings, speeds, and encounter geometries, encouraging the policy to learn a generalizable strategy that can operate across multiple traffic configurations rather than overfitting to a single scenario.

Additional experiments are conducted to investigate whether further improvements can be achieved by continuing the training of an existing policy. The trained policy is subsequently used to stress-test different \ac{colav} implementations, allowing us to assess how effectively it can continue to generate collision-inducing scenarios. The proposed solution to this study case is our \ac{ebast2} algorithm, which is introduced in Section \ref{sec:ebast_v2_desc}.

%% file: Sections/Section_5.tex
    \section{\ac{ebast2}: An Enhanced Event-Based Adaptive Stress-Testing Framework for Co-Simulation}\label{sec:ebast_v2_desc}
Section~\ref{sec:ebast_v2_desc} presents the \ac{ebast2} algorithm, an improved iteration of the original \ac{ebast} method~\cite{sitorus_identifying_2025}. The section describes the proposed algorithm and discusses its key methodological improvements in detail.

\subsection{Proposed Algorithm and Improvements}
\begin{figure}[h!]
\begin{center}
    \includegraphics[width=.8\textwidth]{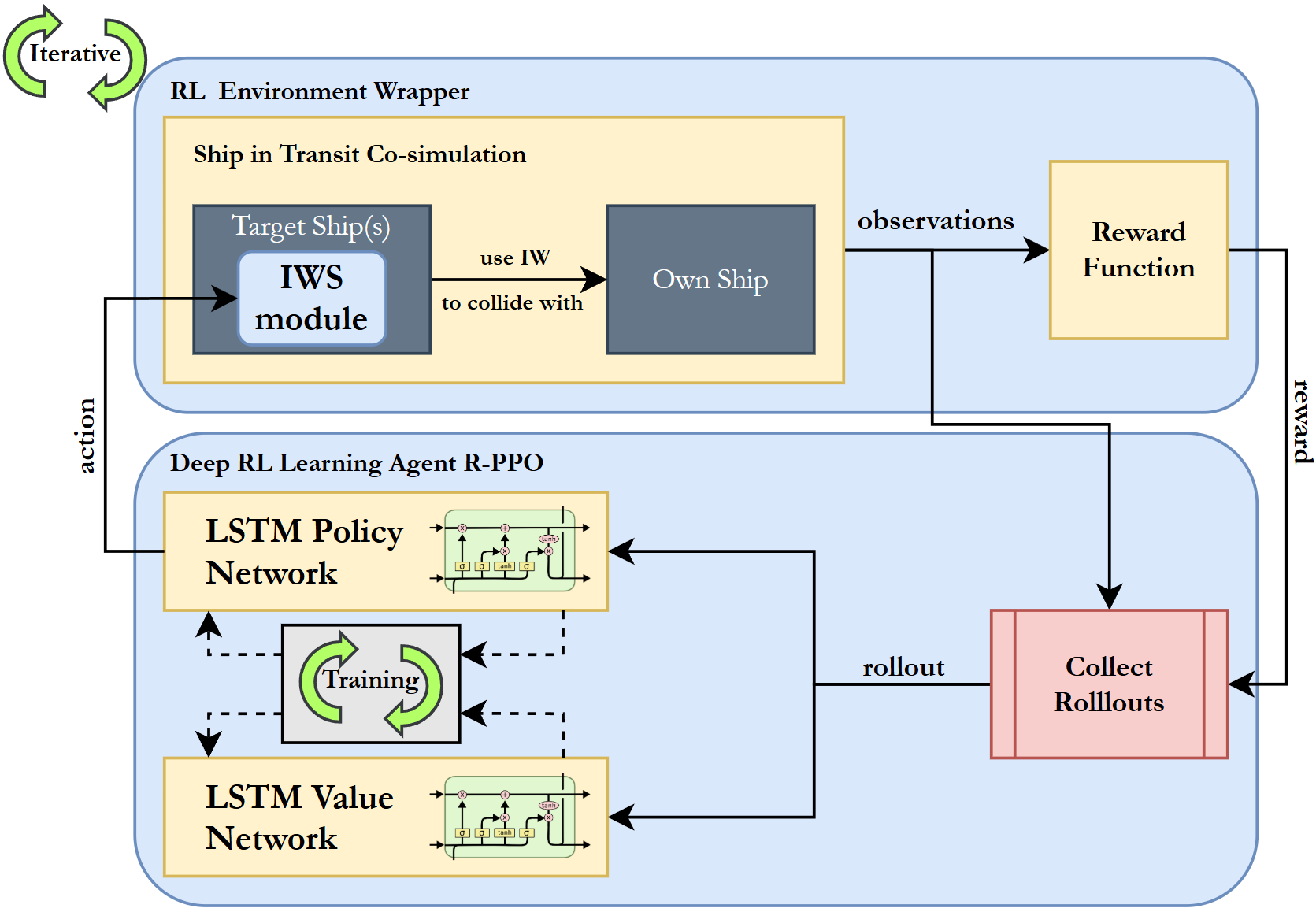}
    \caption{Process diagram of \ac{ebast2} method.}
    \label{fig:EBASTv2_diagram}
\end{center}
\end{figure}
We proposed a novel safety-assessment method specifically tailored for black-box \acp{cps}, where the framework diagram of this method is shown in Figure \ref{fig:EBASTv2_diagram}.  In this paper, we focused on using a maritime Ship-in-Transit co-simulation as a demonstrator. We employ \ac{rppo} algorithm as the primary solver algorithm because of its compatibility with \acp{pomdp}, which are often encountered in black-box simulation like co-simulation. The framework consists of two process blocks that interact with each other iteratively that forms an \ac{ebast2} update cycle process: the \textit{\ac{rl} environment wrapper} and the \textit{\ac{drl} agent}.

\ac{rl} environment wrapper is implemented as a Python wrapper that is built around the Ship-in-Transit co-simulation. The wrapper manages the simulation step function calling, reward signals computation, and relevant \ac{rl} states and termination states compilation for the \ac{drl} agent.  At each cycle, the wrapper accepts actions (or disturbances seen by the own ship) from the \ac{ebast2} policy and forwards it to the corresponding module that handles \ac{iws} mechanism for the target ships. The \ac{iws} mechanism is implemented within the \textit{Mission Manager} \ac{fmu} of each adversarial target ship, which modifies the ship’s trajectory by sampling \acp{iw} to stress-test the own ship’s \ac{colav} system. Once the state-action transition is complete, the wrapper returns observation and its corresponding reward to the \textit{\ac{drl} agent}.

The \ac{drl} agent contains the solver which takes the tuples of observation from the state-action transition along with its associated rewards and process it to improve the \ac{ebast2}'s policy. In this research, we employs the \ac{rppo} algorithm provided by \texttt{stable-baselines3}, an open-source library which supports variety of \ac{rl} algorithms \cite{raffin_stable_2021}. This library offers reliable and easy-to-use \ac{rl} implementations, well suited for research and \ac{rl}-related development. Following the \ac{rppo} algorithm, the \ac{drl} agent first collects rollout trajectories through repeated interactions with the simulation environment. These rollouts are then used to iteratively train the \ac{lstm}-based policy and value networks through the \ac{rppo} policy evaluation and policy improvement.

We specifically adopt the \textit{recurrent policy gradient methods} like \ac{rppo} rather than the \textit{history-based tree search} because of the need to learn a \textit{reusable policy} for a partially observable and continuous scenario space. While history-replay mechanism is effective for online search, it does not directly learn a reusable policy that generalizes across similar histories or scenario instances. In practice, \ac{drl} agent is not strictly tied to \ac{rppo}. It can hosts variety of \ac{rl} algorithms as long as it is compatible with \ac{pomdp}. 

\ac{ebast2} introduces several improvements over its predecessor, \ac{ebast}. In addition to its compatibility with \ac{pomdp} settings, \ac{ebast2} supports multi-adversary control, unlike \ac{ebast} single-adversary control limit. This allows the \ac{ebast2} agent to coordinate multiple target ships and generate more challenging encounter scenarios that may expose weaknesses in the own ship's \ac{colav} system. Furthermore, the \ac{iws} mechanism is enhanced by introducing a variable \ac{iw} sampling \textit{scope-length} as a second action component. The variable \ac{iw} sampling \textit{scope-length} allows the agent to adapt the scale of its intervention. Shorter scope-lengths enable local adjustments to the target ship trajectory, while longer scope-lengths allow larger deviations over a wider portion of the encounter scenario. This additional degree of freedom increases the flexibility of \ac{iw} sampling and enables a more precise exploration of the scenario space.

\subsection{\textit{Gymnasium} Environment for Reinforcement Learning}\label{sec:RL_env_gymnasium}
The \texttt{stable-baselines3} library expected the users to prepare their \ac{rl} environment based on \texttt{Gymnasium} API. \texttt{Gymnasium} API is a common interface for \ac{rl} environments, defining how and agent interacts with an standardized function calls that represents \ac{mdp} (or \ac{pomdp}) \cite{towers_gymnasium_2024}. These main function calls are \texttt{reset()}, \texttt{step()}, and \texttt{isTerminal()}.

Note that although Ship-in-Transit co-simulation are wrapped within the \ac{rl} environment, both entities operate different interpretations of a \textit{time-step}. In the co-simulation, time-step refers to the \textit{time interval} between consecutive numerical integration steps (e.g. Euler integration). In contrasts, a time-step in the \ac{rl} environment represents the \textit{interval between two consecutive events}. Consequently, the number of co-simulation integration steps between two \ac{rl} time steps may vary, as different events may occur after different amounts of simulated time. 

In this work, we define events as the occurence of one of the following conditions:
\begin{itemize}
    \item \textit{\textbf{Event 1}}: the co-simulation reaches the time threshold at which \ac{iws} is enabled;
    \item \textit{\textbf{Event 2}}: a target ship enters the \ac{roa} of its next waypoint;
    \item \textit{\textbf{Event 3}}: the co-simulation reaches a terminal state, either \textit{desired} or \textit{undesired}.
\end{itemize}

The \textit{desired terminal states}, denoted by $\Upsilon$, corresponds to outcomes of interest for the stress-testing process. These includes:
\begin{itemize}
    \item own ship collision between one ore more target ships,
    \item own ship prolonged deviation from its intended trajectories beyond a predefined time limit. The deviation boundary is based on the maximum allowable \textit{cross-track error} of the \ac{los} guidance system.
\end{itemize}

On the contrary, \textit{undesired terminal states} are denoted as $\neg\Upsilon$, correspond to terminal conditions that do not provide relevant information for the stress-testing objective. These include:
\begin{itemize}
    \item own ship reaches its mission endpoint,
    \item collision between target ships.
\end{itemize}

The \texttt{reset()} function are typically called when the environment is initialized. Upon calling this function, the Ship-in-Transit co-simulation is advanced until \textit{Event 1} is encountered for the first time. At this point, the agent receives the initial observation along with additional information regarding the current environment state.

The \texttt{step()} function executes the selected action from the agent's policy and updates the environment accordingly. An action is requested when either \textit{Event 1} or \textit{Event 2} is triggered. Subsequently, the co-simulation advances until another \textit{Event 2} occurs, or until \textit{Event 3} is reached. As a result, the agent receives a new observation along with a reward for taking the action.

The \texttt{isTerminal()} function returns two Boolean values, $\tau$ and $f$. The value $\tau$ is \texttt{TRUE} if the current simulator state $s$ is terminal, and \texttt{FALSE} otherwise. If $s$ is terminal, $f$ indicates whether the terminal state is \textit{desired}: $f=\texttt{TRUE}$ if $s\in\Upsilon$, and $f=\texttt{FALSE}$ if $s\notin\Upsilon$.

Additionally, the \ac{rl} environment also defines the structures such as the \textit{observation space} and \textit{action space}, allowing \ac{rl} libraries like \texttt{stable-baselines3} to work across different environments through the same interface. This will be further explained in the sections below.

\subsubsection{Observation Space}

The observation space $\Omega$ consists of the simulation states that are observable to the agent, as not all states of the Ship-in-Transit co-simulation are exposed to the agent. Throughout this paper, we are going to refer to the ``own ship'' and the ``target ship'' as ``OS'' and ``TS'' respectively. Given the number of adversarial target ships $n_\text{TS}$ and $j = 1, \cdots, n_\text{TS}$, this space consists of:
\begin{itemize}
\item own ship \ac{colav} throttle command: $\%^\text{COLAV}_t$,
\item own ship throttle command: $\%_t$,
\item own ship \ac{colav} rudder angle command: $\psi_{\text{rudder},_t}^\text{COLAV}$,
\item own ship rudder angle command: $\psi_{\text{rudder},t}$,
\item own ship positions: $N_{\text{OS},t}, E_{\text{OS},t}, \psi_{\text{OS},t}$,
\item own ship forward speed: $u_{\text{OS},t}$,
\item own ship cross-track error: $e_{\text{ct,OS},t}$,
\item target ships \textbf{\textit{relative positions}}: $N_\text{TS/OS}^1, E_\text{TS/OS}^1, \psi_\text{TS/OS}^1\cdots, N_\text{TS/OS}^j, E_\text{TS/OS}^j, \psi_\text{TS/OS}^j$,
\item target ships forward speed: $u^1, \cdots, u^j$,
\item \textit{\textbf{actions masks}}: $m_{t}^1, \cdots, m_{t}^j$.
\end{itemize}

The term \textit{relative position} refers to the difference between the position of a target ship and that of the own ship. Representing the target-ship positions relative to the own ship helps provide a more consistent observation representation across different encounter cases.

In addition, the \textit{action masks} $m$ are key observations that enable \textit{multi-adversary control}. Each action mask takes a binary value of either 0 or 1, indicating whether the corresponding target ship can currently be acted upon by the agent. The action-masking mechanism is further explained in Section \ref{sec:mac_through_action_masking}.

\subsubsection{Action Space}
The action space $\mathcal{A}$ comprises of the available actions for the agent to perform. In the \ac{ast} context, action here is also referred to as \textit{disturbance} These actions are the inputs for \ac{iws} mechanism, which are responsible for building the collision-inducing trajectories for the target ships. This space consists of:
\begin{itemize}
    \item scope-angle: $\psi^\text{sc,1}_t, \cdots, \psi^\text{sc,j}_t$,
    \item scope-length: $l^\text{sc,1}_t, \cdots, l^\text{sc,j}_t$.
\end{itemize}
where a sampled action from the policy $\pi$ is expressed as
\begin{equation}
    a_t = \left[
    \psi^\text{sc,1}_t, l^\text{sc,1}_t,
    \psi^\text{sc,2}_t, l^\text{sc,2}_t,
    \cdots, \psi^\text{sc,j}_t, l^\text{sc,j}_t \right].
\end{equation}
Further details on how these actions are used are shown in Section \ref{sec:IWS}.

\subsection{Intermediate Waypoint Sampling}\label{sec:IWS}
\ac{iws} was first developed as a novel action sampling strategy that enables the discovery of realistic collision-inducing trajectories for \ac{ebast} framework through a mechanism referred to as \textit{scoping}. Conceptually, the method can be compared to a marksman try to pinpoint a \textit{target} by adjusting its rifle angle and estimating the target's distance. In this way, \ac{iws} provides lower-resolution control over the adversarial target ship's trajectory, allowing a broader search space to be explored while reducing the agent's aggressiveness and tendency to deliberately pursue collisions.

We first define two helper functions that will be used extensively for \ac{iws}. The first one is \texttt{getHypot(}$A,B$\texttt{)}, used to compute the hypotenuse between point $A$ and $B$, given by
\begin{equation}
    l_{AB} = \texttt{getHypot(}A,B\texttt{)}=\sqrt{(N_A-N_B)^2+(E_A-E_B)^2}.
\end{equation}
The second one is \texttt{getBeta(}$A,B$\texttt{)}, used to compute the angle $\beta$ between the north-axis and the line segment formed by connecting $A$ and $B$. This function is formulated as
\begin{equation}
    \beta_{AB} = \texttt{getBeta(}A,B\texttt{)}= \arctan\left(\frac{E_A-E_B}{N_A-N_B}\right).
\end{equation}

Each target ship is initialized with a waypoint sequence $\zeta$, containing the initial waypoints $w_{\mathrm{start}}$ and $w_{\mathrm{end}}$. From this sequence, a waypoint pair $W=(w_\mathrm{base} , w_\mathrm{head})$ is defined and used by the \ac{los} guidance system to provide a heading reference to the target ship's autopilot. The pair is retrieved using the function \texttt{getWaypointPairs(}$i, \zeta$\texttt{)}, such that
\begin{equation}
W = \left(\zeta(i-1), \zeta(i)\right),
\end{equation}
where $i=1,2,\ldots,n$ denotes the index of the current waypoint in $\zeta$. As the agent encountered a new event, the \texttt{advanceWaypointPairs(}$i$\texttt{)} is called to increment $i$ by 1 to advance the waypoint pairing.  To illustrate how the waypoint pair $W$ relates to the current state of the trajectory, consider the initial waypoint sequence before any \acp{iw} are introduced.
\begin{equation}
    \zeta = (\tikzmark{a}{w_{\text{start}}}, \tikzmark{b}{w_{\text{end}}}),\quad i=1.
\end{equation}
\begin{tikzpicture}[remember picture,overlay]
\draw[<-] 
  ([shift={(2pt,-2pt)}]pic cs:a) |- ([shift={(-10pt,-10pt)}]pic cs:a) 
  node[anchor=east] {$\scriptstyle w_\mathrm{base}$}; 
\draw[<-] 
  ([shift={(2pt,-2pt)}]pic cs:b) |- ([shift={(14pt,-10pt)}]pic cs:b) 
  node[anchor=west] {$\scriptstyle w_\mathrm{head}$}; 
\end{tikzpicture}

The \ac{iws} starts immediately after Event 1 is triggered (see Section \ref{sec:RL_env_gymnasium}), which automatically calls \texttt{initiateIWS()} function. This function insert the \textit{zeroeth} \ac{iw} $w_0$ into $\zeta$ based on the target ship's north-east position when Event 1 is first triggered. Hence the current $\zeta$ becomes

\begin{equation}
    \zeta = (\tikzmark{c}{w_{\text{start}}}, \tikzmark{d}w_0, {w_{\text{end}}}),\quad i=1.
\end{equation}
\begin{tikzpicture}[remember picture,overlay]
\draw[<-] 
  ([shift={(2pt,-2pt)}]pic cs:c) |- ([shift={(-10pt,-10pt)}]pic cs:c) 
  node[anchor=east] {$\scriptstyle w_\mathrm{base}$}; 
\draw[<-] 
  ([shift={(2pt,-2pt)}]pic cs:d) |- ([shift={(14pt,-10pt)}]pic cs:d) 
  node[anchor=west] {$\scriptstyle w_\mathrm{head}$}; 
\end{tikzpicture}

The zeroeth \ac{iw} will then be used as the \textit{scoping base} to compute the next \ac{iw} using \texttt{getNextWaypoint(}$i, \zeta, \psi^\mathrm{sc}_t, l^\mathrm{sc}_t, \beta$\texttt{)}. The algorithm goes as follow:
\begin{enumerate}
    \item Retrieve the-scope length $l^\mathrm{sc}_t$ and scope-angle $\psi^\mathrm{sc}_t$ from the policy, where the scope-angle is constrained so that $\psi^\mathrm{sc}_t \leq |\psi^\mathrm{max}|$.
    \item Evaluate the sampled scope-length using \texttt{evaluateScopeLength()}.
    \item Retrieve the current waypoint pair $W_i$ using \texttt{getWaypointPairs(}$i, \zeta$\texttt{)}.
    \item Compute the $l_{W_i}$ and $\beta_{W_i}$ using \texttt{getHypot(}$W_i$\texttt{)} \texttt{getBeta(}$W_i$\texttt{)}.
    \item Set the $w_{\mathrm{head},i}$ coordinates as the base point for the next waypoint $w_{\mathrm{head},i+1}$.
    \item The next waypoint $w_{i+1} = (N_{i+1}, E_{i+1})$ can then be computed by solving
    \begin{align}
        N_{\mathrm{head},i+1} &= N_{\mathrm{head},i} + l_{W_i}\cos(\beta_{W_i}+\psi^\mathrm{sc}_t) \\ 
        E_{\mathrm{head},i+1} &= E_{\mathrm{head},i} + l_{W_i}\sin(\beta_{W_i}+\psi^\mathrm{sc}_t) .
    \end{align}
    \item Insert $w_{i+1}$ into $\zeta$ at index $i+1$ using \texttt{insertNewWaypoint(}$i, \zeta$\texttt{)} function.
    \item Finally, advance the waypoint pairs using \texttt{advanceWaypointPairs(}$i$\texttt{)}.
\end{enumerate}
Having called this function, the condition of the trajectory $\zeta$ should follows
\begin{equation}
    \zeta = ({w_{\text{start}}}, \tikzmark{g}w_0, \tikzmark{h}w_1,  {w_{\text{end}}}),\quad i=2.
\end{equation}
\begin{tikzpicture}[remember picture,overlay]
\draw[<-] 
  ([shift={(2pt,-2pt)}]pic cs:g) |- ([shift={(-10pt,-10pt)}]pic cs:g) 
  node[anchor=east] {$\scriptstyle w_\mathrm{base}$}; 
\draw[<-] 
  ([shift={(2pt,-2pt)}]pic cs:h) |- ([shift={(14pt,-10pt)}]pic cs:h) 
  node[anchor=west] {$\scriptstyle w_\mathrm{head}$}; 
\end{tikzpicture}

\begin{figure}[htbp]
    \centering
    \begin{minipage}{0.465\textwidth}
        \centering
        \includegraphics[width=\textwidth]{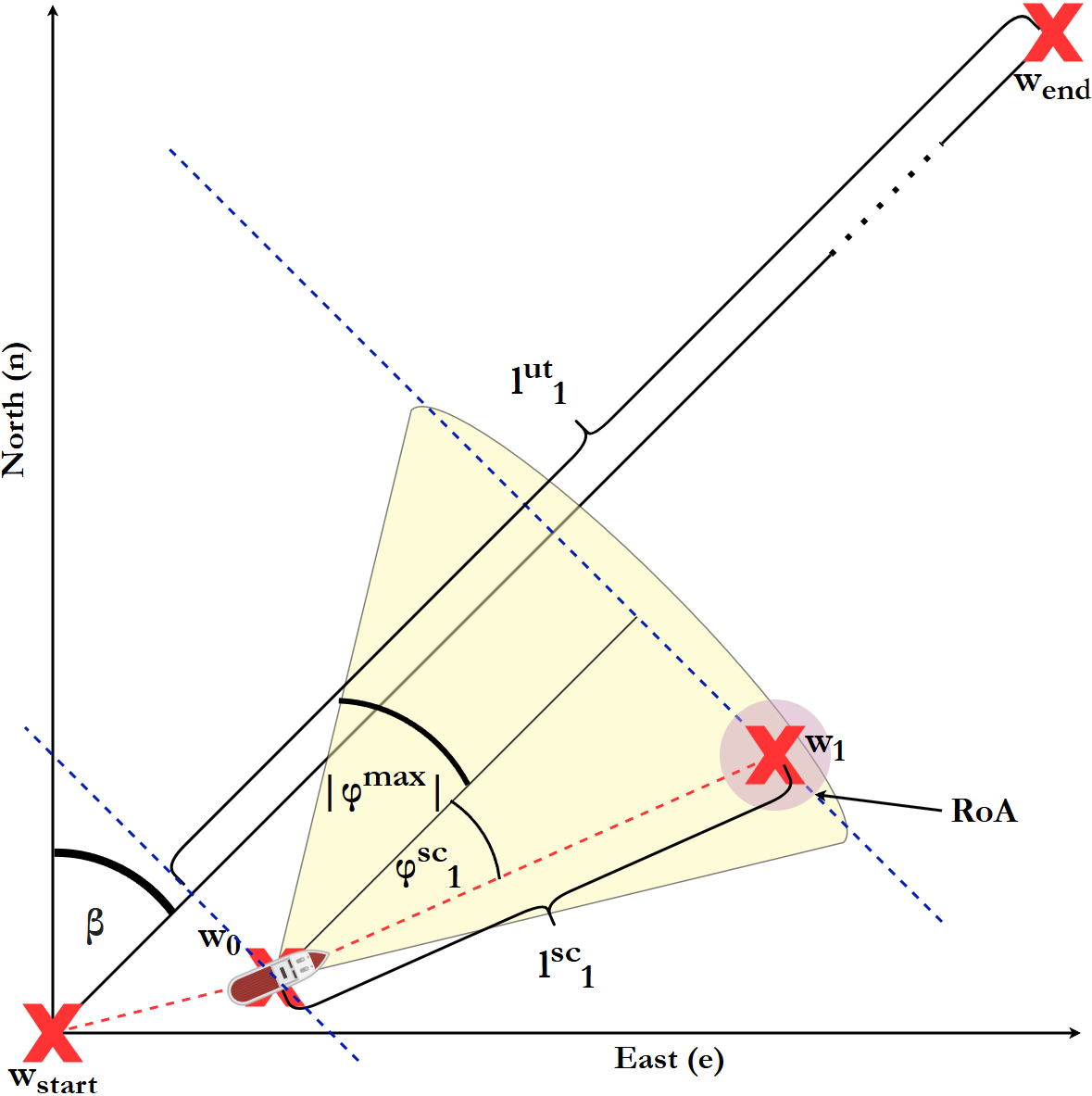}
        \caption{$w_0$ placement \& computing $w_1$.}
        \label{fig:IWS_new_1}
    \end{minipage}
    \hfill
    \begin{minipage}{0.52\textwidth}
        \centering
        \includegraphics[width=\textwidth]{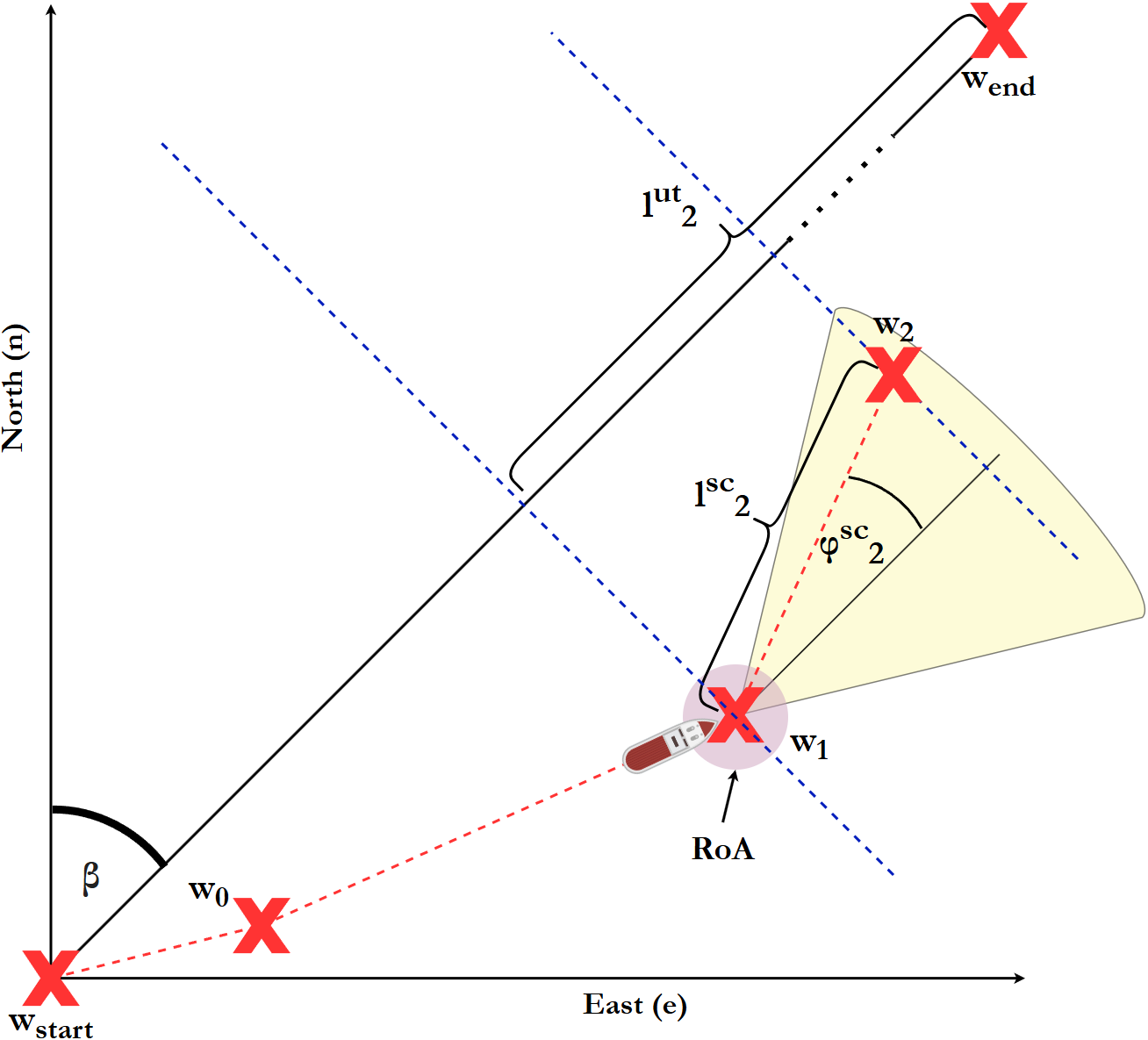}
        \caption{Computing $w_2$.}
        \label{fig:IWS_new_2}
    \end{minipage}
\end{figure}

Furthermore, the \texttt{getNextWaypoint()} function can be called repeatedly as long as Event 2 continues to be triggered by the target ships. This produces the updated trajectory condition defined in Equation \ref{eq:second_IWS}. In addition, the overall process is illustrated in Figures \ref{fig:IWS_new_1} and \ref{fig:IWS_new_2}.

\begin{equation}\label{eq:second_IWS}
    \zeta = ({w_{\text{start}}}, w_0, \tikzmark{k}w_1, \tikzmark{l}w_2,  {w_{\text{end}}}),\quad i=3
\end{equation}
\begin{tikzpicture}[remember picture,overlay]
\draw[<-] 
  ([shift={(2pt,-2pt)}]pic cs:k) |- ([shift={(-10pt,-10pt)}]pic cs:k) 
  node[anchor=east] {$\scriptstyle w_\mathrm{base}$}; 
\draw[<-] 
  ([shift={(2pt,-2pt)}]pic cs:l) |- ([shift={(14pt,-10pt)}]pic cs:l) 
  node[anchor=west] {$\scriptstyle w_\mathrm{head}$}; 
\end{tikzpicture}

Typically, the \ac{iws} mechanism automatically stop once the action-sampling counts reach the predetermined limit of $n_w$. However, there are special cases in which \ac{iws} is forced to stop even before the former condition is not fulfilled. When the sampled scope-length $l^\mathrm{sc}_t$ is greater than the length of the remaining untraversed segment-length $l^\mathrm{ut}_i$, the next waypoint will not be computed and the last sampled \ac{iw} will automatically connect to the end waypoint $w_\mathrm{end}$. Thus, the updated trajectory condition becomes
\begin{equation}
    \zeta = ({w_{\text{start}}}, w_0, w_1, \tikzmark{m}w_2,  \tikzmark{n}{w_{\text{end}}}),\quad i=4
\end{equation}
\begin{tikzpicture}[remember picture,overlay]
\draw[<-] 
  ([shift={(2pt,-2pt)}]pic cs:m) |- ([shift={(-10pt,-10pt)}]pic cs:m) 
  node[anchor=east] {$\scriptstyle w_\mathrm{base}$}; 
\draw[<-] 
  ([shift={(2pt,-2pt)}]pic cs:n) |- ([shift={(14pt,-10pt)}]pic cs:n) 
  node[anchor=west] {$\scriptstyle w_\mathrm{head}$}; 
\end{tikzpicture}
\begin{figure}[h!]
\begin{center}
    \includegraphics[width=.59\textwidth]{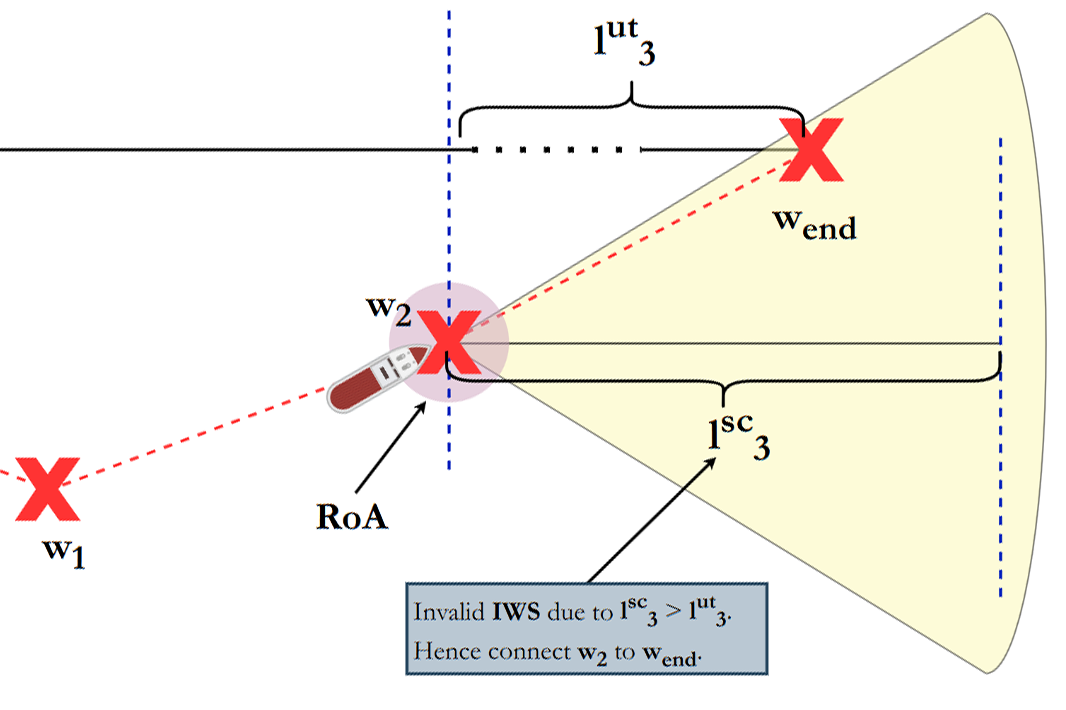}
    \caption{Invalid \ac{iws} for $w_3$. Close the trajectory.}
    \label{fig:IWS_new_3}
\end{center}
\end{figure}

This case takes place only when the \texttt{evaluateScopeLength()} function returns \texttt{TRUE}, as illustrated in Figure \ref{fig:IWS_new_3}. The maximum scope angle, $\psi_\mathrm{max}$, together with the scope-length evaluation, enables the generation of trajectories that remain flexible while maintaining forward progression, therefore supporting the objectives of \ac{ebast2}.

\subsubsection{Multi-Adversary Control through Action Masking}\label{sec:mac_through_action_masking}
\begin{figure}[h!]
\begin{center}
    \includegraphics[width=.75\textwidth]{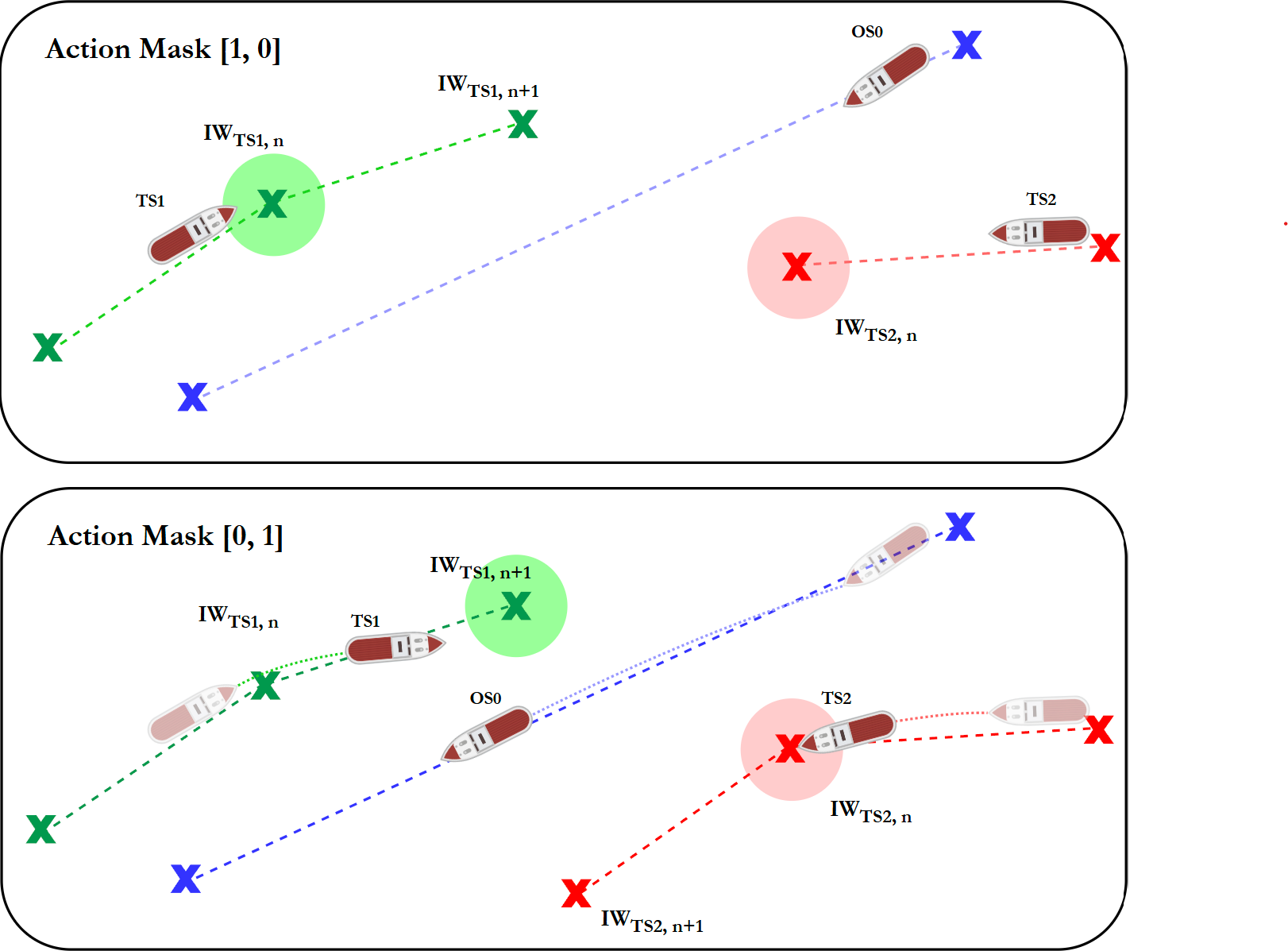}
    \caption{Action masking process. Value of $1$ is given to the Target Ship(s) that is allowed to sample new \ac{iw}.}
    \label{fig:action_masking}
\end{center}
\end{figure}
Due to the structure of the \ac{ebast2} action space $\mathcal{A}$, the policy cannot sample an action exclusively for a particular \ac{iw}-enabled target ship. Instead, an action is sampled simultaneously for all target ships represented in the action space. This design choice is intended because two or more target ships may trigger events at the same time, hence a ready supply of action is required at all decision events. In contrast, the multi-adversary control scheme allows each target ship to trigger a decision event independently of the others. Consequently, an event triggered by one target ship also causes the policy to produce actions for target ships that did not trigger an event. Although these additional actions do not result in trajectory modifications, they are still produced by the policy and may therefore introduce irrelevant action--observation associations during training.

Zouitine in his article \cite{zouitine_masking_2022} discusses three approaches for incorporating action masking: \textit{action-level} masking,  \textit{feature-level} masking and \textit{agent-level} masking. Action-level masking works by altering the value associated to the \textit{impossible action} to $-\infty$ . Feature-level masking controls which information is available to an agent when constructing its internal state representation. Agent-level masking determines which agents contribute to the policy-loss computation in a multi-agent setting.

To address the asynchronous event triggering in \ac{ebast2}, we adopt a mechanism conceptually closer to feature-level masking. Rather than suppressing individual action values or dimensions, an additional observation vector is introduced to indicate which target ships triggered the current decision event. Let
\begin{equation}
    \mathbf{m}_t =
    \left[m_t^{1},m_t^{2},\ldots,m_t^{|\mathcal{I}_t|}\right],
    \qquad
    m_t^j \in \{0,1\},
\end{equation}
denote this event-activity mask, where each entry corresponds to one \ac{iw}-enabled target ship $j$. An entry is set to $m_t^j=1$ when the corresponding target ship triggers an event and to $m_t^j=0$ otherwise. Since target ships trigger events independently, multiple entries may simultaneously take the value $1$ at a given decision event.

The mask $\mathbf{m}_t$ is included in the observation supplied to the policy. It provides an explicit indication of which target-ship-specific action components are active at the current decision event. Through training, the policy can associate entries with $m_t^j=1$ with actions that affect the corresponding target ship, while actions associated with $m_t^j=0$ are inactive and do not modify that ship's trajectory.

\subsection{Reward Function} \label{sec:reward_function}

The reward function provides the learning signal used to guide the \ac{ebast2} policy toward failure-inducing scenarios. Its design follows the general \ac{ast} formulation introduced in Equation~\ref{eq:reward_function_pomdp}, while incorporating additional shaping terms that are specific to the \ac{iws} mechanism.

At each decision event, the reward is determined by whether the episode has terminated and whether the termination corresponds to a desired failure event. Let $\tau_t \in \{0,1\}$ indicate episode termination and $f_t \in \{0,1\}$ indicate whether the desired failure event has occurred. The reward is then expressed as
\begin{equation}
\mathcal{R}_t =
\begin{cases}
R_{\Upsilon},
    & \tau_t \land f_t, \\[2mm]
R_{\neg\Upsilon}(d_f),
    & \tau_t \land \neg f_t, \\[2mm]
R_{\mathrm{NT},t},
    & \neg\tau_t ,
\end{cases}
\label{eq:ebast2_reward}
\end{equation}
where 
\begin{align}
    R_{\neg\Upsilon}(d_f) &=
    \frac{1}{|\mathcal{I}_t|}
    \sum_{j\in\mathcal{I}_t}
    R_{\neg\Upsilon}^{j}\left(d_f^{j}\right)\\
    R_{\mathrm{NT},t} &=
    \frac{1}{|\mathcal{I}_t|}
    \sum_{j\in\mathcal{I}_t}
    R_{\mathrm{NT},t}^{j}\left(d_f^{j},d_{n,t}^{j}, n_w^{j}, c_{w,t}^{j}, \Delta\psi^\text{sc,j}_t, \Delta\psi^\text{sc,j}_\text{ic,t}\right)
\end{align}

$R_{\Upsilon}$ denotes the reward for a desired termination, $R_{\neg\Upsilon}$ denotes the reward for an undesired termination, and $R_{\mathrm{NT},t}$ denotes the non-terminal reward. Here, ``desired'' refers to the objective of the stress-testing agent; from the perspective of the system under test, such an event represents a failure.

Moreover, $d_f$ denotes the final own-ship--target-ship separation at an undesired terminal event, while $d_{n,t}$ denotes the minimum separation observed up to decision event $t$. The variable $n_w$ specifies the maximum number of \acp{iw} that may be sampled for a target ship, and $c_{w,t}\in\{0,\ldots,n_w\}$ denotes the number already sampled. The counter is initialized to zero at the beginning of each episode and incremented after every successful \ac{iw} request. Furthermore, $\Delta\psi^{\mathrm{sc}}_t$ denotes the change between consecutive scope-angle commands, whereas $\Delta\psi^{\mathrm{sc}}_{\mathrm{ic},t}$ denotes the angular difference between the sampled scope angle and the interception scope angle.

Note that both $R_{\neg\Upsilon}$ and $R_\mathrm{NT},t$ are obtained by averaging of reward contributions from the \ac{iw}-enabled target ships contained in the set $\mathcal{I}_t$, hence the superscript $j$. Combined with the action masking protocol, this formulation enables the policy to control multiple adversarial target ships simultaneously. For brevity, the superscript $j$ is omitted from the remaining equations in this section unless otherwise stated.

\begin{figure}[htbp]
    \centering
    \begin{minipage}{0.32\textwidth}
        \centering
        \includegraphics[width=\linewidth]{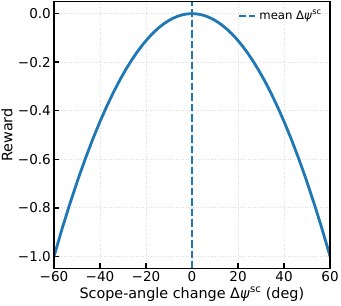}
        \caption{Scope-angle change reward function $F_{\ell\ell}$.}
        \label{fig:s-ac_reward}
    \end{minipage}
    \hfill
    \begin{minipage}{0.32\textwidth}
        \centering
        \includegraphics[width=\linewidth]{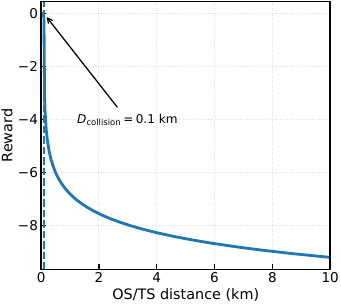}
        \caption{Miss-distance reward function $F_\rho$.}
        \label{fig:md_reward}
    \end{minipage}
    \hfill
    \begin{minipage}{0.32\textwidth}
        \centering
        \includegraphics[width=\linewidth]{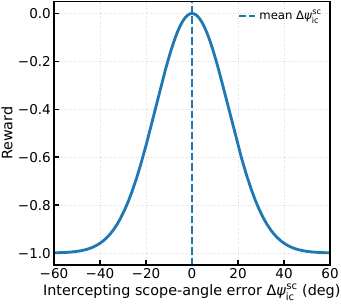}
        \caption{Intercepting scope-angle reward function $F_\Lambda$.}
        \label{fig:is-c_reward}
    \end{minipage}
\end{figure}

For a desired termination, the terminal reward is set to
\begin{equation}
R_{\Upsilon}=0.
\end{equation}
If the episode terminates without reaching the desired failure event, the agent instead receives a distance-based penalty
\begin{equation}
R_{\neg\Upsilon}(d_f)
=
C_{\rho}F_{\rho}(d_f),
\label{eq:undesired_terminal_reward}
\end{equation}
where $C_{\rho}$ is a scaling coefficient. The distance transformation is defined as
\begin{equation}
F_{\rho}(D)
=
-\log\left(1+(D-D_{\mathrm{collision}})\right),
\label{eq:distance_reward}
\end{equation}
where $D_{\mathrm{collision}}$ denotes the separation threshold corresponding to collision. This function is illustrated in Figure \ref{fig:md_reward}.

When the episode remains active, the reward serves two complementary purposes. First, it \textit{discourages large successive changes} in the sampled scope angle, thereby promoting smoother target-ship trajectories. Second, it \textit{guides the sampled scope angle towards an interception course} when such an interception is kinematically feasible. Accordingly, the non-terminal reward is decomposed as
\begin{equation}
R_{\mathrm{NT},t}
=
R_{\Delta\psi,t}
+
R_{\Lambda,t},
\label{eq:nonterminal_reward}
\end{equation}
where $R_{\Delta\psi,t}$ is the \textit{scope-angle-change term} and $R_{\Lambda,t}$ is the \textit{interception-guidance term}.

The scope-angle-change component is defined as
\begin{equation}
R_{\Delta\psi,t}
=
C_{\ell\ell}
\frac{c_{w,t}}{n_w}
F_{\ell\ell}\left(\Delta\psi^{\mathrm{sc}}_t\right),
\label{eq:scope_change_reward}
\end{equation}
where
\begin{equation}
\Delta\psi^{\mathrm{sc}}_t
=
\psi^{\mathrm{sc}}_t
-
\psi^{\mathrm{sc}}_{t-1}
\end{equation}
is the change between two consecutive sampled scope angles. The ratio $c_{w,t}/n_w$ gradually increases the contribution of this term as additional \acp{iw} are sampled. Additionally, $C_{\ell\ell}$ is another scaling factor.

The function $F_{\ell\ell}$, as shown in Figure \ref{fig:s-ac_reward},  penalizes large changes in scope angle. The change is evaluated under a zero-centered truncated normal distribution, such that small changes are assigned higher likelihood than abrupt changes. Let
\begin{equation}
\ell(x)=\log F_{\mathrm{TN}}(x)
\end{equation}
denote the corresponding log-likelihood. The normalized scope-angle change function is then
\begin{equation}
F_{\ell\ell}\left(\Delta\psi^{\mathrm{sc}}\right)
=
\frac{
    \ell\left(\Delta\psi^{\mathrm{sc}}\right)-\ell(0)
}{
    \ell(0)-\ell\left(\Delta\psi^{\mathrm{sc}}_{\max}\right)
},
\label{eq:scope_change_function}
\end{equation}

where $F_{\mathrm{TN}}$ is the probability density function of a doubly truncated normal distribution \cite{johnson_distributions_1994},
\begin{equation}
F_{\mathrm{TN}}(x)
=
\frac{
    \dfrac{1}{\sigma_{\mathrm{TN}}}
    \phi\left(
        \dfrac{x-\mu_{\mathrm{TN}}}{\sigma_{\mathrm{TN}}}
    \right)
}{
    \Phi\left(
        \dfrac{b-\mu_{\mathrm{TN}}}{\sigma_{\mathrm{TN}}}
    \right)
    -
    \Phi\left(
        \dfrac{a-\mu_{\mathrm{TN}}}{\sigma_{\mathrm{TN}}}
    \right)
},
\qquad
a\leq x\leq b,
\label{eq:truncated_normal}
\end{equation}
where $\phi(\cdot)$ and $\Phi(\cdot)$ denote the probability density function and cumulative distribution function of the standard normal distribution, respectively. $F_{\ell\ell}$ plot is given in Figure \ref{fig:s-ac_reward}.

The interception-guidance component is defined as
\begin{equation}
R_{\Lambda,t}
=
\begin{cases}
\displaystyle
\left(
\frac{n_w+1-c_{w,t}}{n_w}
\right)
C_{\Lambda}
F_\Lambda(\Delta\psi^{\mathrm{sc}}_{\mathrm{ic},t}),
&
\text{if interception is feasible},
\\[4mm]
C_{\mathrm{pass}}
F_{\rho}(d_{n,t}),
&
\text{otherwise},
\end{cases}
\label{eq:interception_reward}
\end{equation}
where $C_\Lambda$ and $C_\text{pass}$ scale their respective reward term, and  $F_\Lambda$ acts as a interception-guiding function. $F_\Lambda$ takes inspiration from the first reward design variation based on Goto \textit{et al.} work \cite{goto_design_2023}. This function is illustrated in Figure \ref{fig:is-c_reward} and formulated as
\begin{equation}
F_\Lambda(\Delta\psi^{\mathrm{sc}}_{\mathrm{ic},t}) = \exp\left(
-\frac{
\left|\Delta\psi^{\mathrm{sc}}_{\mathrm{ic},t}\right|
}{
\kappa_{\Lambda}
}
\right).
\end{equation}

This second non-terminal reward term encourages the target ship to select scope angles that are consistent with an interception course through $F_\Lambda(\Delta\psi^{\mathrm{sc}}_{\mathrm{ic},t})$. However, it also \textit{fairly} penalizes the remaining \ac{iw} samplings when any interception attempts are no longer feasible. This is done by giving the target ship a reward based on \textit{how close it was able to get to the own ship} through $F_\rho(d_{n,t})$. At each decision event, a geometrically feasible interception heading is computed from the current relative positions and velocities of the own ship and target ship. The angular discrepancy between the sampled scope angle $\psi^{\mathrm{sc}}_t$ and the corresponding \textbf{interception heading} $\psi^{\mathrm{ic}}_t$ is denoted by
\begin{equation}
\Delta\psi^{\mathrm{sc}}_{\mathrm{ic},t}
=
\psi^{\mathrm{ic}}_t
-
\psi^{\mathrm{sc}}_t.
\end{equation}

The interception heading prediction formula will be presented in Section \ref{sec:interception_heading_prediction}. Unlike the scope-angle-change term, the weighting of the interception component decreases as additional waypoints are sampled. Consequently, interception guidance dominates the early part of an episode, whereas trajectory smoothness becomes increasingly important as the target ship progresses through its available waypoint requests.

%% file: Sections/Section_6.tex
\section{Results}
\subsection{Training Parameters}
For the training process, we includes two \ac{iw}-enabled target ships as the \ac{ast} stressor. Each of target ship is equipped with \ac{iw} \ac{fmu} that allows it to change trajectory mid simulation. The parameter settings of the module for all target ships are shown in Table \ref{tab:IWS_parameters}.
\begin{table}[ht] 
\centering 
\caption{\ac{iws} module parameters, for all target ships.} 
\label{tab:IWS_parameters} 
    \begin{tabular}{lc} 
        \toprule 
        \textbf{Parameter}          & \textbf{Value} \\
        \midrule 
        \textit{\textbf{Event 1}} start time & $250$ seconds \\ 
        \ac{roa}                    & $300$ meters \\
        Maximum \ac{iws} counts     & 5 \\
        Maximum $\psi^\mathrm{sc}$  & $30.0^\circ$\\
        \bottomrule 
    \end{tabular} 
\end{table}

\begin{center}
\begin{minipage}[t]{0.48\textwidth}
    \centering
    \captionof{table}{\ac{rppo} and \ac{ppo} training hyperparameters for \ac{ebast2}.}
    \label{tab:ebast2_rppo_parameters}
    \small
    \begin{tabular*}{\linewidth}{@{\extracolsep{\fill}}ll@{}}
        \hline
        \textbf{Parameter} & \textbf{Value} \\
        \hline
        Learning rate              & $3\times10^{-4}$ \\
        Number of environments     & 64 \\
        Rollout steps              & 512 \\
        Batch size                 & 1024 \\
        Number of epochs           & 5 \\
        Discount factor ($\gamma$) & 1.0 \\
        Clipping range             & 0.2 \\
        Entropy coefficient        & 0.005 \\
        Value function coefficient & 0.5 \\
        Maximum gradient norm      & 0.5 \\
        Advantage normalization    & True \\
        Total timesteps            & $10^7$ \\
        \hline
    \end{tabular*}
\end{minipage}
\hfill
\begin{minipage}[t]{0.48\textwidth}
    \centering
    \captionof{table}{\ac{sac} training hyperparameters for \ac{ebast2}.}
    \label{tab:ebast2_sac_parameters}
    \small
    \begin{tabular*}{\linewidth}{@{\extracolsep{\fill}}ll@{}}
        \hline
        \textbf{Parameter} & \textbf{Value} \\
        \hline
        Learning rate                   & $3\times10^{-4}$ \\
        Number of environments          & 64 \\
        Replay buffer size              & $1.28\times10^{6}$ \\
        Learning starts                 & $1.28\times10^{5}$ \\
        Batch size                      & 1024 \\
        Discount factor ($\gamma$)      & 1.0 \\
        Soft update coefficient ($\tau$)& 0.005 \\
        Training frequency              & 10 \\
        Gradient steps                  & 128 \\
        Entropy coefficient             & Auto \\
        Total timesteps                 & $10^7$ \\
        \hline
    \end{tabular*}
\end{minipage}
\end{center}

The training parameters for each \ac{rl} algorithm are presented in Table \ref{tab:ebast2_rppo_parameters} and Table \ref{tab:ebast2_sac_parameters}. \ac{rppo} and \ac{ppo} share the same training because inherently they are derived from the same algorithm. The difference is that \ac{rppo} uses \texttt{MultiInputLstmPolicy} to enable \ac{pomdp}, whereas both \ac{ppo} and \ac{sac} use \texttt{MultiInputPolicy} due to their \ac{mdp} property. We aim to establish comparable training conditions across all \ac{rl} algorithms to ensure a fair comparison. Specifically, we use the same \textit{learning rate}, \textit{number of environments}, \textit{discount factor} $\gamma$, and \textit{total number of timesteps} across all algorithms. We pick $\gamma$ value of $1.0$ because the failure-discovery problem is considered as a \textit{finite state-action transition} problem. Additionally, we enable parallel training by using \textit{multiple environments} to increase transition data gathering and speed up the training process. We could also identify how many neural network updates for each \ac{rl} core based on the parameters, where the training happens differently for on-policy and off-policy method. 

For \ac{rppo} and \ac{ppo}, training was performed per \textit{rollout}, with each rollout spanning $512$ timesteps across $64$ parallel \textit{environments}. Therefore, each rollout collects $512\times64=32{,}768$ transitions. Given a \textit{batch size} of $1{,}024$, the number of \textit{minibatches} per rollout is
$$\frac{32{,}768}{1024}=32.$$
Since each rollout is optimized for $5$ epochs, the number of neural network updates per rollout is
$$32\times5=160.$$
For a total training budget of $10^7$ transitions, the number of rollouts is approximately
$$\frac{10^7}{32{,}768}\approx305.18.$$
Thus, the total number of neural network updates for each \ac{rppo} and \ac{ppo} training run is approximately
$$160\times305.18\approx48{,}828$$updates. Thus, over $10$ million timesteps, \ac{rppo} and \ac{ppo} performs a total of $\mathbf{48{,}828}$ \textbf{gradient updates}.

For \ac{sac}, training is performed after every \textit{training frequency} of $10$ environment steps once the initial exploration phase, specified by \textit{learning starts}, has ended. With $64$ parallel environments, each training frequency collects $10\times64=640$ transitions, which are stored in the replay buffer. After each collection cycle, \ac{sac} performs $128$ gradient updates, with each update using a randomly sampled minibatch of $1024$ transitions from the replay buffer.
Over a total of $10$ million timesteps, the number of collection cycles is
$$\frac{10^7}{640}=15{,}625.$$
The exploration phase lasts for $128{,}000$ timesteps, corresponding to
$$\frac{128{,}000}{640}=200$$
collection cycles during which no gradient updates are performed. Therefore, the total number of training cycles is
$$15{,}625-200=15{,}425.$$
Since each training cycle consists of $128$ gradient updates, the total number of gradient updates is
$$15{,}425\times128=1{,}974{,}400.$$
Thus, over $10$ million timesteps, \ac{sac} performs a total of $\mathbf{1{,}974{,}400}$ \textbf{gradient updates}.

We train the policy using a 100 traffic cases generated with \ac{stg}. Each traffic case centered on the own ship, whose initial position is located at north-east coordinate of (0, 0). For each generated case, \ac{stg} parameters are sampled within their respective allowed range. which are specified in a JSON file named \texttt{encounter\_settings}. The file's contents are provided in Appendix Section \ref{sec:stg_parameters}.  At each call to \texttt{reset()}, the \ac{rl} environment wrapper randomly selects one of the 100 traffic cases. The cases are indexed from $0$ to $99$, and the selection is performed using a uniform distribution, such that each case has an equal probability of being selected.

\subsection{\ac{rl} Cores' Performance Comparisons}\label{sec:cores_comparison}
In this section, we discuss each \ac{rl} core's performances by comparing their reward collection and episode length trends over the $10$ million training timesteps. We evaluate \ac{rppo}'s performance by comparing its results directly with \ac{ppo} and \ac{sac} in the following sections. Note that the maximum episode length will be $10$ given that each target ship only allowed to sample $5$ \acp{iw} per episode.

\begin{table}[htbp] 
\centering 
\caption{\ac{rppo} policy's failure scenario generation training score.} 
\label{tab:eval_rppo} 
    \begin{tabular}{lccc} 
        \toprule 
        \textbf{Outcome} & \textbf{Count} & \textbf{Percentage} & \textbf{Mean Episode Length} \\ 
        \midrule 
        No Collision & 32 / 100 & 32.00\% & 6.6 \\ 
        Target Ship Collision & 0 / 100 & 0.00\% & 0.0 \\ 
        \textbf{Own Ship Collision} & \textbf{68 / 100} & \textbf{68.00\%} & 5.6 \\ 
        \bottomrule 
    \end{tabular} 
\end{table}

\begin{table}[htbp] 
\centering 
\caption{\ac{ppo} policy's failure scenario generation training score.}
\label{tab:eval_ppo} 
    \begin{tabular}{lccc} 
        \toprule
        \textbf{Outcome} & \textbf{Count} & \textbf{Percentage} & \textbf{Mean Episode Length} \\
        \midrule
        No Collision                  & 42 / 100 & 42.00\% & 6.8 \\
        Target Ship Collision         &  0 / 100 &  0.00\% & 0.0 \\
        \textbf{Own Ship Collision}   & \textbf{58 / 100} & \textbf{58.00\%} & 5.3 \\
        \bottomrule
    \end{tabular} 
\end{table}

\begin{table}[htbp] 
\centering 
\caption{\ac{sac} policy's failure scenario generation training score.}
\label{tab:eval_sac} 
    \begin{tabular}{lccc} 
        \toprule
        \textbf{Outcome} & \textbf{Count} & \textbf{Percentage} & \textbf{Mean Episode Length} \\
        \midrule
        No Collision                  & 46 / 100 & 46.00\% & 7.9 \\
        Target Ship Collision         &  0 / 100 &  0.00\% & 0.0 \\
        \textbf{Own Ship Collision}   & \textbf{54 / 100} & \textbf{54.00\%} & 6.9 \\
        \bottomrule
    \end{tabular} 
\end{table}

\subsubsection{\ac{rppo} vs \ac{ppo}}
\begin{figure}[htbp]
    \centering
    \begin{minipage}{0.48\textwidth}
        \centering
        \includegraphics[width=\textwidth]{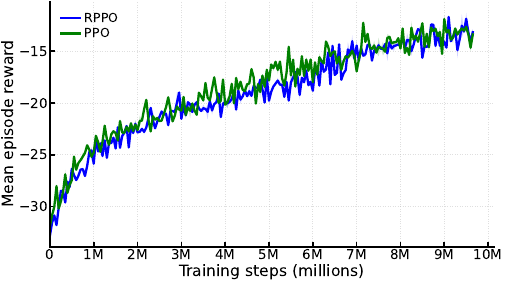}
        \caption{Reward comparison.}
        \label{fig:rppo_v_ppo_reward}
    \end{minipage}
    \hfill
    \begin{minipage}{0.48\textwidth}
        \centering
        \includegraphics[width=\textwidth]{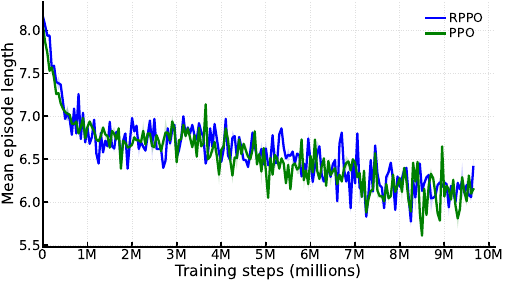}
        \caption{Episode length comparison.}
        \label{fig:rppo_v_ppo_eps_len}
    \end{minipage}
\end{figure}
Based on Figure \ref{fig:rppo_v_ppo_reward}, we see that both core yields a nearly-identical trends, where \ac{ppo} reward is only negligibly higher compared to the \ac{rppo}. Nevertheless, both grow from a low episodic reward of $-30$ to a value above $-15$. This signifies that both implementation are steadily improved overtime, owing to the same objective architecture of \ac{rppo} and \ac{ppo}. The same goes with how the episode length decrease overtime, showed in Figure \ref{fig:rppo_v_ppo_eps_len}. Both implementation manage to decrease the episode length to nearly 6 \ac{iw} samplings per episode before a termination is discovered. 

Looking more closely at the results in Table \ref{tab:eval_rppo} and Table \ref{tab:eval_ppo}, \ac{rppo} policy generates a larger number of failure scenarios, with 68 out of 100 cases resulting in own ship collision, compared with 58 out of 100 for the \ac{ppo} policy. Additionally, the average episode length of the \textit{no collision} and \textit{own ship collision} differ slightly between the two methods: \ac{rppo} policy tends to produce shorter \textit{no collision} episodes of $6.6$ and  longer \textit{own ship collision} episodes of $5.6$, whereas \ac{ppo} results exhibit the opposite trend with $6.8$ long \textit{no collision} episodes and $5.3$ long \textit{own ship collision} episodes.


\subsubsection{\ac{rppo} vs \ac{sac}}
\begin{figure}[htbp]
    \centering
    \begin{minipage}{0.48\textwidth}
        \centering
        \includegraphics[width=\textwidth]{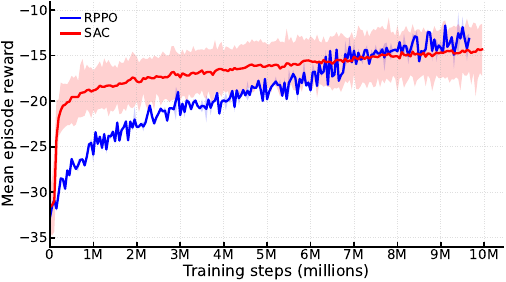}
        \caption{Reward comparison.}
        \label{fig:rppo_v_sac_reward}
    \end{minipage}
    \hfill
    \begin{minipage}{0.48\textwidth}
        \centering
        \includegraphics[width=\textwidth]{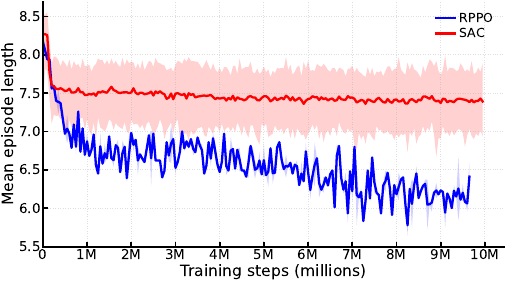}
        \caption{Episode length comparison.}
        \label{fig:rppo_v_sac_eps_len}
    \end{minipage}
\end{figure}

The results reveal a notable difference between the behavior of the \ac{rppo} and \ac{sac} methods. Due to its off-policy learning scheme and substantially higher update frequency, the \ac{sac} results contain a denser set of training observations with greater variance. For visual clarity, the \ac{sac} results are therefore presented as mean curves with shaded variance bands.

As shown in Figure \ref{fig:rppo_v_sac_reward}, the mean reward achieved by \ac{sac} increased rapidly for the first million training timesteps, until the increase slows down considerably for the remainder of the training phase. Eventually, the \ac{rppo}'s mean reward catches up around the $7\times10^6$ timesteps and stay higher compared to the \ac{sac}'s mean reward. However, the variability in the \ac{sac} results is considerably larger, with the upper range of the shaded bands frequently extending beyond the reward achieved by \ac{rppo}. This indicates that although \ac{sac} performs worse on average, some training outcomes are comparable to or higher than those of \ac{rppo}.

Figure \ref{fig:rppo_v_sac_eps_len} further shows that \ac{sac} produces longer episodes than \ac{rppo}, averaging approximately $7$--$8$ \ac{iw} samplings per episode. This suggests that \ac{sac} generally requires more trajectory modifications before reaching a terminal state compared with \ac{rppo} and \ac{ppo}. This observation is also reflected in the evaluation results presented in Table \ref{tab:eval_sac}. In particular, the \ac{sac} policy produces longer \textit{no collision} and \textit{own ship collision} episodes, with average lengths of $7.9$ and $6.9$, respectively. In comparison, the corresponding values are $6.6$ and $5.6$ for \ac{rppo}, and $6.8$ and $5.3$ for \ac{ppo}. Overall, these results indicate that \ac{sac} tends to require more \ac{iw} sampling steps to reach a terminal outcome than either \ac{rppo} or \ac{ppo}.

\subsection{Performance Variability Across Independent Training Runs}
We then investigate how the performance of each policy varies across independent training runs using identical hyperparameters and training settings. This experiment assesses whether the observed performance is consistent across runs, rather than being the result of particular training converging to a suboptimal local optimum. For each assessment, we compare the \textit{collision-inducing score} and \textit{training time} of each method. The score is defined as the percentage of 100 generated scenarios that result in a collision. All runs were performed on \ac{ntnu}'s IDUN High-Performance Computing systems \cite{sjalander_epic_2024}, using 64 CPU cores and 256 GB of RAM per run. We limit the experiment to five runs per policy due to the substantial computational costs, as each training run generally requires more than five days.

\begin{table}[htbp]
    \centering
    \caption{Training time and performance across independent runs.}
    \label{tab:training_time_steps}

    \setlength{\tabcolsep}{5pt}

    \begin{tabular}{crrrrrr}
        \toprule
        & \multicolumn{2}{c}{\ac{rppo}}
        & \multicolumn{2}{c}{\ac{ppo}}
        & \multicolumn{2}{c}{\ac{sac}} \\
        
        
        \cmidrule(lr){2-3}
        \cmidrule(lr){4-5}
        \cmidrule(lr){6-7}

        Trial
        & Score (\%) & Time (days)
        & Score (\%) & Time (days)
        & Score (\%) & Time (days) \\
        \midrule

        1 & 68.0 & 6.15
          & 58.0 & 6.30
          & 54.0 & 7.26 \\

        2 & 64.0 & 4.97
          & 68.0 & 6.10
          & 57.0 & 8.31 \\

        3 & 65.0 & 4.95
          & 66.0 & 6.08
          & 51.0 & 5.55 \\

        4 & 68.0 & 4.96
          & 63.0 & 6.77
          & 43.0 & 7.01 \\

        5 & 69.0 & 6.03
          & 71.0 & 4.83
          & 54.0 & 7.33 \\

        \midrule
        Mean
          & 66.80 & 5.41
          & 65.20 & 6.02
          & 51.80 & 7.09 \\

        \bottomrule
    \end{tabular}
\end{table}

Based on Table \ref{tab:training_time_steps}, we observe that \ac{rppo} policy achieves the highest average score of $66.80\%$, compared to \ac{ppo} and \ac{sac} scores of $65.20\%$ and $51.80\%$ respectively. \ac{rppo} policy also requires the shortest average training time of $5.41$ days, while \ac{ppo} and \ac{sac} policies requires averagely $6.02$ days and $7.09$ days. However, \ac{ppo} policy outperforms \ac{rppo} policy in several runs, specifically in Trial 2, 3, and 5, and achieves the highest single-run score of $71.0\%$ compared with \ac{rppo}'s highest score of $69.0\%$.

\begin{figure}[!h]
\begin{center}
    \includegraphics[width=\textwidth]{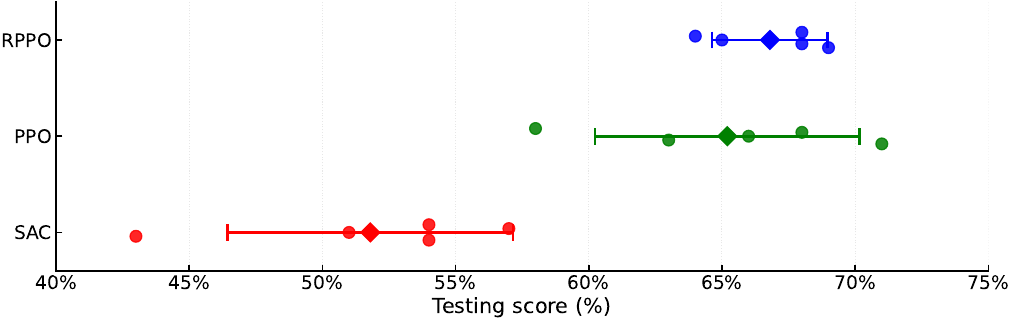}
    \caption{Collision-inducing performance comparison of all \ac{ebast2} policies with 5 repeated trainings for each method.}
    \label{fig:performance_variance}
\end{center}
\end{figure}

To examine this variability more closely, the results' statistics are summarized in Table \ref{tab:performance_variability} and visualized in Figure \ref{fig:performance_variance}. Although \ac{ppo} policy occasionally achieves higher score than \ac{rppo} policy, its scores are more dispersed across the independent runs. In contrasts, \ac{rppo} scores are more tightly clustered around the mean, indicating more consistent performance. On the other hand, \ac{sac} policy's score are generally lower and show and even wider spread than those of \ac{ppo}.

\begin{table}[!htbp]
    \centering
    \caption{Performance statistics across independent test runs.}
    \label{tab:performance_variability}
    \begin{tabular}{lrrrrrrrr}
        \toprule
        Method & $n$ & Mean & Std. & Median & Q25 & Q75 & Min & Max \\
        \midrule
        \ac{rppo} & 5 & 66.80 & 2.17 & 68.00 & 65.00 & 68.00 & 64.00 & 69.00 \\
        \ac{ppo}  & 5 & 65.20 & 4.97 & 66.00 & 63.00 & 68.00 & 58.00 & 71.00 \\
        \ac{sac}  & 5 & 51.80 & 5.36 & 54.00 & 51.00 & 54.00 & 43.00 & 57.00 \\
        \bottomrule
    \end{tabular}
\end{table}

\begin{figure}[!h]
\begin{center}
    \includegraphics[width=\textwidth]{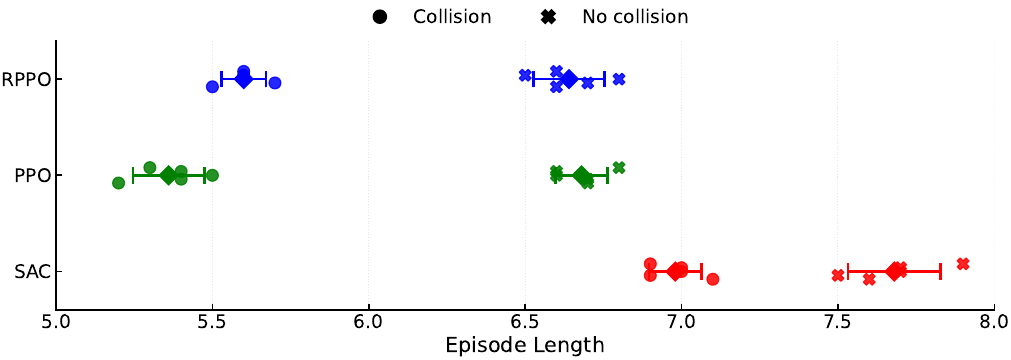}
    \caption{Collision-inducing episode lengths comparison of all \ac{ebast2} policies with 5 repeated trainings for each method.}
    \label{fig:eps_length_variance}
\end{center}
\end{figure}

We can also learn the scenario generation behavior of each policy by comparing the length of the generated collision and non-collision scenarios. Episode length is defined as the number of state-action transitions within an episode. The corresponding statistics are presented in Table \ref{tab:episode_length_variability} with the distribution visualized in Figure \ref{fig:eps_length_variance}. The \ac{rppo} policy tends to generate a longer collision scenarios, with episode length of $5.60$, but a shorter non-collision scenarios, with a mean of $6.64$. The opposite applies to \ac{ppo} policy, where it tends to generate a shorter collision scenarios, with a mean of $5.36$, and slightly longer non-collision scenarios, with a mean of $6.64$. The \ac{sac} policy generates the longest scenarios for both outcomes, with a mean episode lengths of $6.98$ and $7.68$ for collision and non-collision respectively. These patterns are consistent with those observed in the initial-run comparisons in Section \ref{sec:cores_comparison}, suggesting that the scenario-generation characteristics of the \ac{ebast2} policies remain similar across independent training runs.

\begin{table}[!htbp]
    \centering
    \caption{Scenario lengths statistics across independent test runs.}
    \label{tab:episode_length_variability}
    \begin{tabular}{lrlrrrrrrr}
        \toprule
        Method & $n$ & Outcome & Mean & Std. & Median & Q25 & Q75 & Min & Max \\
        \midrule

        \multirow{2}{*}{\ac{rppo}}
            & \multirow{2}{*}{5}
            & \textbf{Collision}    & \textbf{5.60} & \textbf{0.07} & \textbf{5.6} & \textbf{5.6} & \textbf{5.6} & \textbf{5.5} & \textbf{5.7} \\
            & 
            & No collision          & 6.64 & 0.11 & 6.6 & 6.6 & 6.7 & 6.5 & 6.8 \\
        \addlinespace

        \multirow{2}{*}{\ac{ppo}}
            & \multirow{2}{*}{5}
            & \textbf{Collision}    & \textbf{5.36} & \textbf{0.11} & \textbf{5.4} & \textbf{5.3} & \textbf{5.4} & \textbf{5.2} & \textbf{5.5} \\
            & 
            & No collision          & 6.68 & 0.08 & 6.7 & 6.6 & 6.7 & 6.6 & 6.8 \\
        \addlinespace

        \multirow{2}{*}{\ac{sac}}
            & \multirow{2}{*}{5}
            & \textbf{Collision}    & \textbf{6.98} & \textbf{0.08} & \textbf{7.0} & \textbf{6.9} & \textbf{7.0} & \textbf{6.9} & \textbf{7.1} \\
            & 
            & No collision          & 7.68 & 0.15 & 7.7 & 7.6 & 7.7 & 7.5 & 7.9 \\

        \bottomrule
    \end{tabular}
\end{table}

These differences in episode length also suggest distinct scenario-generation behavior across the policies. \ac{ppo} policy tend to be more conservative when generating collision-inducing scenarios, which reflects its slow-and-stable learning behavior influenced by the clipped surrogate objective. \ac{sac} policy tend to generate a more complex failure-inducing scenarios, where it tends to have more \acp{iw} that enables pronounced heading change and more flexible course. This behavior may be attributed to the higher variability during training, associated with \ac{sac}'s off-policy nature and exploration strategy. \ac{rppo} policy lies between \ac{ppo} and \ac{sac} policies in terms of its failure-inducing scenario generation tendency. This may be related to its recurrent architecture, which can better exploit state-action transition information while retaining the relatively stable learning characteristics of \ac{ppo}. This ultimately allows \ac{rppo} policy to handle better the limited observations from the co-simulation modules.

\subsection{Scenario Generation Comparisons}

\begin{figure}[htbp]
    \centering

    \begin{minipage}{0.48\textwidth}
        \centering
        \includegraphics[width=\textwidth]{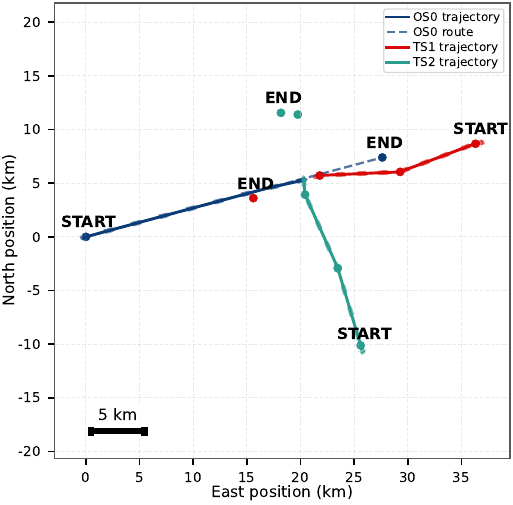}
        \caption{\ac{rppo} trajectory from Case 50 (TS2-collision).}
        \label{fig:50_rppo}
    \end{minipage}
    \hfill
    \begin{minipage}{0.48\textwidth}
        \centering
        \includegraphics[width=\textwidth]{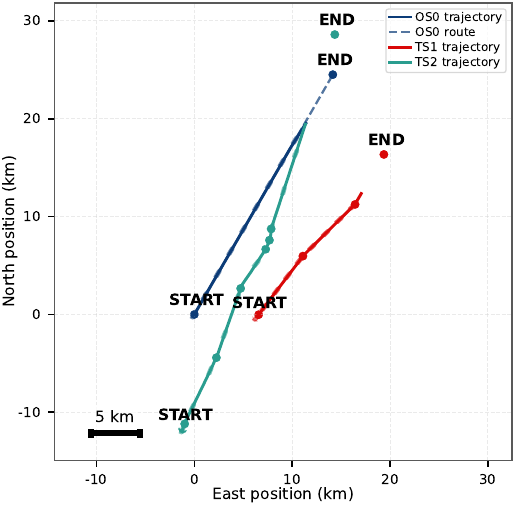}
        \caption{\ac{rppo} trajectory from Case 66 (TS2-collision).}
        \label{fig:96_rppo}
    \end{minipage}

    \vspace{0.5cm}

    \begin{minipage}{0.48\textwidth}
        \centering
        \includegraphics[width=\textwidth]{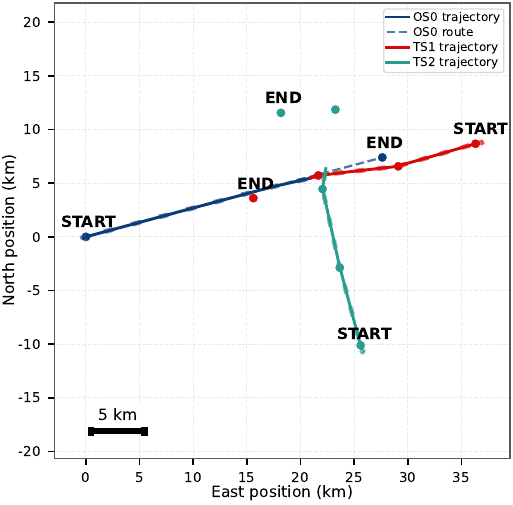}
        \caption{\ac{ppo} trajectory from Case 50 (TS1-collision).}
        \label{fig:50_ppo}
    \end{minipage}
    \hfill
    \begin{minipage}{0.48\textwidth}
        \centering
        \includegraphics[width=\textwidth]{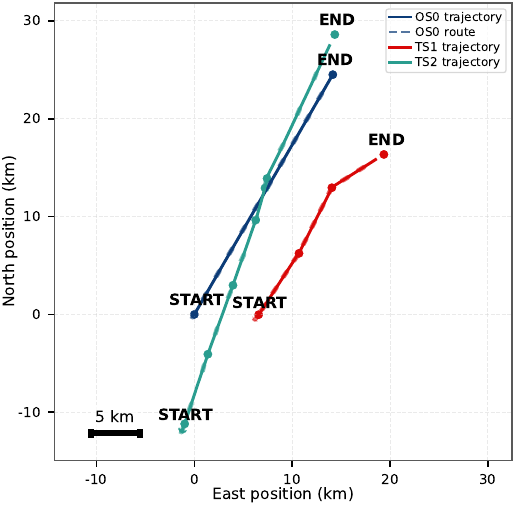}
        \caption{\ac{ppo} trajectory from Case 66 (No collision).}
        \label{fig:96_ppo}
    \end{minipage}

    \begin{minipage}{0.48\textwidth}
        \centering
        \includegraphics[width=\textwidth]{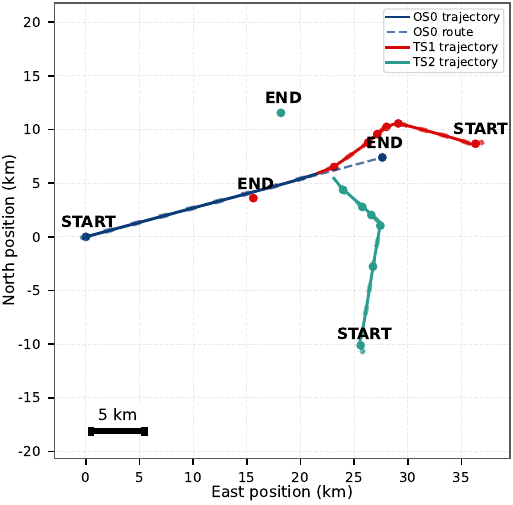}
        \caption{\ac{sac} trajectory from Case 50 (TS1-collision).}
        \label{fig:50_sac}
    \end{minipage}
    \hfill
    \begin{minipage}{0.48\textwidth}
        \centering
        \includegraphics[width=\textwidth]{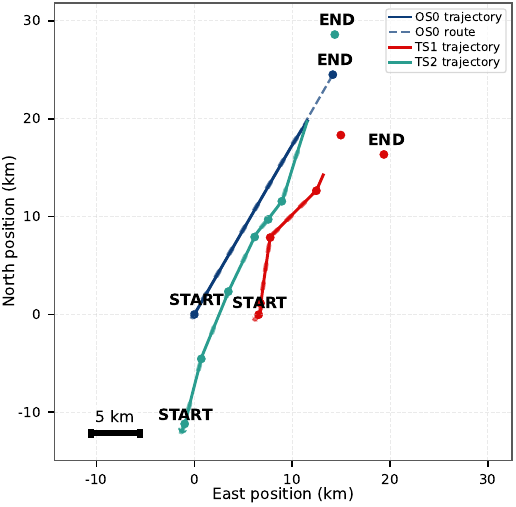}
        \caption{\ac{sac} trajectory from Case 66 (TS2-collision).}
        \label{fig:96_sac}
    \end{minipage}

    \vspace{0.5cm}

\end{figure}

\begin{figure}[htbp]
    \centering

    \begin{minipage}{0.48\textwidth}
        \centering
        \includegraphics[width=\textwidth]{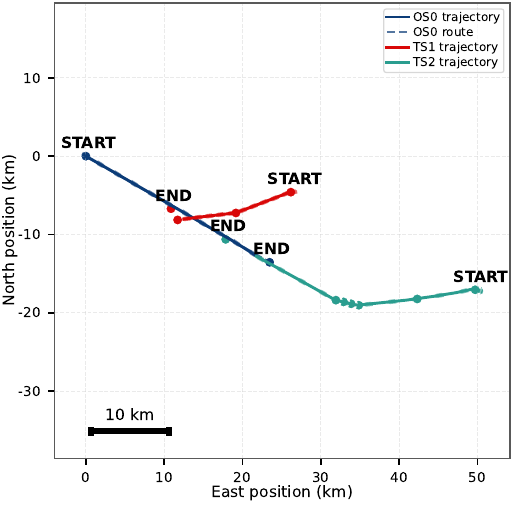}
        \caption{\ac{rppo} trajectory from Case 96 (TS2-collision).}
        \label{fig:77_rppo}
    \end{minipage}
    \hfill
    \begin{minipage}{0.48\textwidth}
        \centering
        \includegraphics[width=\textwidth]{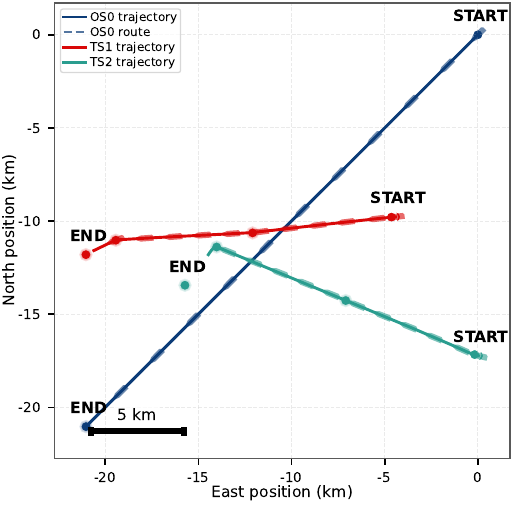}
        \caption{\ac{rppo} trajectory from Case 77 (No collision).}
        \label{fig:44_rppo}
    \end{minipage}

    \vspace{0.3cm}

    \begin{minipage}{0.48\textwidth}
        \centering
        \includegraphics[width=\textwidth]{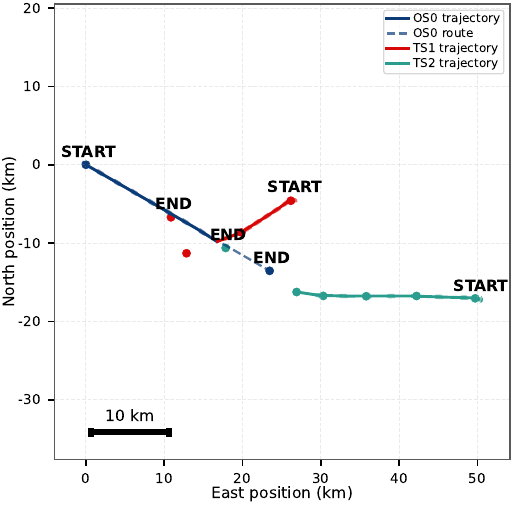}
        \caption{\ac{ppo} trajectory from Case 96 (TS1-collision).}
        \label{fig:77_ppo}
    \end{minipage}
    \hfill
    \begin{minipage}{0.48\textwidth}
        \centering
        \includegraphics[width=\textwidth]{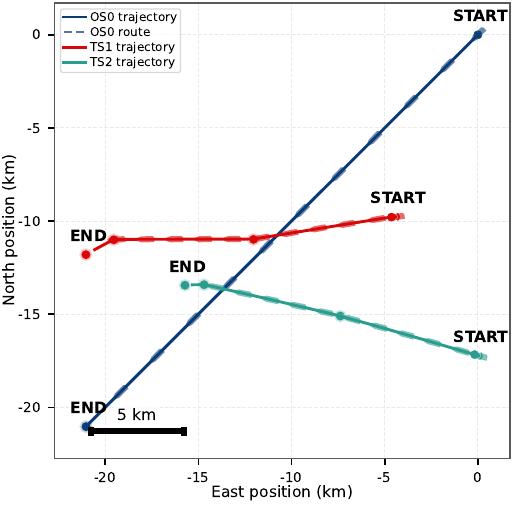}
        \caption{\ac{ppo} trajectory from Case 77 (No collision).}
        \label{fig:44_ppo}
    \end{minipage}

    \begin{minipage}{0.48\textwidth}
        \centering
        \includegraphics[width=\textwidth]{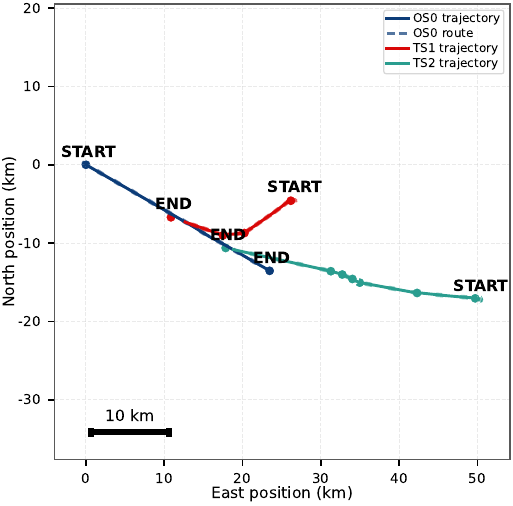}
        \caption{\ac{sac} trajectory from Case 96 (No collision).}
        \label{fig:77_sac}
    \end{minipage}
    \hfill
    \begin{minipage}{0.48\textwidth}
        \centering
        \includegraphics[width=\textwidth]{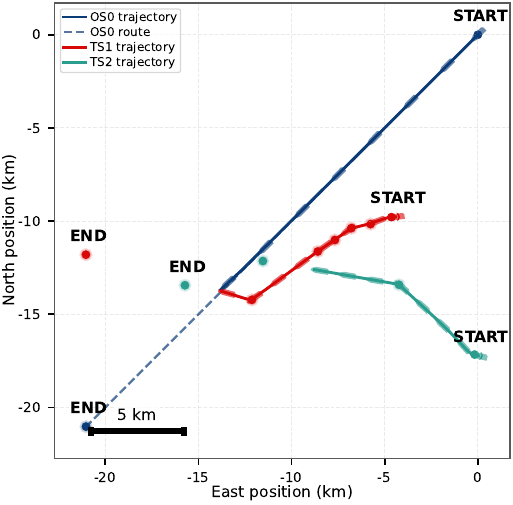}
        \caption{\ac{sac} trajectory from Case 77 (TS1-collision).}
        \label{fig:44_sac}
    \end{minipage}

    \vspace{0.5cm}

\end{figure}

\begin{figure}[htbp]
    \centering

    \begin{minipage}{0.48\textwidth}
        \centering
        \includegraphics[width=\textwidth]{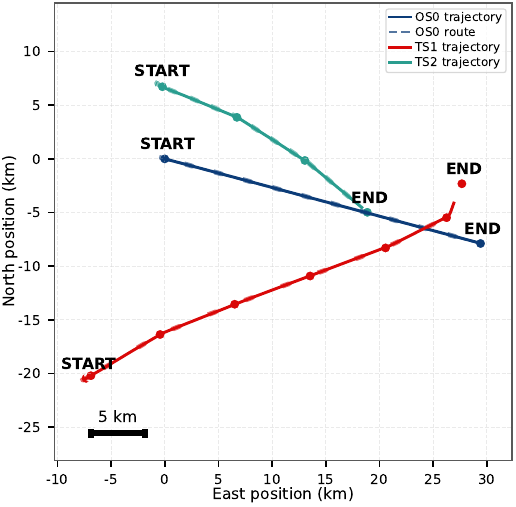}
        \caption{\ac{rppo} trajectory from Case 89 (No collision).}
        \label{fig:77_rppo}
    \end{minipage}
    \hfill
    \begin{minipage}{0.48\textwidth}
        \centering
        \includegraphics[width=\textwidth]{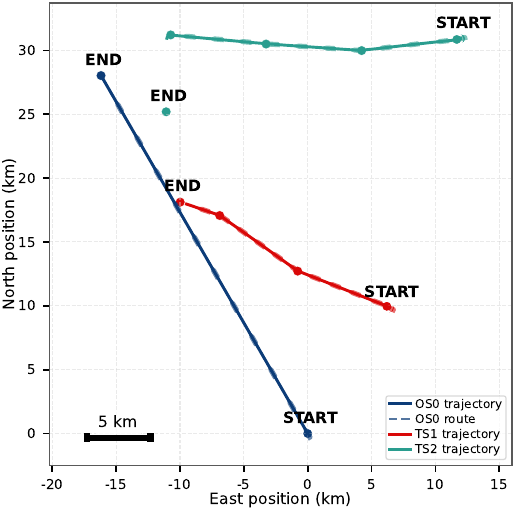}
        \caption{\ac{rppo} trajectory from Case 44 (No collision).}
        \label{fig:44_rppo}
    \end{minipage}

    \vspace{0.3cm}

    \begin{minipage}{0.48\textwidth}
        \centering
        \includegraphics[width=\textwidth]{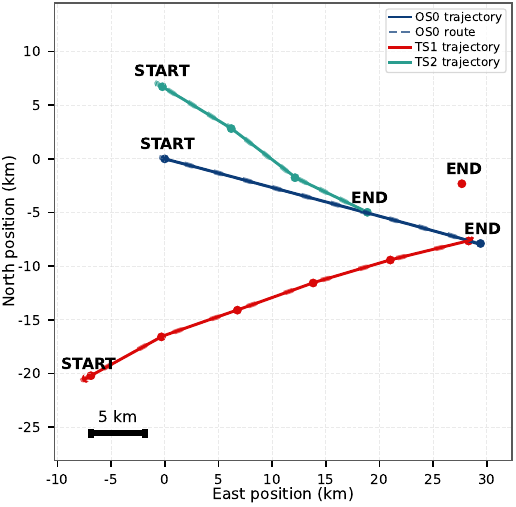}
        \caption{\ac{ppo} trajectory from Case 89 (No collision).}
        \label{fig:77_ppo}
    \end{minipage}
    \hfill
    \begin{minipage}{0.48\textwidth}
        \centering
        \includegraphics[width=\textwidth]{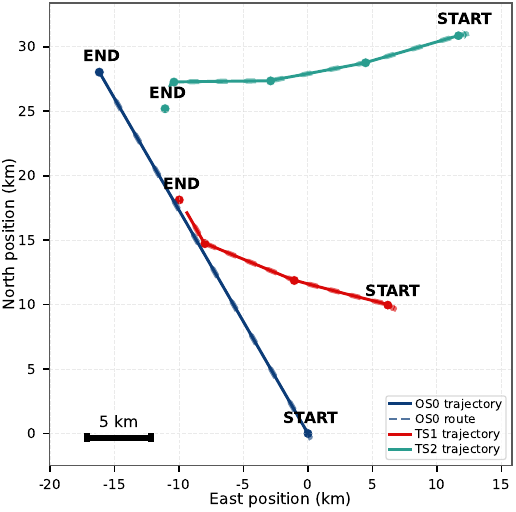}
        \caption{\ac{ppo} trajectory from Case 44 (No collision).}
        \label{fig:44_ppo}
    \end{minipage}

    \begin{minipage}{0.48\textwidth}
        \centering
        \includegraphics[width=\textwidth]{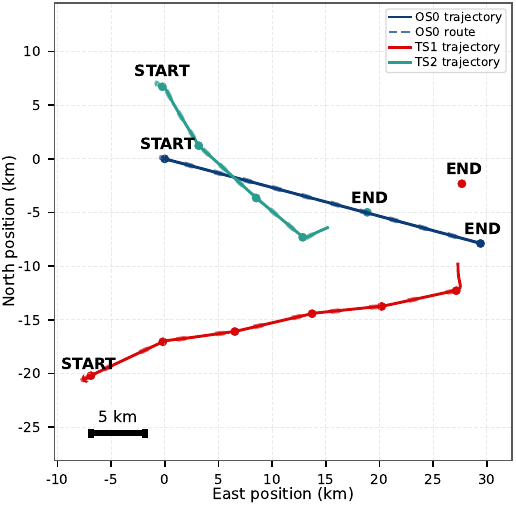}
        \caption{\ac{sac} trajectory from Case 89 (No collision).}
        \label{fig:77_sac}
    \end{minipage}
    \hfill
    \begin{minipage}{0.48\textwidth}
        \centering
        \includegraphics[width=\textwidth]{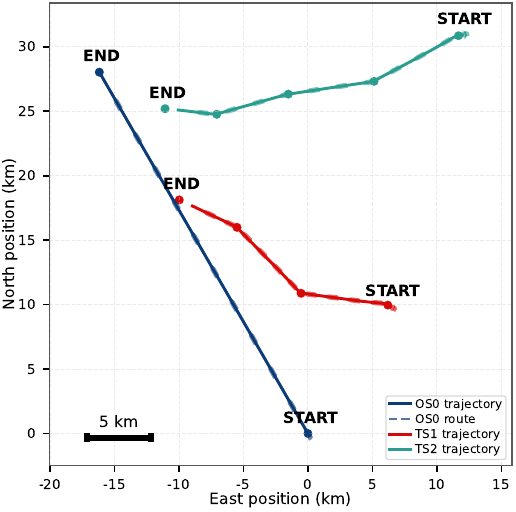}
        \caption{\ac{sac} trajectory from Case 44 (No collision).}
        \label{fig:44_sac}
    \end{minipage}

    \vspace{0.5cm}

\end{figure}

In this section, we compare different generated scenarios by \ac{rppo}, \ac{ppo}, and \ac{sac} policies from a selected number of traffic cases. More specifically, we compare the generated scenarios from Case $50$, $66$, $96$, $77$, $89$, and $44$.

In Case $50$, all three policies successfully generate an own ship collision scenario. For the \ac{rppo} policy, the collision is triggered by target ship 2, whereas for both \ac{ppo} and \ac{sac}, the collision involves target ship 1. In all three scenarios, the target ships maneuver toward the own ship and eventually overwhelm its \ac{colav} system, resulting in a collision. However, the trajectories generated by \ac{sac} contain more \acp{iw} and exhibit more pronounced heading changes than those generated by \ac{rppo} and \ac{ppo}. This indicates that \ac{sac} tends to produce more complex and highly maneuvering trajectories when searching for collision-inducing scenarios.

In some cases, the \ac{ppo} policy is unable to generate a true own ship collision scenario. This is demonstrated in Case $66$, where both \ac{rppo} and \ac{sac} successfully identify collision-inducing scenarios, whereas \ac{ppo} does not. The scenarios generated by \ac{rppo} and \ac{sac} share a notable similarity: both contain additional turns and maneuvers before eventually leading to an own ship collision. On the contrary, the trajectory generated by \ac{ppo} remains comparatively straight, suggesting that its more conservative maneuvering limits its ability to discover a collision scenario in this case.

Conversely, there are cases in which the \ac{sac} policy fails to generate an own ship collision scenario. In Case $96$, \ac{rppo} generates a scenario in which the own ship collides with target ship 2, while \ac{ppo} generates a collision involving target ship 1. The \ac{rppo} trajectories, particularly that of target ship 2, contain more \acp{iw} and therefore exhibit more frequent and pronounced heading changes although still maintaining a rather simple course. The \ac{ppo} trajectories are considerably simpler and succeed in triggering a collision relatively early by maneuvering target ship 1 toward the own ship. The trajectories generated by \ac{sac}, on the other hand, are comparatively complex but ultimately fail to produce an own ship collision.

Nevertheless, there are specific cases in which \ac{sac} performs better than its \ac{rppo} and \ac{ppo} counterparts. In Case $77$, both \ac{rppo} and \ac{ppo} struggle to identify collision-inducing scenarios because the own ship travels significantly faster than both target ship 1 and target ship 2. Consequently, a substantially larger interception angle is required for either target ship to reach the own ship. The trajectories generated by \ac{rppo} and \ac{ppo} remain relatively conservative, with both policies maintaining comparatively straight courses. In contrast, \ac{sac} overcomes this limitation by selecting a trajectory with a much larger interception angle, followed by a sharp turn that ultimately results in an own ship collision. This case illustrates how the greater maneuvering flexibility exhibited by \ac{sac} can be advantageous when more aggressive trajectory changes are required to induce a collision.

Furthermore, none of the policies is guaranteed to identify a failure scenario for every initial condition. This is illustrated by Cases $89$ and $44$, in which none of the three policies succeeds in generating a collision-inducing trajectory. In Case $89$, target ship 1 is initially located sufficiently far from the own ship that interception is infeasible given its speed, while target ship 2 travels too slowly to intercept the own ship. Similarly, in Case $44$, interception is infeasible because the feasible trajectories of the two target ships do not intersect the own ship's trajectory in a way that allows a collision to occur. Nevertheless, even in these unsuccessful cases, the generated trajectories still exhibit behavior consistent with an attempt to induce a collision: the policies maneuver the target ships toward the own ship and thereby increase the collision risk, despite ultimately being unable to produce an actual collision.

\subsection{Continued \ac{ebast2} Training for Improving the Policy}
\begin{figure}[htbp]
    \centering
    \begin{minipage}{0.48\textwidth}
        \centering
        \includegraphics[width=\textwidth]{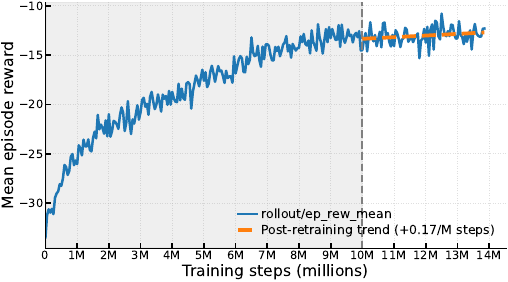}
        \caption{Reward comparison.}
        \label{fig:rppo_continue_reward}
    \end{minipage}
    \hfill
    \begin{minipage}{0.48\textwidth}
        \centering
        \includegraphics[width=\textwidth]{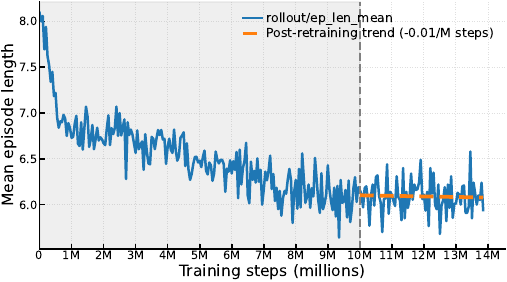}
        \caption{Episode length comparison.}
        \label{fig:rppo_continue_eps_len}
    \end{minipage}
\end{figure}
\begin{table}[!htbp] 
\centering 
\caption{Second-trained \ac{rppo} policy's training score.} 
\label{tab:eval_rppo_continue} 
    \begin{tabular}{lccc} 
        \toprule 
        \textbf{Outcome} & \textbf{Count} & \textbf{Percentage} & \textbf{Mean Episode Length} \\ 
        \midrule 
        No Collision & 30 / 100 & 30.00\% & 6.5 \\ 
        Target Ship Collision & 0 / 100 & 0.00\% & 0.0 \\ 
        \textbf{Own Ship Collision} & \textbf{70 / 100} & \textbf{70.00\%} & 5.6 \\ 
        \bottomrule 
    \end{tabular} 
\end{table}

We then evaluate whether the \ac{rppo} policy can be further improved through continued training. The previously trained policy is trained for an additional $4$ million timesteps, and its performance during this second training phase is evaluated. As shown in Table \ref{tab:eval_rppo_continue}, the retrained \ac{rppo} policy achieves a $2\%$ improvement in generating own-ship collision scenarios. This is consistent with the mean reward trend in Figure \ref{fig:rppo_continue_reward}. The gray-shaded region represents the previous training phase, while the orange dashed line shows an ordinary least-squares linear fit to the continued training phase. The mean reward increase are quite modest, whereas the mean episode length remains relatively stable. Although small, this improvement demonstrates the potential for further refinement of an already trained \ac{ebast2} policy.

\subsection{Continued \ac{ebast2} Training with Different \ac{colav} System}
\begin{table}[!htbp] 
\centering 
\caption{First-trained \ac{rppo} policy's training score using different \ac{colav} system.} 
\label{tab:eval_rppo_multi_configs_before} 
    \begin{tabular}{lccc} 
        \toprule 
        \textbf{Outcome} & \textbf{Count} & \textbf{Percentage} & \textbf{Mean Episode Length} \\ 
        \midrule 
        No Collision & 81 / 100 & 81.00\% & 6.6 \\ 
        Target Ship Collision & 0 / 100 & 0.00\% & 0.0 \\ 
        \textbf{Own Ship Collision} & \textbf{19 / 100} & \textbf{19.00\%} & 6.4 \\ 
        \bottomrule 
    \end{tabular} 
\end{table}

\begin{table}[!htbp] 
\centering 
\caption{Second-trained \ac{rppo} policy's training score using different \ac{colav} system.} 
\label{tab:eval_rppo_multi_configs_after} 
    \begin{tabular}{lccc} 
        \toprule 
        \textbf{Outcome} & \textbf{Count} & \textbf{Percentage} & \textbf{Mean Episode Length} \\ 
        \midrule 
        No Collision & 45 / 100 & 45.00\% & 6.8 \\ 
        Target Ship Collision & 0 / 100 & 0.00\% & 0.0 \\ 
        \textbf{Own Ship Collision} & \textbf{55 / 100} & \textbf{55.00\%} & 5.5 \\ 
        \bottomrule 
    \end{tabular} 
\end{table}

\begin{figure}[!htbp]
    \centering
    \begin{minipage}{0.48\textwidth}
        \centering
        \includegraphics[width=\textwidth]{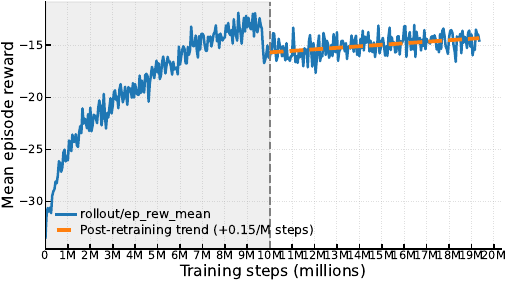}
        \caption{Reward comparison.}
        \label{fig:rppo_continue_multi_configs_reward}
    \end{minipage}
    \hfill
    \begin{minipage}{0.48\textwidth}
        \centering
        \includegraphics[width=\textwidth]{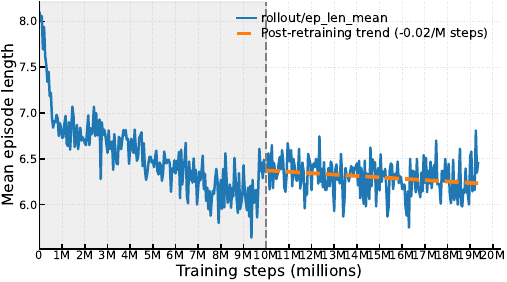}
        \caption{Episode length comparison.}
        \label{fig:rppo_continue_multi_configs_eps_len}
    \end{minipage}
\end{figure}

We also evaluate whether the trained \ac{rppo} policy can generalize to a different \ac{colav} system during stress testing. For simplicity, the alternative system is based on the implementation in Section \ref{sec:simple_colav}, but its response is changed from \textit{turning to starboard} to \textit{turning to port}. When the trained \ac{rppo} policy is applied to this modified system, its success rate in generating own ship collision scenarios drops significantly from  $68\%$ to $19\%$, as shown in Table \ref{tab:eval_rppo_multi_configs_before}. This is consistent with Figure \ref{fig:rppo_continue_multi_configs_reward}, where the mean reward drops sharply after $10$ million timesteps. At the same time, the mean episode length increases, indicating that the policy struggles to find effective collision-inducing scenarios.

We therefore retrain the \ac{rppo} policy for another $10$ million timesteps. During this training phase, each episode has a $50\%$ chance of using either \textit{turning to starboard} or \textit{turning to port} \ac{colav} system, exposing the policy to both behaviors during training. After retraining, the policy achieves a $55\%$ collision generation rate for either configuration. This improvement is also reflected in Figure \ref{fig:rppo_continue_multi_configs_reward}, where the mean reward increases throughout the second training phase, while the mean episode length decreases in Figure \ref{fig:rppo_continue_multi_configs_eps_len}. These results indicate that the retrained policy generalizes better across different \ac{colav} behaviors while remaining effective at generating collision-inducing scenarios consistent with the \ac{ast} objective.

%% file: Sections/Section_7.tex
\section{Conclusion}

Through our experiment, the proposed \ac{ebast2} framework demonstrates its capability as both verification and safety validation methods for \ac{cps} by fulfilling these three goals:
\begin{itemize}
    \item Providing counter-examples across a wide range of ship traffic scenarios by adopting \ac{ast} as the core algorithm.
    \item Supports realistic scenario discovery by generating advancing adversarial trajectories and maintaining a balance between trajectory smoothness, interception intent through \ac{iws}.
    \item Demonstrates an increasing ability of the trained policies to generate collision-inducing scenarios, quantified by the failure rate across training cases.
\end{itemize}

We also compared different \ac{rl}-algorithm cores, namely \ac{rppo}, \ac{ppo} and \ac{sac}, to demonstrate how the \ac{pomdp} formulation might be advantageous, compared to its \ac{mdp} alternatives, as the core of the proposed \ac{ebast2} method to stress test black-box simulation like Co-simulation. The proposed \ac{rppo} policy adoption proves a higher average failure-discovery performance of $66.80\%$ compared to \ac{ppo} policy ($65.20\%$) and \ac{sac} policy ($51.80\%$). Despite this, different \ac{rl} policy cores exhibit distinct scenario-generation characteristics. Overall, the three \ac{rl} cores exhibit distinct scenario-generation characteristics. \ac{ppo} shows more conservative behavior, \ac{sac} generates more variable and complex failure scenarios, while \ac{rppo} provides a balance between these characteristics. The recurrent structure of \ac{rppo} appears particularly beneficial in handling the limited observations provided by the co-simulation modules.

Additionally, we demonstrate that a trained policy can be further improved through continued training. The retrained \ac{rppo} policy achieves a higher own-ship collision generation rate, suggesting the potential for an iterative safety-assessment cycle: after a system is stress tested and improved, the \ac{ebast2} policy can be retrained to evaluate the updated system. We also show that the policy can adapt to different system behaviors by exposing it to both the original and modified systems during retraining phase. This is achieved by uniformly sampling the system configuration at each episode \texttt{reset()}. In our experiment, varying the \ac{colav} response enables the retrained policy to generalize across both configurations while still generating meaningful failure scenarios. Together, these results demonstrate the potential of the \ac{ebast2} framework as a safety-assessment method for black-box maritime \ac{cps}.
\vspace{-1em}
\begin{ack}
\vspace{-1em}
    This work was supported by the Research Council of Norway under the project System optimisation and stress testing in co-simulations (OptiStress) (grant no.\ 344238). The authors would also like to thank Børge Rokseth and Stian Skjong for their valuable support in facilitating this research collaboration between \ac{ntnu}, SINTEF, and \ac{dnv}
\end{ack}
